\documentclass[twocolumn,floatfix,prb,showpacs,a4paper]{revtex4}

\usepackage{graphicx}
\usepackage{color}
\usepackage{amsmath}
\usepackage{bm}

\begin{document}

\title{Universal spatial correlations in the anisotropic Kondo screening cloud:
analytical insights and numerically exact results from a coherent state expansion}

\author{Serge Florens}
\affiliation{Institut N\'{e}el, CNRS and Universit\'e Grenoble Alpes, F-38042 Grenoble, France}
\author{Izak Snyman}
\affiliation{Mandelstam Institute for Theoretical Physics, School of Physics, University 
of the Witwatersrand, Wits, 2050, South Africa}

\date{July 2015}
\begin{abstract} 
We analyze the spatial correlation structure of the spin density of an
electron gas in the vicinity of an antiferromagnetically-coupled Kondo impurity. Our analysis extends to the regime of
spin-anisotropic couplings, where there are no quantitative results for spatial correlations in the literature.
We use an original and numerically exact method, based on a systematic
coherent-state expansion of the ground state of the underlying spin-boson Hamiltonian. 
It has not yet been applied to the computation of observables that are
specific to the fermionic Kondo model.
We also present an important technical improvement to the method, that 
obviates the need to discretize modes of the Fermi sea, and allows one to tackle the
problem in the thermodynamic limit. As a result, one can obtain
excellent spatial resolution over arbitrary length scales, for a relatively low 
computational cost, a feature that gives the method an advantage over popular techniques 
such as the Numerical and Density-Matrix Renormalization Groups.
We find that the anisotropic Kondo model shows rich universal scaling
behavior in the spatial structure of the entanglement cloud. First, SU(2)
spin-symmetry is dynamically restored in a finite domain in parameter space in vicinity of the 
isotropic line, as expected from poor man's scaling. 
More surprisingly, we are able to obtain in closed analytical form a set of different, yet 
universal, scaling curves for strong exchange asymmetry, which are parametrized by the 
longitudinal exchange coupling.
Deep inside the cloud, i.e. for distances smaller than the Kondo length, the 
correlation between the electron spin density and the impurity spin oscillates 
between ferromagnetic and antiferromagnetic values at the scale of the Fermi
wavelength, an effect that is drastically enhanced at strongly anisotropic couplings. 
Our results also provide further numerical checks and alternative analytical
approximations for the Kondo overlaps that were recently computed by
 Lukyanov, Saleur, Jacobsen, and Vasseur [Phys. Rev. 
Lett. {\bf 114}, 080601 (2015)] .
\end{abstract}
\pacs{73.40.Gk, 72.10.Fk}
\maketitle
\section{Introduction}
\label{secintro}
The spin $1/2$ Kondo model describes a localized magnetic moment interacting
with an electron gas, via an antiferromagnetic exchange coupling.\cite{Hewson}
Despite a long history, and even an exact solution, it has not yet surrendered
all its secrets. It is well established that the ground state is a spin singlet 
in which the impurity spin is quenched by the electron gas, and the
spatial region where the electron gas is correlated with the impurity
is referred to as the Kondo screening cloud.\cite{Affleck1} Even for an
isotropic system, its precise spatial profile is not known analytically, except
asymptotically,\cite{Affleck2} and it is only in the past few years that it has been
calculated numerically.\cite{Gubernatis,Borda,Lechtenberg} Despite wide-ranging
proposals,\cite{Affleck3,Cornaglia,Hand,Pereira,Park,Nishida,Bauer} it has
eluded direct measurement, partly because of the difficulty in measuring spin
correlations. 

In the isotropic case, the screening cloud is characterized by the ground state correlation function
$X(x)=4\left<\vec{S}^{\rm imp}\cdot\vec{\mathcal S}^{\rm el}(x)\right>$ where $\vec{S}^{\rm imp}$
is the impurity spin operator, and $\vec{\mathcal S}^{\rm el}(x)$ is the electron spin density at
$x$. When the Kondo temperature is much lower than the Fermi energy, 
the screening cloud can be decomposed into a forward scattering
contribution $X_0(x)$ and a backscattering contribution $X_{2k_F}(x)$, so 
that $X(x)= X_0(x)+\cos(2k_F x)X_{2k_F}(x)$, where the two functions $X_0(x)$ and 
$X_{2k_F}(x)$ vary slowly on the scale of the Fermi wavelength $2\pi/k_F$.
In the scaling regime and for spin-isotropic Kondo exchange, the profile of the 
screening cloud displays a universal line shape,\cite{Affleck4,Mitchell} which is dependent on 
the value of Kondo coupling $J$ only through an emergent 
length $\xi$ that is inversely proportional to the Kondo 
temperature.\cite{Holzner,Busser} To be specific, if $X_k(x)$ and $X_k'(x)$, with $k\in\{0,2k_F\}$, 
are correlation functions corresponding to different values $J$ and
$J'$ of the Kondo coupling, and $\xi$ and $\xi'$ are the corresponding Kondo
lengths, then
\begin{equation}
X_k'(x)=\frac{\xi}{\xi'}X_k(\xi' x/\xi),
\end{equation} 
for all $x\gg 2\pi/k_F$.

Recently, we have realized that it may be possible to measure the longitudinal
forward scattering ($k=0$) component of the screening cloud (a precise
definition is given below) in a chain of tunnel-coupled superconducting
islands.\cite{Snyman} It turns out that the Hamiltonian describing the charge
and phase degrees of freedom in this system is equivalent to the {\em
spin-anisotropic} Kondo model in one dimension,\cite{Goldstein} described
by two coupling constants: The $z$-components of the impurity and electron spins 
couple with a strength $J_\parallel$, while the components perpendicular to 
the $z$-axis couple with a different strength $J_\perp$. 
A detailed study of the screening cloud in the anisotropic Kondo model
has to our knowledge not been performed before, and is 
therefore timely.

In the anisotropic case, there are four correlation functions of 
interest, namely $X^\parallel_k(x)$, that measures the correlation between 
$z$-components of the impurity and electron spins (both in the forward $k=0k_F$ and 
backward $k=2k_F$ scattering channels), and $X^\perp_k(x)$, that measures correlations 
between components perpendicular to the $z$ axis (in each channel). Clearly, an 
anisotropic Kondo interaction, characterized by two coupling constants
$J_\parallel$ and $J_\perp$, will affect the universal scaling picture 
non-trivially. 

The universal scaling of the isotropic model has been confirmed
using the Numerical Renormalization Group (NRG), a popular method to study the 
Kondo model.\cite{Borda,Lechtenberg} However, there is still room to improve the
accuracy of existing results.
For instance, it is predicted analytically~\cite{Affleck1} that backscattering dominates 
forward scattering inside the screening cloud. This produces
oscillations from ferromagnetic to antiferromagnetic correlations between the impurity and the electron gas, on the
scale of half the Fermi wavelength.
The first numerical calculation of the universal scaling
functions were reported in Ref.~\onlinecite{Borda}.  However, 
the predominance of the backscattering
component over the forward scattering component inside the cloud was not resolved. This was
probably due to the fact that results were obtained at relatively large Kondo
temperatures, so that there was not a sufficient separation of scales between 
the size $\xi$ of the cloud, and the short distance ultraviolet cut-off scale. 
In a more recent work,\cite{Lechtenberg} the NRG method of
Ref.~\onlinecite{Borda} was refined, and the alternation of ferro- and antiferromagnetic 
correlations inside the cloud clearly be seen. However, results were only presented for
distances up to $10$ Fermi wavelengths from the impurity. At these scales, the
correlation functions show non-universal modulations.

In this Article, we perform numerical calculations that are sufficiently accurate
to investigate the universal scaling behavior in regimes
ranging from isotropic to strongly anisotropic Kondo couplings. 
At the same time, we can clearly resolve the predominace of the backscattering 
component over the forward scattering component inside the cloud.
In precise terms, we investigate the the following questions: Let $X^j_k(x)$ and $(X^j_k)'(x)$, with 
$j\in\{\parallel,\perp\}$ and $k\in\{0,2k_F\}$, be correlation functions at 
distinct values $(J_\perp,J_\parallel)$ and $(J_\perp',J_\parallel' )$. Under
which conditions are there constants $\lambda_1$ and $\lambda_2$ such that
$(X_k^j)'(x)=\lambda_1 X_k^j(\lambda_2 x)$, for all $x\gg 2\pi/k_F$, and what
are the line shapes of these universal scaling curves?
Our general numerical findings show that scaling is typically obeyed for
fixed values of $J_\parallel$.
In other words, correlators at the same $J_\parallel$ but different $J_\perp$ can 
be scaled onto each other. This statement holds as long as $J_\perp\lesssim 1$ 
(in dimensionless units of the inverse density of states), while for 
transverse couplings of order one, the Kondo temperature becomes comparable to 
the Fermi energy, and universality is lost. 
For $J_\parallel \gg J_\perp$, we find that the line shapes of the cloud correlation 
functions acquire a $J_\parallel$-dependence that cannot be scaled away. However, as $J_\parallel$ 
approaches $J_\perp$, the $J_\parallel$ dependence rapidly becomes very weak. 
One thus find a sizable region of parameter space, located around the
strict spin-isotropic line $J_\parallel=J_\perp \lesssim 1$, in which the universal 
curves have little discernible $J_\parallel$ dependence. In this region, correlators 
calculated at $J_\perp\not=J_\perp'$ {\em and} $J_\parallel\not=J_\parallel'$ can 
to a very good approximation be scaled onto the unique universal curve of the 
isotropic model. These conclusions are consistent with poor man's scaling 
arguments~\cite{Chakravarty} at small $J_\perp$, as we will explain in detail.

Our work also elaborates on a different but related topic, that was raised 
very recently by Lukyanov {\it et al.} in Ref. \onlinecite{luk}, where the overlap between two
Kondo wave functions with different Kondo couplings was computed analytically using
integrability techniques, and numerically with the Density Matrix Renormalization Group. 
We complement these interesting results, by providing a simpler 
(approximate) analytical expression for the Kondo overlaps in the spin-anisotropic 
limit. We also show that our numerical method reproduces the full analytical result, as it should.

Regarding methodology, our approach combines both analytics and numerically 
exact calculations. In the case of strong spin-anisotropy, $J_\parallel\gg
J_\perp$, we provide asymptotically exact formulas, that were not discussed in previous literature,
for the fermionic Kondo cloud. 
For a general choice of parameters, we employ a recent approach using a coherent state
expansion of the wave function, \cite{Bera1,Bera2} that has not yet been applied to 
the computation of purely fermionic observables that pertain to the Kondo model.
This method relies on a variational Ansatz, formulated in the
language of the spin-boson model, which describes a two-level system coupled to
a multi-mode ohmic bosonic bath.\cite{Leggett,Weiss} 
This spin-boson model is known to be equivalent to the anisotropic Kondo 
model,\cite{Guinea} with bosonization providing an exact mapping between the 
two.\cite{Kotliar,Costi} We exploit this mapping to apply the
coherent-state expansion to the Kondo model.
In addition, the flexible structure of the Ansatz allows one to add progressively 
more contributions to it, so that the method converges rapidly to the exact ground 
state with arbitrary accuracy.

Historically, the systematic coherent state expansion proposed by Bera {\it et
al.}~\cite{Bera1,Bera2} builds on seminal works of Emery and 
Luther,\cite{EmeryLuther} and of Silbey and Harris,\cite{Silbey,Harris} 
in the language of the spin-boson model, and on independent works by 
Anderson,\cite{Anderson1,Anderson2} and Bergmann and Zhang\cite{Bergmann} in the 
language of the Kondo model. The approach of Emery and Luther, Silbey and Harris, and 
Anderson corresponds to the lowest order approximation, involving a single
coherent state, and provides {\em quantitatively} accurate results only in the limit 
$J_\parallel\gg J_\perp$. Bergmann and Zhang took a step in the direction of a
two coherent-state Ansatz by adding an extra term to the lowest order approximation,
but constrained its form sub-optimally compared to the full solution with two
coherent states.
We stress that this type of approach differs from the well-known method pioneered by
Yosida,\cite{Yosida1} that takes as starting point a state in which a single
electron binds into a singlet with the impurity, while the rest of the Fermi sea
is unaffected. The wave function of the bound electron is chosen to maximize the binding energy. 
The relationship between the Yosida approximation and 
the single-coherent state Ansatz is reminiscent of the relationship in superconductivity
between the single Cooper pair on top of a Fermi sea on the one hand, and the
full BCS wave function on the other. Yosida's approach gives qualitatively
correct results, but even when the effect of additional particle-hole
excitations are included,\cite{Okiji,Yosida2} it cannot yield arbitrarily accurate
results for the correlations inside the Kondo screening cloud.

In this Article, we also contribute an important innovation to the coherent-state
expansion methodology. In previous implementations, the bosonic bath was limited to a finite
number of modes, because the number of variational parameters in the Ansatz scaled
linearly with the number of bath modes. Here, we first partially solve the
variational problem analytically, so that the number of remaining variational parameters
that have to be determined numerically is independent of the number of bath
oscillators. This allows us to take the thermodynamic limit before we optimize
the energy numerically, and probe arbitrarily large distances. 
In the Kondo language, this means that we work directly with an infinite conductor. 

With NRG, probing large length scales comes at the cost of losing 
information about shorter lengths scales. In order to maintain good spatial
resolution, one has to introduce a fictitious impurity at the position where 
correlators are evaluated. Each position considered then requires the solution of 
a given two-channel Kondo impurity problem, so that a high-resolution
computation of the Kondo cloud with NRG is quite expensive numerically.
Our method remarkably deals with all length scales on an equal footing, so that 
the Kondo cloud can be calculated in a single step, once the many-body ground state 
is known. The coherent state expansion thus nicely complements the existing tools 
for studying the Kondo model.

The rest of the Article is structured as follows. In Sec.~\ref{secham} we define
the anisotropic Kondo Hamiltonian. We also state the equivalent spin-boson
model, and the relation between the parameters of the two models. The precise mapping
between them is reviewed in Appendix~\ref{appmap}. In Sec.~\ref{sco} we define
the spatial correlation functions, that contain information 
on the entanglement cloud around the spin impurity, and that we will study. 
We first consider the
correlations in the fermionic language of the Kondo model and then in the bosonic 
language of the spin-boson model. 
The bosonic observables are derived from the fermionic ones in Appendix~\ref{appcloud}. 
Sec.~\ref{secmpa} introduces the systematic coherent state expansion. 
The innovation we mentioned in the previous paragraph, that reduces the number of
variational parameters to solve for numerically, is presented in
Sec.~\ref{secreduce}. In Sec.~\ref{secen} we take the thermodynamic limit for
the energy, and express it in terms of the remaining variational parameters. The
correlation functions that we study are expressed in terms of the remaining
variational parameters in Sec.~\ref{cloudcalc}. Next, we find the optimal values
for the variational parameters by minimizing the energy, and we present our results
in Sec.~\ref{secresults}. We first confirm the accuracy of the method and then
study in great detail the various correlation functions describing the Kondo cloud, as well as
the Kondo wavefunction overlaps. Finally, in Sec.~\ref{secdisc} we summarize our
main findings and, where applicable, compare them to results in the literature.

\section{Kondo and spin-boson Hamiltonians}

\label{secham}
In this section we write down the two related models that we study, and make explicit
the deep connection between them.
The first Hamiltonian is the anisotropic spin-$1/2$ Kondo model in one dimension,
that describes a spin-1/2 impurity coupled to a Fermi sea.
The Fermi sea does not necessarily represent electrons confined to one dimension. If the 
electron gas is higher dimensional, but the impurity is a point-like scatterer, it only 
interacts with electronic $s$-waves. Such $s$-waves in the electron gas can then be described 
by a one-dimensional Kondo Hamiltonian, where the spatial coordinate is the radial distance 
from the impurity.
The second Hamiltonian is the ohmic spin-boson model, that appears in many area of
physics, from superconducting nano-circuits to biological systems.
Below an ultraviolet energy scale set by the Fermi energy, the two models are 
equivalent.\cite{Guinea,Kotliar,Costi} The full mapping is reviewed in 
Appendix \ref{appmap}.

The anisotropic spin-$1/2$ Kondo model in one dimension reads 
$H=H_0+H_\parallel+H_\perp$, where:
\begin{eqnarray}
H_0&=&\sum_{k\sigma} (\varepsilon_k-\mu)\tilde c^\dagger_{k\sigma}\tilde c_{k\sigma},\nonumber\\
H_\parallel&=&\frac{J_\parallel^B}{4}\sigma_z\left[\tilde \psi_{\uparrow}^\dagger(0)\tilde \psi_{\uparrow}(0)-\tilde \psi_{\downarrow}^\dagger(0)\tilde \psi_{\downarrow}(0)\right],\nonumber\\
H_\perp&=&\frac{J_\perp^B}{2}\left[\tilde \psi^\dagger_\uparrow(0)\tilde \psi_\downarrow(0)\sigma^-
+\tilde \psi^\dagger_\downarrow(0)\tilde
\psi_\uparrow(0)\sigma^+\right]\label{eq1}.
\end{eqnarray}
Here $\tilde c_{k\sigma}$ annihilates an electron of wave number $k$ and spin
direction $\sigma$ on a ring of length $L$, and $\tilde \psi_\sigma(x)=\sum_k
e^{ikx}\tilde c_{k\sigma}/\sqrt{L}$. We denote these operators with tildes,
because we will denote the useful slow modes (defined below) without tildes.
We assume $\varepsilon_k=\varepsilon_{-k}$, so that there are both left and 
right-movers at the Fermi energy. The superscript $B$ indicates the 
bare values of the coupling constants $J^B_\parallel$ and $J^B_\perp$.

Integrating out high energy degrees of freedom towards the Fermi surface, we 
end up with a linear dispersion relation, and an effective low energy
Hamiltonian:
\begin{eqnarray}
H_0&=&\sum_{k\sigma} k \left(c^\dagger_{k\sigma} c_{k\sigma}+\bar c_{k\sigma}^\dagger \bar c_{k\sigma}\right),\nonumber\\
H_\parallel&=&\frac{J_\parallel}{2}\sigma_z\left[\psi_{\uparrow}^\dagger(0) \psi_{\uparrow}(0)- \psi_{\downarrow}^\dagger(0) \psi_{\downarrow}(0)\right],\nonumber\\
H_\perp&=&J_\perp\left[\psi^\dagger_\uparrow(0)\psi_\downarrow(0)\sigma^-
+\psi^\dagger_\downarrow(0) \psi_\uparrow(0)\sigma^+\right],\label{h2}
\end{eqnarray}
in units where the Fermi velocity $v_F=1$. We have defined slow modes
\begin{eqnarray}
c_{k\sigma}&=&\frac{1}{\sqrt{2}}\left(\tilde c_{k_F+k,\sigma}+\tilde c_{-k_F-k,\sigma}\right),\nonumber\\
\bar c_{k\sigma}&=&\frac{1}{\sqrt{2}}\left(\tilde c_{k_F+k,\sigma}-\tilde c_{-k_F-k,\sigma}\right),
\nonumber\\
\psi_\sigma(x)&=&\frac{1}{\sqrt{L}}\sum_k e^{ikx}e^{-a|k|/2}c_{k\sigma},\label{psi}
\end{eqnarray}
and a similar relation between $\bar \psi_\sigma(x)$ and $\bar c_{k\sigma}$.
Here $1/a$ is the ultraviolet cut-off energy. Operators without over-bars are
associated with even-parity single particle wave functions, so that
$c_{k\sigma}^\dagger$ creates an electron in the state
$\sqrt{2/L}\cos[(k_F+k)x]$. In $D>1$ dimensions, an equivalent role is played by
$s$-waves. For negative $x$, $\psi^\dagger_\sigma(x)$ creates an electron in an
even-parity state consisting of two wave packets centered around $\pm x$ and
propagating in opposite directions towards the impurity. For positive $x$ on
the other hand, it creates an electron in a state consisting of two wave packets
propagating away from the impurity. Because $\psi^\dagger_\sigma(x)$ and
$\psi^\dagger_\sigma(-x)$ create electrons localized to the same physical
regions in space, we refer to (\ref{h2}) as the unfolded representation.
Operators with over-bars are associated with odd-parity single particle wave
functions, so that $\bar c_{k\sigma}^\dagger$ creates an electron in the state
$i\sqrt{2/L}\sin[(k_F+k)x]$. Since the even and odd modes decouple, and only
the even modes couple to the impurity, we focus on electrons in even orbitals
and drop the $\bar c^\dagger \bar c$ terms in $H_0$.
 
In the process of linearizing the spectrum (integrating out fast modes), the
bare couplings $J^B_\parallel$ and $J^B_\perp$ are renormalized to new values 
$J_\parallel$ and $J_\perp$, that depend on the ultraviolet scale $1/a$. 
Note however that SU(2) symmetry, if present, is preserved under renormalization. 
Thus, if $J^B_\perp=J^B_\parallel$ (isotropic limit), then $J_\perp=J_\parallel$. 

In this work, we will make extensive use of the well-known fact, reviewed in 
Appendix \ref{appmap}, that the Kondo Hamiltonian (\ref{h2}) can be mapped onto an 
ohmic spin-boson model:
\begin{equation}
H_{\rm SB}=\sum_{q>0} q b_q^\dagger b_q - \sum_{q>0} 
\frac{g_q}{2} (b_q^\dagger+b_q)s_z+\frac{\Delta}{2}s_x.\label{sbham}
\end{equation}
Here $b_q$ are bosonic operators such that $[b_q,b_{q'}]=0$ and
$[b_q,b_{q'}^\dagger]=\delta_{q,q'}$, while $s_x$ and $s_z$ are Pauli matrices.
These pseudo-spin operators are not the physical angular momentum operators of the Kondo 
impurity, but are related to them. 
The parameters of the spin-boson model are related to
those of the Kondo model by 
\begin{align}
& g_q=2\sqrt{\frac{\alpha\pi q}{L}}e^{-aq/2},
~\alpha=\left(1-\frac{J_\parallel}{2\pi}\right)^2,
~\Delta=\frac{J_\perp}{\pi a}.\label{mappar}
\end{align}
 
\section{Screening cloud observables}
\label{sco}
Having defined the system we investigate, we now write down the observables that 
we will study.
We are interested here in ground state correlation functions between the
impurity spin and the electron spin density at a distance $x$ from the impurity,
which we refer to collectively as the Kondo screening cloud (or cloud for 
short).\cite{Affleck1} We give equivalent expressions for the cloud correlators
as ground state expectation values in both the fermionic and the bosonic
pictures. In Appendix \ref{appcloud} the bosonic expressions are derived
starting from the fermionic expressions. 

In the fermionic picture, the cloud correlation functions is defined as:
\begin{align}
&X^\parallel(x)=4\left<S^{\rm imp}_z \mathcal S^{\rm el}_z(x)\right>_{\rm K},\nonumber\\
&X^\perp(x)=4\left<S^{\rm imp}_x \mathcal S^{\rm el}_x(x)+S^{\rm imp}_y \mathcal 
S^{\rm el}_y(x)\right>_{\rm K},\label{clouddef}
\end{align}
where $S_j^{\rm imp}=\sigma_j/2$ is the $j\in \{x,y,z\}$ component of the
impurity spin operator, and
\begin{equation}
\mathcal S_j^{\rm el}(x)=\frac{1}{2}\sum_{\sigma\sigma'}\tilde 
\psi_\sigma^\dagger(x)[\sigma_j]_{\sigma\sigma'}
\tilde \psi_{\sigma'}(x)
\end{equation}
is the $j$ component of the electron spin density at $x$. 
The subscript K indicates that the expectation
value is with respect to the fermionic Kondo ground state. We refer to $X^\parallel$ as the longitudinal
and $X^\perp$ as the transverse cloud. Both consist of a component $X^{j}_{0}$ that varies slowly
on the scale of the Fermi wavelength, and a component that oscillates with wave vector $2k_F$, and has 
an amplitude $X^{j}_{2k_F}$ . 
The former is the result of scattering events that change the electron momentum by an
amount that is small compared to $k_F$, while the latter results from scattering between the 
Fermi points at $\pm k_F$. Explicitly, one has:
\begin{widetext}
\begin{align}
&X^\perp(x)=X^\perp_0(x)+\cos(2k_Fx)X^\perp_{2k_F}(x),\nonumber\\
&X^\parallel(x)=X^\parallel_0(x)+\left[\cos(2k_Fx)-\frac{a}{2x}\sin(2k_F x)\right]
X^\parallel_{2k_F}(x).
\label{cloud1}
\end{align}
For $x\gg a$, both the transverse and longitudinal $2k_F$ components are proportional to $\cos(2k_Fx)$. 
In the bosonic language, and in the thermodynamic limit $|x|\ll L$, one finds
using standard bosonization identities:
\begin{align}
&X^\perp_0(x)=\frac{a}{\pi(x^2+a^2)}{\rm Re}\left<s^-e^{\varphi^\dagger(x)-i\varrho^\dagger(x)}
e^{-\varphi(x)+i\varrho(x)}\right>+(x\to-x),\nonumber\\
&X^\perp_{2k_F}=\frac{2a}{\pi(x^2+a^2)}{\rm Re}\left<s^-e^{\varphi^\dagger(x)}
e^{-\varphi(x)}\right>,\nonumber\\
&X^\parallel_0(x)=\frac{\partial_x}{2\pi}\left<s_z\left[\varrho^\dagger(x)+\varrho(x)\right]\right>
-\frac{a}{\pi(x^2+a^2)},\nonumber\\
&X^\parallel_{2k_F}(x)=\frac{2x}{\pi(a^2+4x^2)}
{\rm Im}\left<s_z\left(\frac{a+ix}{a-ix}\right)^{s_z}e^{i\varrho^\dagger(x)}e^{i\varrho(x)}\right>,\label{bosoncloud}
\end{align}
with the bosonic fields:
\begin{align}
&\varphi(x)=2\sum_{q>0}\sqrt{\frac{\pi}{Lq}}e^{-aq/2}\left[\cos(qx)-1\right]b_q,~~~\varrho(x)=2\sum_{q>0}\sqrt{\frac{\pi}{Lq}}e^{-aq/2}\sin(qx)b_q.
\end{align}
\end{widetext}

At $x<a$, the behavior of these expressions depend strongly on ultraviolet
physics that our model does not attempt to represent accurately. The regime of
physical significance is $x\gg a$ where the behavior of the cloud is insensitive
to ultraviolet details of the model. None the less, we delay taking the $a/x\to
0$ limit until the end of the calculation, reasoning that it is of some interest
to see how a particular cut-off scheme regularizes ultraviolet singularities.

We presented expressions for
a one-dimensional cloud. However, the generalization to a higher dimensional
electron gas is trivial, assuming a point impurity that therefore only scatters $s$-waves.\cite{Affleck1}
In this case, one interprets the coordinate $x$ as a radial distance and 
divides the one-dimensional correlator by the area of a spherical shell of radius $x$. One also 
reverses the sign of the $2k_F$ components, because the radial part of the higher dimensional problem is 
defined on the half-line, with a $\pi$ phase shift between left- and right movers.
    
To make further progress in the computation of the cloud observables, we need an 
accurate approximation for the ground state of the spin-boson model. That is the 
topic of the next section.

\section{Coherent-state expansion of the spin-boson ground state}
\label{secmpa}

We will use here a systematic coherent-state decomposition~\cite{Bera1,Bera2} of
the many-body ground state of the Ohmic spin-boson model.
Physically, coherent states appear as natural degrees of freedom, because, in the absence of spin-tunneling
$\Delta$, the 
two spin configurations are associated with a displacement of the bath modes from $0$ to 
$\pm f_q = \pm \sqrt{\pi\alpha/Lq}$.
Indeed, at $\Delta=0$, there are two degenerate ground states
\begin{equation}
\left|f_+\right>\otimes\left|\uparrow\right>,~~~\left|f_-\right>\otimes\left|\downarrow\right>,
\end{equation}
where
\begin{equation}
\left|f_\pm\right>=\exp\left[\pm\sum_{q>0}f_q(b_q^\dagger-b_q)\right]\left|0\right>,
\end{equation}
$\left|0\right>$ is the bosonic vacuum, and $\{\left|\uparrow\right>
,\left|\uparrow\right>\}$ are the spin eigenstates of $s_z$.
The single coherent-state Ansatz (usually dubbed the Silbey-Harris state) includes 
the effect of tunneling by promoting $f_q$ to a variational parameter and taking as 
trial state the linear combination $\left|f\right>=(\left|f_+\right>\otimes\left|\uparrow\right>-\left|f_-\right>\otimes\left|\downarrow\right>)/\sqrt{2}$. 
This results readily in the simple expression $f_q^\mathrm{SH} = \sqrt{\pi\alpha q/L} g_q/(q+\Delta_R)$ with
the self-consistency condition $\Delta_R=\Delta
\left<f_+|f_-\right>=\Delta\left<f|-f\right>$.
This trial wavefunction provides an excellent approximation to the true ground state
provided the shifted equilibrium positions $f_q^\mathrm{SH}$ 
of most oscillators are not too far apart, {\it i.e.} 
if $\alpha$ is small compared to unity.

For larger $\alpha$, the overlap $\left<f\right|\left.-f\right>$, with $f_q$ 
determined variationally, becomes exponentially small, and this strongly 
underestimates the tunneling energy $\Delta\left<\sigma_x\right>/2$. 
The coherent-state expansion addresses this issue by extending the
Silbey-Harris form to a more general linear superposition of coherent states:
\begin{equation}
\left|\psi\right>=\sum_{m=1}^M 
c_m\frac{\left|f^{(m)}\right> \otimes \left|\uparrow\right>
- \left|-f^{(m)}\right> \otimes \left|\downarrow\right>}{\sqrt{2}},
\label{superpose}
\end{equation}
with the set of displacement $f_q^{(m)}$ parametrizing a family of Silbey-Harris
states:
\begin{equation}
\left|f^{(m)}\right>=\exp\left[\sum_{q>0}f^{(m)}_q(b_q^\dagger-b_q)\right]
\left|0\right>.
\label{family}
\end{equation}
Here $M$ is the maximal number of allowed coherent states in the
decomposition~(\ref{superpose}), and sets the level of
approximation.
The displacements $f^{(m)}_q$ and coefficients $c_m$ are determined by minimizing
\begin{equation}
\mathcal E=\left<\psi\right|H_{\rm SB}\left|\psi\right>
-\lambda\left<\psi\right|\left.\psi\right>,\label{ener}
\end{equation}
where $\lambda$ is a Lagrange multiplier that is used to enforce 
normalization. These parameters are generically found to be real in
the ground state.

The coherent-state expansion already dramatically improves the 
estimate for the tunneling energy $\Delta\left<\psi\right|\sigma_x\left|\psi\right>/2$
for $M=2$, by allowing for cross-terms 
$\left<f^{(m)}\right|\sigma_x\left|f^{(n)}\right>$, in which the 
displacements $f_q^{(m)}$ and $f_q^{(n)}$ are anti-correlated at low $q$.
Such cross-terms are therefore not exponentially small in $\alpha$, and allow
a sizable energy gain compared to the Silbey-Harris approximation.

It is important to note that an arbitrary state of the spin-boson
Hamiltonian can be written as an infinite but {\em discrete} sum of the form
(\ref{superpose}), at least if the bosonic bath contains a finite number of
modes.  This follows from a theorem, proved by Cahill,\cite{Cahill} that for a
single bosonic mode, countable sets
$\left\{\left|f^{(m)}\right>,\,m=1,\,2,\,\ldots\right\}$ of (real) coherent
states exist, that form a complete basis.  For a finite number of bosonic modes,
a general state can therefore be approximated to any required accuracy, by
making $M$ sufficiently large. It is in this sense that the coherent state
expansion is numerically exact, provided in practice that good convergence 
to the true many-body ground state occurs. Previous
investigations~\cite{Bera2}, as well as the extensive comparisons made in 
Sec.~\ref{secresults}, demonstrate indeed 
that the expansion~(\ref{superpose}) rapidly approaches the exact ground state 
of the spin-boson model as $M$ is increased to moderate values, also for an
infinite bath.

\section{Reducing the number of variational parameters}
\label{secreduce}
In previous implementations of the coherent expansion~(\ref{superpose}-\ref{family}), 
the bosonic bath was restricted to a large but finite number $N$ of modes, and
each of the $M\times N$ displacements $f_q^{(m)}$ was treated as an independent
variational parameter. In this work, we want to calculate the Kondo screening
cloud, which requires a high spatial resolution from short to possibly
exponentially large distances, so that a huge number of bath modes needs to
be included. It is therefore desirable to have an implementation of the
method in which the number of variational parameters does not depend on the
number of included modes. Such an implementation would also allow for
computations for a continuous spectral density (thermodynamic limit), which
numerical techniques like NRG are not able to perform.
In this section, we derive such an implementation, which is the the main technical 
innovation of our work.

We show that the algebraic structure of the variational equations 
constrains the functional dependence of the displacements $f^{(m)}_q$ with respect 
to momentum $q$ to such an extent that there are in fact only $M^2+M-1$ variational 
parameters, independently of the number of modes $N$ of the bosonic bath.
To stress that the argument does not require a linear bath spectrum, we replace the term
$\sum_{q>0} qb^\dagger_q b_q$ in the spin-boson Hamiltonian, with the more general 
kinetic energy expression $\sum_{q>0} \omega_q b_q^\dagger b_q$ for the remainder of 
this section. In subsequent sections, we will specialize again to the linear bath spectrum. 
 
Exploiting the coherent state structure of the Ansatz, we can express
the total energy $\mathcal E$ in terms of $f^{(m)}_q$ and $c_m$ 
as:
\begin{widetext}
\begin{equation}
\mathcal E=\sum_{m,n=1}^M c_m c_n\left\{\left<f^{(m)}\right|\left.f^{(n)}\right>\sum_{q>0}\left[\omega_q f_q^{(m)}f_q^{(n)}
-\frac{g_q}{2}\left(f_q^{(m)}+f_q^{(n)}\right)-\lambda\right]-\frac{\Delta}{2}\left<f^{(m)}\right|
\left.-f^{(n)}\right>\right\},\label{eF}
\end{equation}
\end{widetext}
where
\begin{equation}
\left<f^{(m)}\right|\left.\pm f^{(n)}\right>=\exp\left[-\frac{1}{2}\sum_{q>0}\left(f_q^{(m)}\mp f_q^{(n)}
\right)^2\right].\label{ol}
\end{equation}

Considering the minimization condition $\partial \mathcal E/\partial f_q^{(m)}=0$
with the use of (\ref{eF}) and (\ref{ol}), this leads to 
\begin{equation}
\sum_{n=1}^M\left\{U_{mn}\omega_q+V_{mn}\right\}f_q^{(n)}=W_m g_q,\label{sys}
\end{equation}
where the entries of the $M\times M$ matrices $U$ and $V$, and the $M$-dimensional column vector $W$ 
are:
\begin{align}
&U_{lm}=2c_m\left<f^{(l)}\right|\left. f^{(m)}\right>,\nonumber\\
&V_{lm}=P_{lm}+Q_{lm}-\delta_{lm}\sum_n\left(P_{ln}-Q_{ln}\right),\nonumber\\
&P_{lm}=c_m\left<f^{(l)}\right|\left.f^{(m)}\right>\nonumber\\
&~~~~~~~~~~~~\times\sum_{k>0}\left[\omega_k f_k^{(l)}f_k^{(m)}-\frac{g_k}{2}\left(f^{(l)}_q-f^{(m)}_q\right)-\lambda\right],\nonumber\\
&Q_{lm}=c_m\frac{\Delta}{2}\left<f^{(l)}\right|\left. -f^{(m)}\right>,\nonumber\\
&W_l=\frac{1}{2}\sum_m U_{lm}.\label{mat}
\end{align} 

It is crucial to note that $U$, $V$ and $W$ in (\ref{sys})
do not depend on the value of the momentum $q$ labelling the displacement $f_q^{(m)}$ 
that we consider within the variational equation. 
These matrices and vectors however depend non-linearly on the complete set of 
displacements, through summations over a dummy momentum
index as in Eq.~(\ref{ol}).
From (\ref{sys}), it readily follows that $f_q^{(m)}$ is given by:
\begin{equation}
f_q^{(m)}=g_q\sum_{n=1}^M \left[(U\omega_q+V)^{-1}\right]_{mn}W_n.
\end{equation}
Since $U\omega_q+V$ is linear in $\omega_q$, 
\begin{equation}
\sum_{n=1}^M
\left[(U\omega_q+V)^{-1}\right]_{mn}W_n=\frac{N^{(m)}(\omega_q)}{D(\omega_q)},
\end{equation}
where $N^{(m)}(z)$ is a polynomial of order $M-1$ in $z$ and $D(z)={\rm
det}(Uz+V)$ is a polynomial of order $M$ in $z$. Note that the
denominator $D(z)$ is common to all sets of displacements $f^{(m)}$ for
$m=1,\ldots,M$.
Using the expression (\ref{mat}) for $W$ in terms of $U$, we see that for 
large $z$, $\frac{N^{(m)}(z)}{D(z)}\to 1/2z$ independent of $m$. 
Thus we arrive at the form:
\begin{align}
&f_q^{(m)}=\frac{g_q}{2}h_m(q),\nonumber\\
&h_m(q)=\frac{\sum_{n=0}^{M-1}\mu_{mn}\omega_q^n}{\prod_{n=1}^M(\omega_q-\omega_n)},\label{e7}
\end{align}
where $\mu_{m\,M-1}=1$, and in general $\mu_{mn}$ is real. The poles $\omega_n$ that are not real
come in complex conjugate pairs.
The optimal state $\left|\psi\right>$ can be obtained by numerically minimizing the energy with
respect to the $M\times (M-1)$ 
unknown coefficients $\mu_{mn}$, the $M$ poles $\omega_n$, and the $M-1$
weights $c_m$ (accounting for wavefunction normalization), so that there are 
$M^2+M-1$ parameters to be found in total, independently of the number of bosonic modes 
in the problem.

Since the dimension of the search space does not depend on the number of bath
modes, the main obstacle in considering a bath with a very large number of
modes has been removed. The next question is whether the energy, which involves
sums over all bath modes, can be calculated efficiently for a large number of
modes. Only single sums over the momentum have to be computed in
the energy functional, and in the worst case, the total numerical cost is linear in
the number of modes. This is a major improvement with respect to a brute force
diagonalization of the model, where scaling of the Hilbert space dimension is 
exponential with the number of degrees of freedom. In order to reach the 
continuum limit, the discrete momentum sums are replaced by integrals, which
fortunately can be performed analytically rather than numerically, as we now 
demonstrate.

\section{Thermodynamic limit and the analytical evaluation of momentum integrals}

\label{secen}
In this section we specialize again to the case of a linear bath spectrum
$\omega_q=q$, and analytically perform the momentum integrals involved in the
calculating the energy for given set of variational parameters $\mu_{mn}$,
$\omega_m$ and $c_m$. We do so for the smooth ultraviolet cut-off
$g_q\propto e^{- aq/2}$ that the bosonization of the Kondo model naturally introduces.
Although we do not work out the details here, we note that the momentum integrals
can also be performed analytically in the case of a sharp cut-off $g_q\propto\theta(1/a-q)$.
In the $L\to\infty$ (thermodynamic) limit, momentum sums are replaced by integrals according to
\begin{equation}
\frac{2\pi}{L}\sum_{q>0}\to\int_0^\infty dq.
\end{equation}
With $f^{(m)}_q$ of the form (\ref{e7}), and $g_q$ of the form (\ref{mappar}) dictated by the 
mapping from the Kondo model, the energy 
\begin{equation}
E=\frac{\left<\psi\right|H_{\rm SB}\left|\psi\right>}{\left<\psi\right|\left.\psi\right>}
\end{equation}
can be written as a functional of the variational parameters as:
\begin{equation}
E=\frac{\sum_{mn=1}^M c_mc_n\left(\alpha I_{mn}^0e^{-\frac{\alpha}{4}I^{-1}_{mn}}
-\Delta e^{-\frac{\alpha}{4}I_{mn}^{+1}}\right)}{2\sum_{mn=1}^M c_mc_ne^{-\frac{\alpha}{4}I^{-1}_{mn}}},
\label{en}
\end{equation}
which involves the Laplace transforms 
\begin{equation}
I^\lambda_{mn}=\int_0^\infty dk\, e^{-ak}J^\lambda_{mn}(k)
\end{equation}
of the rational functions
\begin{eqnarray}
J_{mn}^{\pm 1}(k)&=&k\left[h_m(k)\pm h_n(k)\right]^2,\nonumber\\
J_{mn}^0(k)&=&\left[kh_m(k)-1\right]\left[kh_n(k)-1\right].
\end{eqnarray}
Note that we have omitted here a constant term
\begin{equation}
E_0=-\frac{\alpha}{2a},\label{econst}
\end{equation}
that does not depend on the variational parameters.
The functions $J_{mn}^{\lambda}(k)$ have $M$ second order poles of
the type $(k-\omega_l)^{-2}$. For large $k$, the behavior is
\begin{equation}
J_{mn}^\lambda(k)\simeq k^{-2+\lambda}.
\end{equation}

The Laplace transforms can be performed analytically, using the 
following method. In general, we consider integrals of the form
\begin{equation}
I=\int_0^\infty dk\,e^{-ak}R(k),\label{e1}
\end{equation}
where the function $R(k)$ is a ratio of
polynomials, and all poles are of second order, i.e.
\begin{equation}
R(k)=\frac{p(k)}{\prod_{n=1}^M(k-\omega_n)^2}.
\end{equation}
This excludes the possibility of poles on the positive real line, which yields $E=+\infty$.
Furthermore, the numerator $p(k)$ is a polynomial of at most order $2M-1$.

\begin{figure}
\begin{center}
\includegraphics[width=.8\columnwidth]{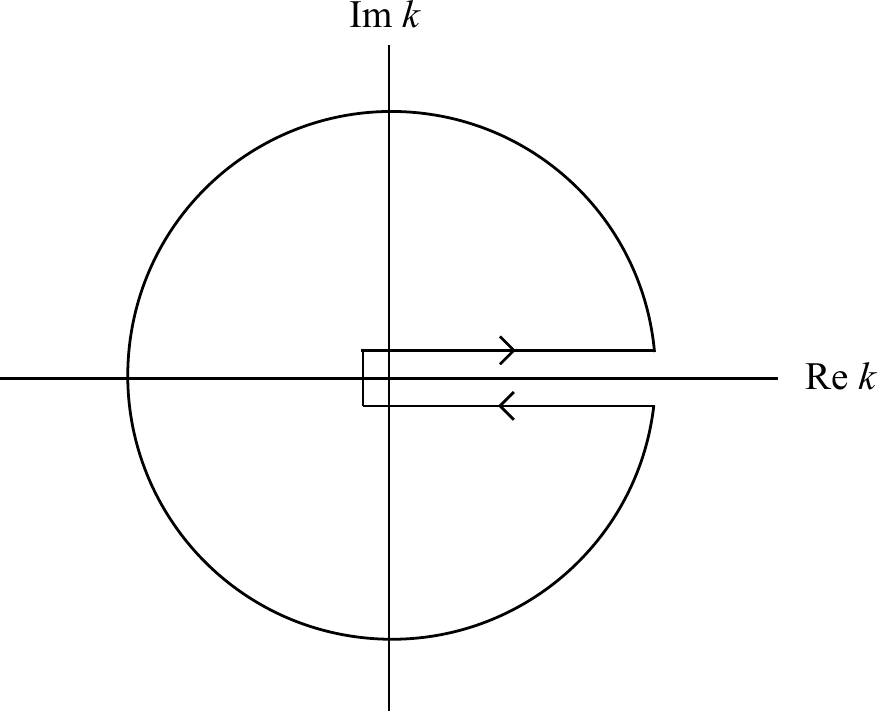}
\caption{Chosen deformation of the contour for the integral~(\ref{e1}).\label{f1}}
\end{center}
\end{figure}

It would be useful if we could replace the integration contour in (\ref{e1}) by a closed contour
such as the one in Figure \ref{f1}. We can do so, provided we multiply the integrand by a function
$\mathcal F(k)$ with a branch cut along the positive real line such that 
$\mathcal F(k+i0^+)-\mathcal F(k-i0^+)\propto\theta(k)$. Furthermore, $\mathcal F$ must be such that the contribution to the integral
that comes from closing the contour at large $k$ is negligible. A function that exactly meets these requirements
is $\Gamma(0,-ak)$, where $\Gamma(0,z)$ is the incomplete Gamma function of order zero.\cite{nist} 
It is defined as 
\begin{equation}
\Gamma(0,z)=\int_z^\infty dt\,\frac{e^{-t}}{t},
\end{equation}
where the integration path does not intersect the negative real line and excludes the origin.
Its main relevant properties are
\begin{equation}
\Gamma(0,-(x+i0^+))-\Gamma(0,-(x-i0^+))=2\pi i,
\end{equation}
for $x>0$, and 
\begin{equation}
\Gamma(0,-z)\sim -\frac{e^{z}}{z},
\end{equation}
for large $|z|$.
Furthermore $\Gamma(0,-z)$ is analytical except in the neighborhood of the positive real line.
Thus we find
\begin{align}
&\int_0^\infty dk\,e^{-ak}R(k)=\int_C \frac{dk}{2\pi i}\, R(k)e^{-ak}\Gamma(0,-ak)\nonumber\\
&=\sum_n{\rm Res}\left\{R(k)e^{-ak}\Gamma(0,-ak),\omega_n\right\}.
\end{align} 
In the first line, $C$ is the contour depicted in Figure \ref{f1}.

Returning to the integrals $I^\lambda_{mn}$ ,
one then has
\begin{equation}
I_{mn}^\lambda=\sum_{l=1}^M{\rm Res}\left\{J_{mn}^\lambda(k)e^{-ak}\Gamma(0,-ak),\omega_l\right\}.
\label{resexp}
\end{equation}
Setting
\begin{equation}
F(z)=e^{-z}\Gamma(0,-z),\label{efl}
\end{equation}
and noting that 
\begin{equation}
\frac{d}{dz}F(z)=-\left[F(z)+\frac{1}{z}\right],\label{res1}
\end{equation}
evaluation of the residues yield 
\begin{equation}
I_{mn}^\lambda=\sum_l \dot{K}_{mnl}^\lambda F(\omega_l a)
-K_{mnl}^\lambda\left(aF(\omega_l a)+\frac{1}{\omega_l}\right),\label{e4}
\end{equation}
where
\begin{eqnarray}
K_{mnl}^\lambda&=&\lim_{k\to \omega_l}(k-\omega_l)^2J_{mn}^\lambda(k),\nonumber\\
\dot{K}_{mnl}^\lambda&=&\lim_{k\to \omega_l}\frac{d}{dk}\left[(k-\omega_l)^2J_{mn}^\lambda(k)\right].
\end{eqnarray}
To relate the $K$ parameters to the coherent state displacements explicitly, we define 
\begin{eqnarray}
h_{mn}&=&\lim_{k\to\omega_n}(k-\omega_n)h_m(k)\nonumber\\
&=&\frac{\sum_{l=0}^{M-1}\mu_{ml}\omega_n^{~l}}{\prod_{l=1\not=n}^M(\omega_n-\omega_l)}\label{defh}
\end{eqnarray}
and 
\begin{eqnarray}
\dot{h}_{mn}&=&\lim_{k\to\omega_n}\frac{d}{dk}\left[(k-\omega_n)h_m(k)\right]\nonumber\\
&=&\frac{\sum_{l=1}^{M-1}l\mu_{ml}\omega_n^{~l-1}}{\prod_{l=1\not=n}^M(\omega_n-\omega_l)}
-h_{mn}\sum_{l=1\not=n}\frac{1}{\omega_n-\omega_l}.\nonumber\\\label{defhd}
\end{eqnarray}
This gives
\begin{eqnarray}
K_{mnl}^{\pm1}&=&\omega_l(h_{ml}\pm h_{nl})^2,\nonumber\\
K_{mnl}^0&=&\omega_l^2h_{ml}h_{nl},\nonumber\\
\dot{K}^{\pm1}_{mnl}&=&(h_{ml}\pm h_{nl})^2+2\omega_l(h_{ml} \pm h_{nl})(\dot{h}_{ml}\pm \dot{h}_{nl}),\nonumber\\
\dot{K}^0_{mnl}&=&\omega_l(h_{ml}+\omega_l\dot{h}_{ml}-1)h_{nl}\nonumber\\
&&~~+\omega_lh_{ml}(h_{nl}+\omega_l\dot{h}_{nl}-1).\label{ee5}
\end{eqnarray}
Substitution of (\ref{ee5}) into (\ref{e4}), and (\ref{e4}) into (\ref{en})
yields an explicit expression for the energy in terms of the variational parameters.

As a consistency check, we can work out the single coherent state theory ($M=1$) to make 
sure that we recover the results of Silbey and Harris. This approximation is accurate 
for $\alpha\ll 1$, which, in the Kondo language, corresponds to
the strongly anisotropic $J_\parallel\gg J_\perp$ regime. 
For $M=1$, there is only one variational parameter, which is conventionally defined as 
$\Delta_R=-\omega_1$. This translates to $h_{11}=1$ and $\dot h_{11}=0$
[see Eqs. (\ref{defh}) and (\ref{defhd})]. The energy of the Silbey-Harris state then reads
\begin{equation}
E=-\frac{1}{2}\left[\alpha\Delta_R^2\partial_{\Delta_R}F(-\Delta_R a)+\Delta \left<f\right|\left.-f
\right>\right],
\end{equation}
with
\begin{equation}
\left<f\right|\left.-f\right>=\exp\left\{-\alpha\left[F(-\Delta_R a)+\Delta_R \partial_{\Delta_R}
F(-\Delta_R a)\right]\right\},
\end{equation}
and we have used the identity (\ref{res1}) to relate $F$ and its derivative.
Minimizing $E$ with respect to $\Delta_R$, we recover the known self-consistency condition
\begin{equation}
\Delta_R=\Delta\left<f\right|\left.-f\right>.
\end{equation}
Assuming than $\Delta_R\ll 1/a$, and using the fact that for small $z$,
\begin{equation}
F(-z)=-{\rm ln}(z)-\gamma_E+\mathcal O(z),
\end{equation}
where $\gamma_E=0.577$ is the Euler-Mascheroni constant, one can solve the self-consistency
condition for $\Delta_R$ to obtain
\begin{equation}
\Delta_R=\Delta \left(e^{1+\gamma_E}a\Delta\right)^\frac{\alpha}{1-\alpha}.
\end{equation}
Up to a pre-factor, $\Delta_R$ corresponds to the Kondo temperature in the regime of small $\alpha$,
where the single coherent state Ansatz is accurate.

In general, it is far more efficient to evaluate the energy analytically, as was done in
this section, than to do the integrals in (\ref{en}) numerically. There are
however small regions of the search space where the analytical evaluation of the
energy is not stable. These are regions in which two or more of the poles
$\omega_m$ lie close to each other. In these regions, there are large
cancellations between individual terms in the sum over residues in
(\ref{resexp}), and this leads to large numerical errors when individual
residues are calculated separately before they are summed. 
We circumvent the problem as follows. When the minimization algorithm searches a
dangerous region of the search space, it does not try to evaluate the residues
at the offending poles individually. It rather takes the slow but safe option
of numerically integrating around a loop that circles all closely spaced poles
at a safe distance. Fortunately, one does not have to fall back on this
contingency plan too often, as the problematic regions of the search space are
small and do not seem to be particularly favored in the actual optimal solution.

\section{Expressing the cloud with coherent states}
\label{cloudcalc}

In the previous section, we obtained an analytical expression for the energy in
terms of $M^2+M-1$ variational parameters. This result allows for a significant
speed-up of the numerical minimization of the energy. In this section, we apply
the same analytical technique to evaluate the momentum integrals involved in
the calculation of the Kondo screening cloud, in terms of bosonic displacements.
Evaluating the four different cloud correlators (\ref{bosoncloud}) for the 
$M$-coherent state wavefunction can be done using straight-forward
coherent-state algebra:
\begin{widetext} 
\begin{align}
X_0^\perp(x)&=-\frac{a}{\pi(x^2+a^2)}\sum_{m,n=1}^Me^{\sqrt{\alpha}\left[A(0)_{mn}-A(x)_{mn}\right]}\cos(\sqrt{\alpha}B(x)_{mn})c_mc_n
\left<f_m\right|\left.-f_n\right>,\nonumber\\
X_{2k_F}^\perp(x)&=-\frac{a}{\pi(x^2+a^2)}\sum_{m,n=1}^Me^{\sqrt{\alpha}[A(0)_{mn}-A(x)_{mn}]}c_mc_n
\left<f_m\right|\left.-f_n\right>,\nonumber\\
X^\parallel_0(x)&=-(1-\sqrt{\alpha})\frac{a}{\pi(x^2+a^2)}+\frac{\sqrt{\alpha}}{2\pi}\sum_{m,n=1}^M C(x)_{mn}c_mc_n\left<f_m\right|\left.f_n\right>,\nonumber\\
X^\parallel_{2k_F}(x)&=\frac{2x}{\pi(4x^2+a^2)}{\rm Im}\left[\frac{a+ix}{a-ix}\sum_{m,n=1}^M e^{i\sqrt{\alpha}D(x)_{mn}}c_mc_n\left<f_m\right|\left.f_n\right>\right],
\label{cloudeqs}
\end{align}
\end{widetext}
together with the normalization condition 
\begin{equation}
\sum_{m,n=1}^M c_m c_n \left<f_m\right|\left. f_n\right>=1.
\end{equation}
The matrices $A$, $B$, $C$, and $D$ are defined as:
\begin{align}
A(x)_{mn}&=\int_0^\infty dq\, \cos(qx) e^{-aq}[h_m(q)+h_n(q)],\nonumber\\
B(x)_{mn}&=\int_0^\infty dq\, \sin(qx)e^{-aq}[h_m(q)-h_n(q)],\nonumber\\
C(x)_{mn}&=\int_0^\infty dq\, \cos(qx)e^{-aq}[qh_m(q)+qh_n(q)-2],\nonumber\\
D(x)_{mn}&=\int_0^\infty dq\, \sin(qx)e^{-aq}[h_m(q)+h_n(q)].\label{cint}
\end{align}
Note that $B$, $C$ and $D$ remain finite if the limit $a\to 0$ followed by $x\to 0$ is taken, while $A$ 
diverges logarithmically.
The method we used to evaluate $\left<f_m\right|\left.\pm f_n\right>=\exp(-\alpha I_{mn}^{\mp1}/4)$ analytically can be extended
to evaluate the above integrals as well. The detail of the calculation can be found in
Appendix \ref{appci}. The resulting expressions are:
\begin{align}
A(x)_{mn}&={\rm Re}\sum_{l=1}^M(h_{ml}+h_{nl})F(\omega_l(a-ix)),\nonumber\\
B(x)_{mn}&={\rm Im}\sum_{l=1}^M(h_{ml}-h_{nl})F(\omega_l(a-ix)),\nonumber\\
C(x)_{mn}&={\rm Re}\sum_{l=1}^M \omega_l(h_{ml}+h_{nl})F(\omega_l(a-ix)),\nonumber\\
D(x)_{mn}&={\rm Im}\sum_{l=1}^M(h_{ml}+h_{nl})F(\omega_l(a-ix)),\label{abcd}
\end{align}
where $F(z)$ is defined in (\ref{efl}). 
 
For future reference, we now consider the large (compared to the Kondo
length) $x$ asymptotic behavior of the expressions (\ref{cloudeqs}) for the
screening cloud. At $|z|\gg 1$, 
\begin{equation}
F(z)\simeq\frac{-1}{z}-\frac{1}{z^2}.
\end{equation}
The residue theorem can be used to derive the identities 
\begin{align}
&\sum_{l=1}^Mh_{ml}=1,~~~~\sum_{l=1}^M\frac{h_{ml}}{\omega_l}=-h_m(0).
\end{align}
With the aid of the above equations, the cloud correlators (\ref{cloudeqs}) are found 
to decay like $1/x^2$ at large $x$. Explicitly, one finds for
$x\gg\xi_\parallel$:
\begin{align}
&X_{2k_F}^\perp(x)=X_0^\perp(x)=-\frac{2\xi_\perp}{x^2},\nonumber\\
&X_{2k_F}^\parallel(x)=X_0^\parallel(x)+ \frac{a}{\pi x^2}=-\frac{\xi_\parallel}{x^2}
\label{asymptotics}
\end{align}
where the lengths $\xi_\perp$ and $\xi_\parallel$ are given by
\begin{align}
& \xi_\perp=\frac{a}{2\pi}\sum_{m,n=1}^Me^{\sqrt{\alpha}
\sum_{l=1}^M(h_{ml}+h_{nl})F_l(a)}c_mc_n
\left<f_m\right|\left.-f_n\right>,\nonumber\\
& \xi_\parallel=\frac{\sqrt{\alpha}}{2\pi}\sum_{m,n=1}^M[h_m(0)+h_n(0)]c_mc_n
\left<f_m\right|\left.f_n\right>.\label{ea1}
\end{align}
 
The above expression establishes a very direct connection between the $q=0$ behavior
of the coherent state displacements and the large $x$ behavior of the longitudinal cloud.
The small contribution $\mathcal O (a/ x^2)$ to $X^\parallel_0$ is a feature
of the smooth ultraviolet cut-off $\sim e^{-aq}$, and will be ignored in what
follows. Note that the $0k_F$ and $2k_F$ components of the cloud become equal
at large $x$. Furthermore, there are two emergent length scales $\xi_\perp$ and
$\xi_\parallel$ in the problem, which are in general not equal, unless
spin-isotropy is restored.

In the single coherent-state approximation, which is accurate for small $\alpha$, the 
cloud correlators can be explicitly computed:
\begin{widetext}
\begin{align}
&X_0^\perp(x)=X_{2k_F}^\perp(x)=-\frac{a\Delta_R}{\pi\Delta(x^2+a^2)}
e^{2\sqrt{\alpha}\left[F(-\Delta_R a)-{\rm Re}\,F(-\Delta_R(a-ix))\right]},\nonumber\\
&X_0^\parallel(x)=-(1-\sqrt{\alpha})\frac{a}{\pi(x^2+a^2)}-\frac{\sqrt{\alpha}\Delta_R}{\pi}{\rm Re}\,F(-\Delta_R(a-ix)),\nonumber\\
&X_{2k_F}^\parallel(x)=\frac{2x}{\pi(a^2+4x^2)}{\rm Im}\left[\frac{a+ix}{x-ix}
e^{2i\sqrt{\alpha}{\rm Im}\,F(-\Delta_R(a-ix))}\right].\label{shcloud}
\end{align}
\end{widetext}
where, as noted in Sec.~\ref{secen}, $\Delta_R=-\omega_1$ is the Kondo energy scale. 
The two correlation lengths $\xi_\perp$ and $\xi_\parallel$ 
[cf. (\ref{ea1})] are in this anisotropic limit:
\begin{align}
&\xi_\perp=\frac{a \Delta_R}{2\pi\Delta}e^{2\sqrt \alpha F(-\Delta_R a)},
~~~\xi_\parallel=\frac{\sqrt{\alpha}}{\pi\Delta_R}.\label{shxi}
\end{align}
These simple analytical expressions for the Kondo cloud, valid in the 
limit of strong spin-anisotropy, have not appeared in the literature before. 
We will analyze them further in the next section, along with numerical results 
obtained for larger $\alpha$ values from the systematic coherent
state expansion.

\section{Results}
\label{secresults}
In Sec.~\ref{secen} and Sec.~\ref{cloudcalc}, we have collected the tools to
calculate the average energy and the components of the screening cloud, for an
$M$ coherent-state wavefunction, in terms of only $\mathcal O(M^2)$ variational parameters. 
In order to find the variational parameters, the energy must be minimized, and
for this purpose we use a standard simulated annealing algorithm.\cite{Corana,Goffe} 
Clearly, the quality of the approximation is limited by the maximum number of
coherent states that can be handled with the available computational resources.
Using a single personal computer and simulated annealing minimization, we have found 
it possible to go up to $M=7$ coherent states, which is enough for our purposes
here, although an improved algorithm combining global and local optimization~\cite{Bera2} 
can reach values as large as $M=24$. In fact, the required number of numerical operations
is not a limitation {\it per se} here. Rather, the main difficulty is that the energy
landscape in the space of variational parameters is very shallow and contains
several low lying minima.

This section is divided into three extended sub-parts.
First, we benchmark the coherent state method, both against
previous results, and also by studying the convergence properties of the coherent state
expansion, establishing its domain of validity for the computation of the
screening cloud. Then, we consider the Kondo overlaps proposed in
Ref.~\onlinecite{luk}, which we compute both analytically in the
spin-anisotropic regime, and numerically for the nearly  spin-isotropic case. Our
results quantitatively agree with recent field theoretic calculations.\cite{luk}
We finally perform an extensive study of the Kondo cloud correlators over
a wide range of parameter values  by a combination of analytical arguments and
extensive coherent state simulations. This allows us to uncover the precise 
universal features of the screening cloud of the anisotropic Kondo
model.

\begin{figure}[t]
\begin{center}
\includegraphics[width=\columnwidth]{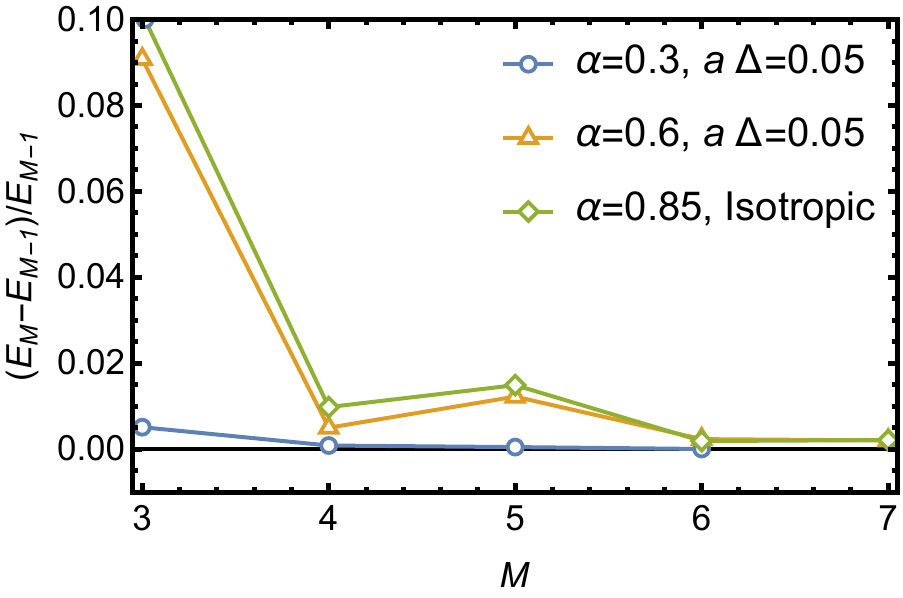}
\caption{(color online) Fractional improvement $(E_M-E_{M-1})/E_{M-1}$ in the minimum energy
$E_M$ as the number of coherent states $M$ is increased, at three different points in
parameter space. Circles correspond to a strongly spin-anisotropic point
$(\alpha=0.3,\Delta=0.05/a)$, triangles to a moderately spin-anisotropic point
$(\alpha=0.6,\Delta=0.05/a)$, and diamonds to a perfectly spin-isotropic point
$(\alpha=0.85,\Delta=0.156/a)$, associated to equal Kondo exchange couplings
$J_\perp=J_\parallel=0.49$.}\label{f2}
\end{center}
\end{figure}

\begin{figure}[t]
\begin{center}
\includegraphics[width=\columnwidth]{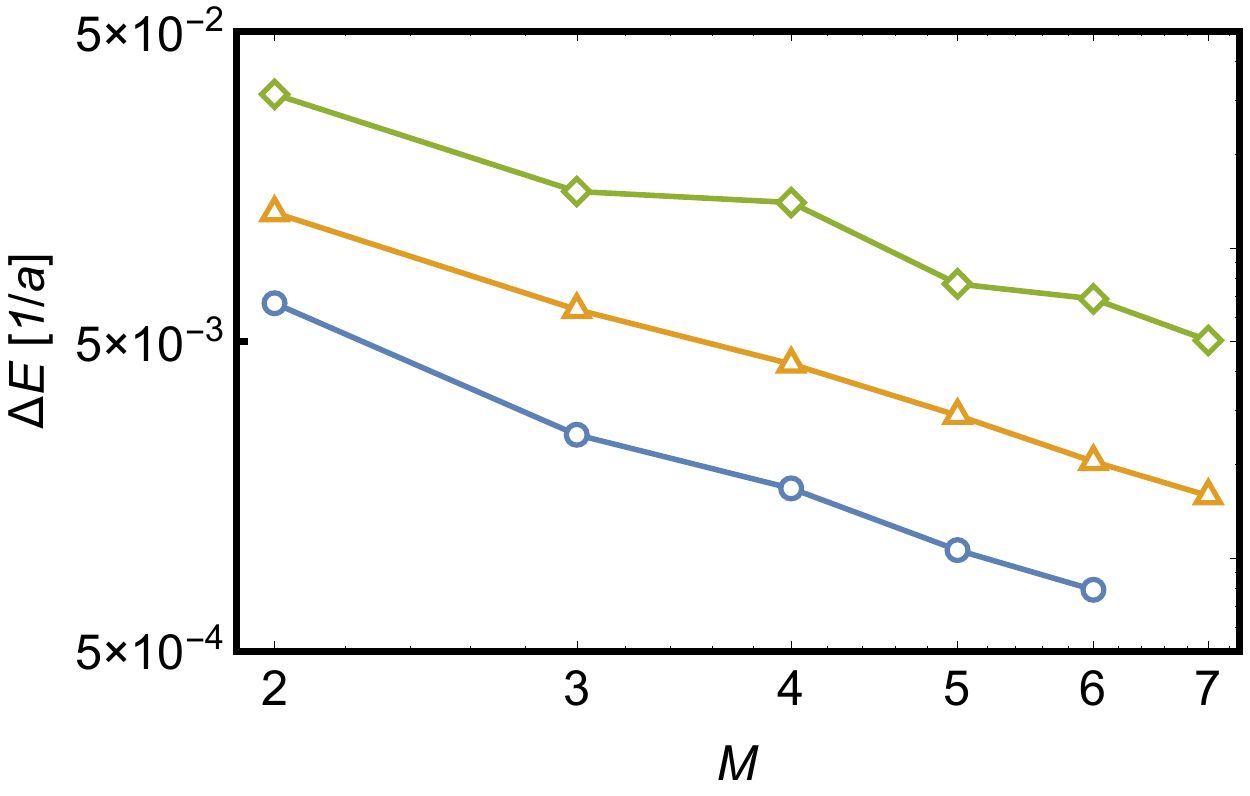}
\caption{(color online) Energy uncertainty $\Delta E$ as a function of the number $M$ of
coherent states, for the same three points in parameter space as the data in
Figure~\ref{f2}.
\label{fenvar}}
\end{center}
\end{figure}

\subsection{Convergence properties of the coherent state expansion}
In this subsection, we present strong evidence for the rapid convergence of the
wavefunction~(\ref{superpose}) as the number $M$ of coherent states increases,
using various observable quantities.
In Refs.~\onlinecite{Bera1} and \onlinecite{Bera2}, convergence was established only for a
spin-boson model containing a large but finite number of modes, and we demonstrate here
that fast convergence also occurs for a continuous bath spectrum. 

We start by plotting in Figure \ref{f2} the fractional improvement
$(E_M-E_{M-1})/E_{M-1}$ in the minimum energy $E_M$ obtained by adding an extra
coherent state in the Ansatz, as a function of $M$. Note that, as in Sec.~\ref{secen}, 
we do not include the constant $E_0=-\alpha/{2a}$ in the definition
of $E_M$. This convention leads to a denominator in $(E_M-E_{M-1})/E_{M-1}$
that is closer to zero, and therefore to a more stringent measure of
convergence. Figure~\ref{f2} shows results for three points in parameter
space. One of the points, $(\alpha=0.3, \Delta=0.05/a)$, corresponds to a
strongly anisotropic situation where convergence is very rapid. Another curve
$(\alpha=0.6, \Delta=0.05/a)$ corresponds to a less anisotropic situation,
where the convergence is slower. The remaining point $(\alpha=0.85,
\Delta=0.156/a)$ corresponds to the isotropic coupling
$J_\perp=J_\parallel=0.49$, or a Kondo temperature $T_K=\exp(-\pi/J)/a\simeq1.6\times 10^{-3}/a$. 
For $(\alpha=0.3,\Delta=0.05/a)$, the minimum
energy changes by an amount comparable to the accuracy goal of the minimization
module, by the time that $M$ reaches $6$. For the other two points in parameter
space, the marginal change in the minimum energy is a fraction of a percent 
at $M=7$. All the results that we present below are at least as converged as these 
last two cases.

As a complementary test, we can also verify that the multiple coherent 
state $\left|\psi\right>$ converges to
an eigenstate of $H$ as the number of coherent states increases. To do so, we calculate
the energy uncertainty 
\begin{equation}
\Delta E=\sqrt{\left<\psi\right|H^2\left|\psi\right>-\left<\psi\right|H\left|\psi\right>^2},
\end{equation}
for the optimal coherent states Ansatz $\left|\psi\right>$, as a function of $M$.
For this purpose, an expression for the overlap
$\left<f^{(m)}\right|H^2\left|f^{(n)}\right>$ is needed. It turns out that the
same integrals $I^\lambda_{mn}$ as in Sect.~\ref{secen} are involved. In terms
of these integrals, the overlap is expressed analytically as:
\begin{align}
&\left<f^{(m)}\right|H^2\left|f^{(n)}\right>\nonumber\\
&=\left\{\left(\frac{\alpha}{2}\right)^2\left(I_{mn}^0-\frac{1}{a}\right)^2+\left(\frac{\Delta}{2}\right)^2-\frac{\alpha}{2}\partial_aI_{mn}^0\right\}
e^{-\frac{\alpha}{4}I_{mn}^{-1}}\nonumber\\
&+\frac{\alpha\Delta}{8}e^{-\frac{\alpha}{4}I_{mn}^{+1}}\partial_a\left(I_{mn}^{+1}-I_{mn}^{-1}\right).
\end{align}
The rest of the calculation can then proceed with the technology developed in
Sect.~\ref{secen} for evaluating the integrals $I_{mn}^\lambda$. Note that
derivatives $\partial_a I_{mn}^\lambda$, refer to partial derivatives with
respect to the explicit $a$ dependence of the integrals, and not to total
derivatives involving the implicit $a$-dependence of the variational parameters.
The energy uncertainty $\Delta E$ is showed in Figure~\ref{fenvar}, 
as a function of coherent state number $M$, for the same three points in parameter 
space for which we have investigated the energy convergence above. In all three cases 
the energy uncertainty decreases monotonically, and roughly as a power law $\propto
M^{-1.7}$. The fact that the uncertainty clearly tends to zero as $M$ increases,
shows that the trial wavefunction~(\ref{superpose}) converges to the true ground state 
as the number of coherent states is increased.
 
\begin{figure}[t]
\begin{center}
\includegraphics[width=.9\columnwidth]{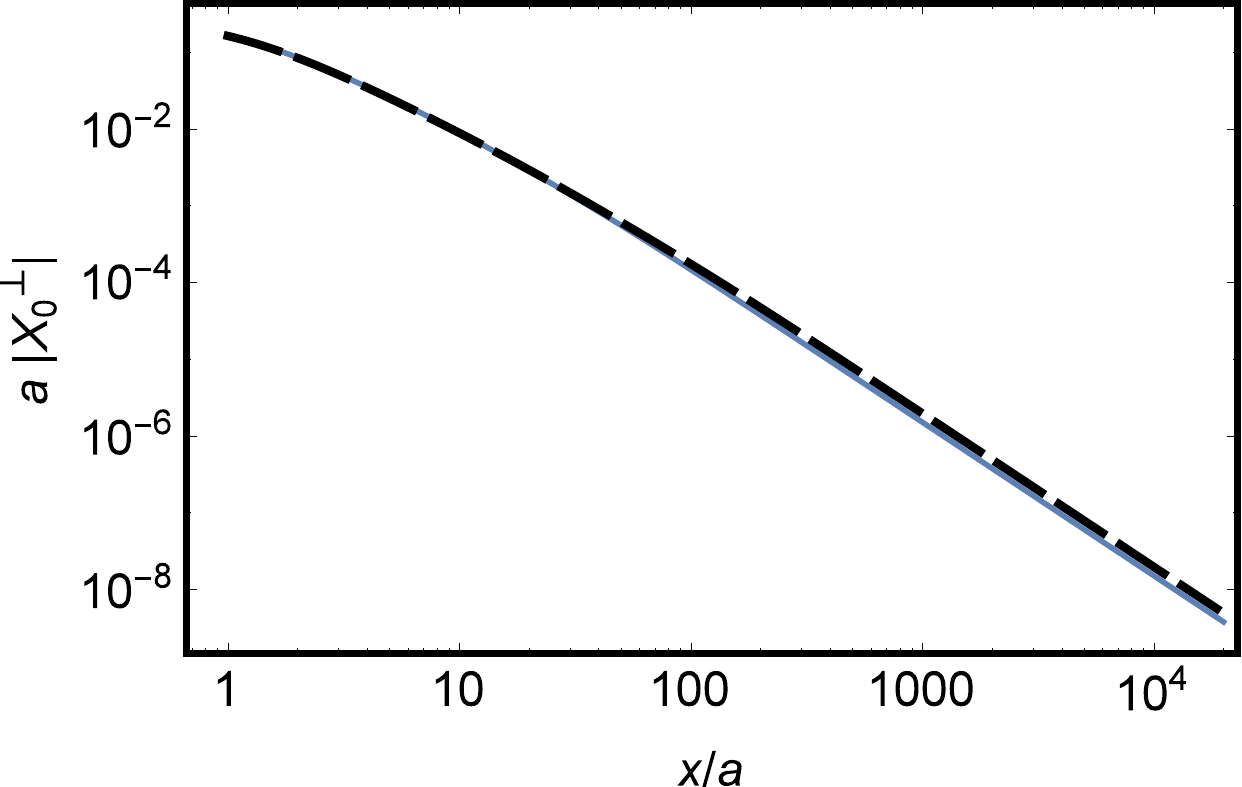}
\caption{(color online) The $0k_F$ transverse component $X_0^\perp$ of the screening cloud,
versus distance $x$ from the impurity, for $\alpha=0.1$, $\Delta=0.05/a$. The
black dashed curve is the (fully converged) result of a $6$ coherent state
calculation. The solid blue curve corresponds to the single coherent state
result (\ref{shcloud}). In the spin-anisotropic limit $\alpha\ll 1$, the
single coherent state approximation is thus very accurate. \label{f3}}
\end{center}
\end{figure}
 
\begin{figure}
\begin{center}
\includegraphics[width=.9\columnwidth]{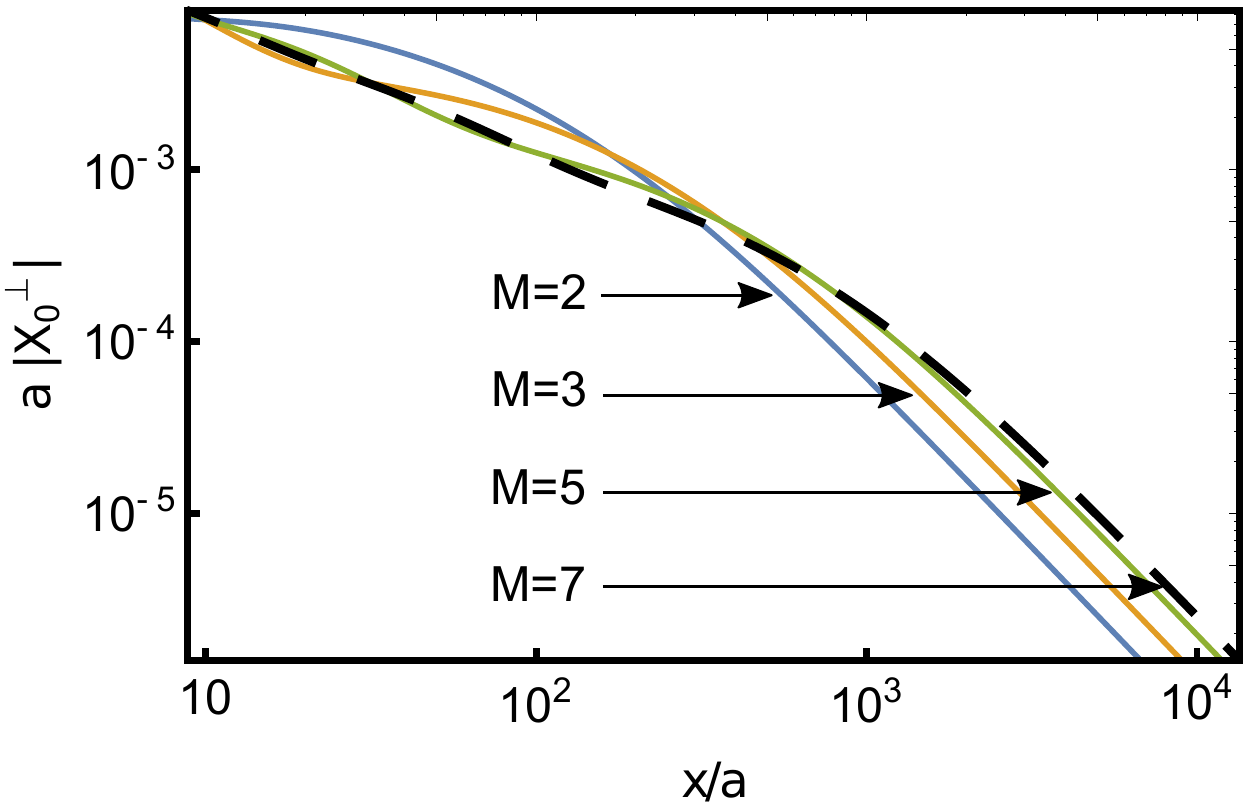}
\caption{(color online) Convergence of the cloud for the $0k_F$ transverse component $X_0^\perp$
versus distance $x$, for $\alpha=0.85$, $\Delta=0.156/a$. This
parameter choice corresponds to an isotropic Kondo coupling
$J_\perp=J_\parallel=0.49$. The black dashed line represents the result of a
$M=7$ coherent state calculation, while solid curves represent results for 
$M=2,\,3$ and $5$ coherent states. \label{f4}}
\end{center}
\end{figure}

Next, we investigate the convergence of the screening cloud itself. In Figure
\ref{f3} we show an instance of very rapid convergence in the case of strong
spin-anisotropy, and in Figure \ref{f4} an example where there is a noticeable 
change between $M=1$ and $M=7$ coherent states (note also that this data set 
is one of the least converged ones included in this work).
In both figures we plot the $0k_F$ transverse correlator $X_0^\perp(x)$, as 
we found this component to show the most dramatic change as $M$ is increased. Figure
\ref{f3} corresponds to  $\alpha=0.1$  and
$\Delta=0.05$ (strong spin-anisotropy), and the comparison to a $M=6$ coherent state calculation shows
that the single coherent state (Silbey-Harris Ansatz) is nearly exact in this limit.
The results in Figure \ref{f4} were obtained at $\alpha=0.85$ and
$\Delta=0.156/a$, which corresponds to an isotropic Kondo coupling
$J_\perp=J_\parallel=0.49$, for $M=2,\,3,\,5$ and $7$ coherent states. 
Good convergence is clearly ensured by the computation with $M=7$ coherent
states.

\subsection{Comparison to exact results based on integrability: ground state 
energy and Kondo overlaps}

In this subsection, we give further evidence that ground state properties of the
Kondo model can be calculated accurately using the coherent state expansion.
In particular, we focus here on several physical quantities that can be computed
exactly via the Bethe Ansatz or related integrability techniques.

\begin{figure}
\begin{center}
\includegraphics[width=\columnwidth]{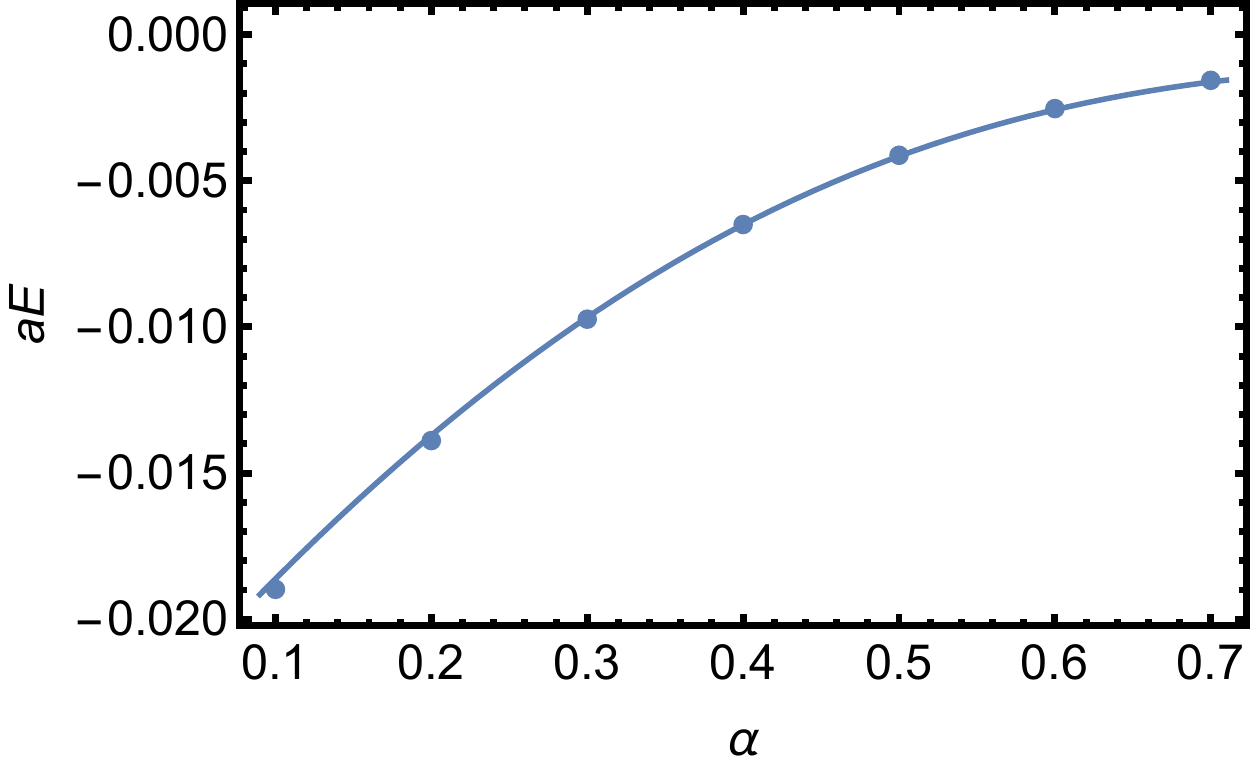}
\caption{(color online) Ground state energy versus $\alpha$ as obtained via the
multiple coherent state approximation (dots) and the exact Bethe Ansatz solution
(solid line), for $\Delta=0.05/a$. \label{f6}}
\end{center}
\end{figure}

In Figure \ref{f6}, we compare the multiple coherent state estimate for the ground state
energy to results obtained via the exact Bethe Ansatz~\cite{hur}, at $\Delta=0.05/a$ and 
various $\alpha$ values. The high energy cutoff $1/a$ corresponds to the
parameter $\omega_c$ in Ref.~\onlinecite{hur}, and we used from this reference 
expression (C.9) for the ground state energy and equation (8) for the Kondo
temperature $T_K$, with the relationship between $D$ and $\omega_c$ given in (C.8). 
Although our calculation is variational, the energy is typically converged
to about 0.1\% (see Figure~\ref{f2}). One has to bear in mind however that,
in the ultraviolet, the model for which the Bethe Ansatz yields the exact solution 
differs from the model we consider here. The Bethe Ansatz result is only valid if all
relevant energy scales in the problem are much smaller than the ultraviolet
cut-off scale. 
For finite $\Delta a$, the Bethe Ansatz expressions even present spurious 
divergences around $\alpha=(2n+1)/(2n+2)$, for $n=1,\, 2,\, \ldots.$
As a result, exact agreement can only be expected at $|Ea|\ll1$, 
and this explains why our variational result is slightly off the Bethe Ansatz 
result for small $\alpha$, although the variational approach presents better 
convergence in this regime.
In general, the numerical data agrees so closely with the Bethe Ansatz result that it is
impossible to distinguish between errors resulting from truncation of the coherent state 
expansion at $M=7$ and errors due to ultraviolet differences between the models. 

\begin{figure}
\begin{center}
\includegraphics[width=\columnwidth]{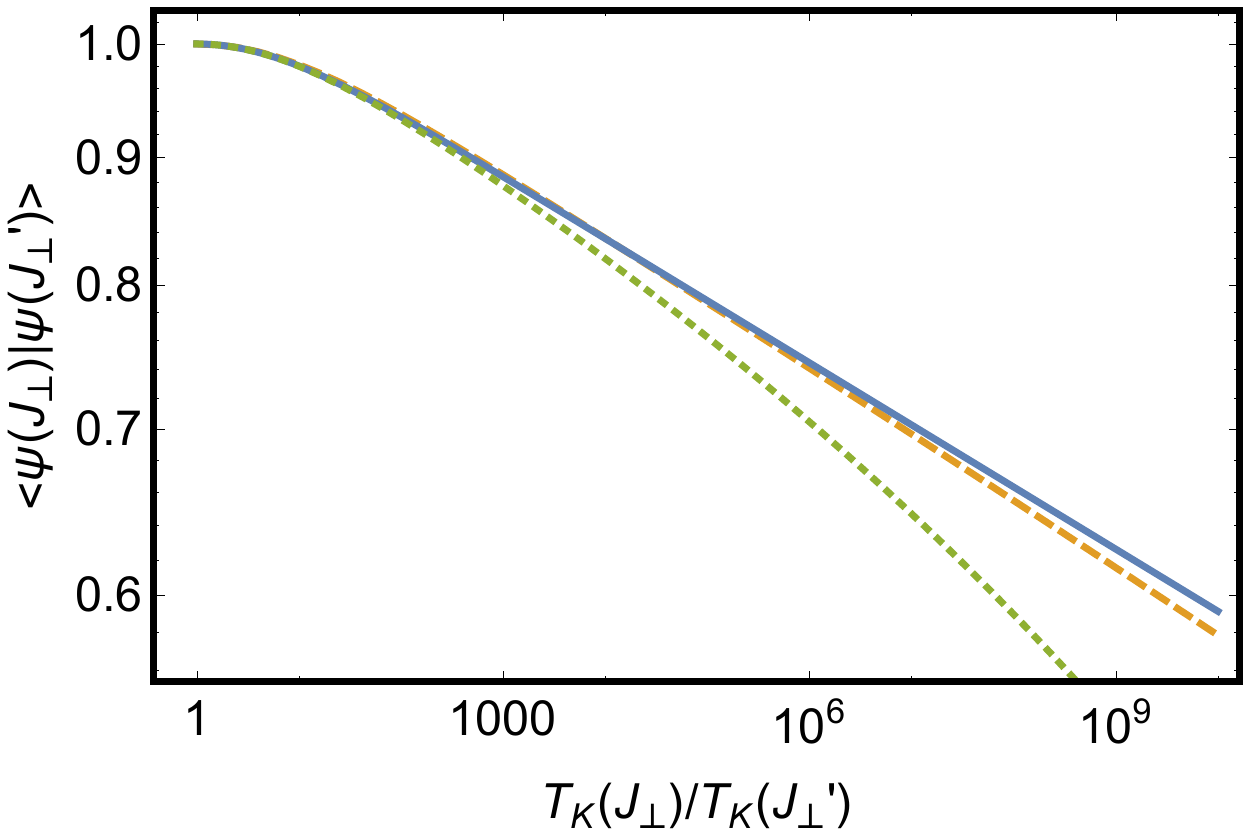}
\caption{(color online) Kondo overlap $\left<\psi(J_\perp)\right|\left.\psi(J_\perp')\right>$ 
between ground states at different and varying values of $J_\perp$, but with the same
fixed value of $J_\parallel$, versus the ratio $T_K(J_\perp)/T_K(J_\perp')$. 
Here, the strongly spin-anisotropic regime is considered with $\alpha=0.1$, i.e. 
$J_\parallel=4.30$. The solid line shows the single coherent state 
formula~(\ref{rpa}), whereas the dashed line is the exact analytical result~(\ref{luk1}). 
The dotted line is the small $\alpha$, small $|z|$ approximation~(\ref{luk2}).\label{f7}}
\end{center}
\end{figure}

\begin{figure}
\begin{center}
\includegraphics[width=\columnwidth]{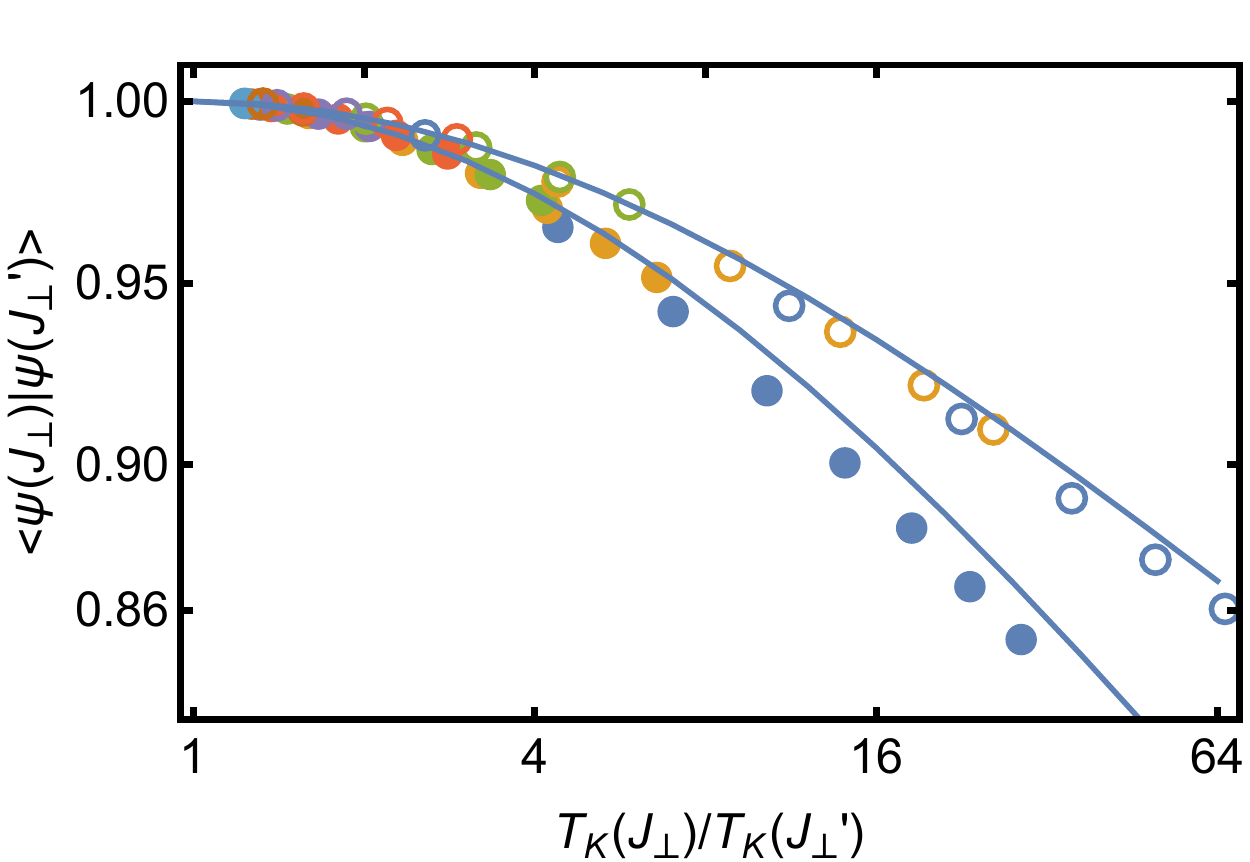}
\caption{(color online) Kondo overlap $\left<\psi(J_\perp)\right|\left.\psi(J_\perp')\right>$ at
intermediate $\alpha$ values. Open symbols in various colors show the $M=6$ coherent state
results for $J_\parallel=0.905$ (i.e. $\alpha=0.3$) for several
values of $J_\perp$ and $J_\perp'\in[0.016,0.346]$ (i.e. $\Delta\in[0.005/a,0.110/a]$). 
Closed symbols show the $M=7$ coherent state results
for $J_\parallel=0.451$ (i.e. $\alpha=0.6$) for several values of 
$J_\perp$ and $J_\perp'\in[0.016,0.346]$ (i.e. $\Delta\in[0.236/a,0.785/a]$). 
Solid curves represent the exact analytical result (\ref{luk1}). 
The Kondo temperature was taken from the approximate
identification $T_K=1/\xi_\parallel$.
\label{f7b}}
\end{center}
\end{figure}

\begin{figure}
\begin{center}
\includegraphics[width=\columnwidth]{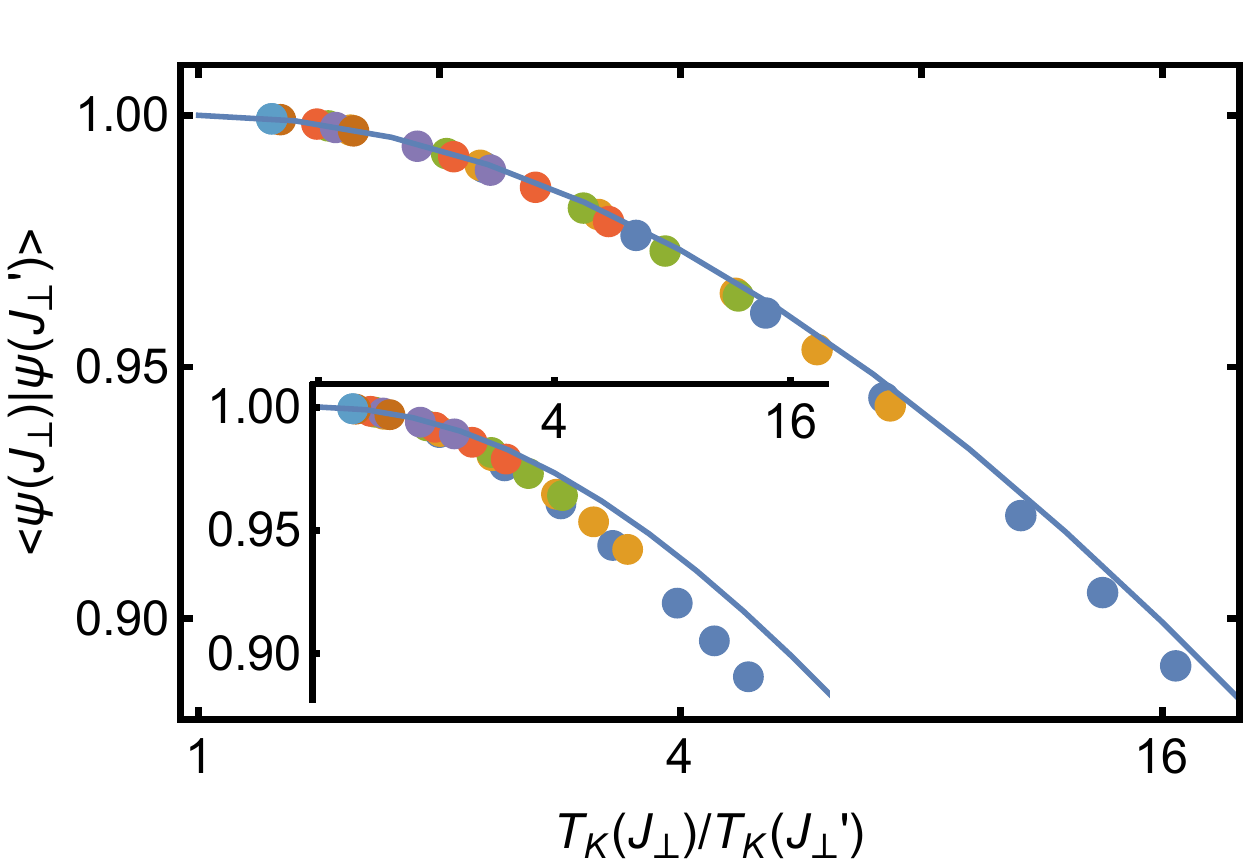}
\caption{(color online) Kondo overlap $\left<\psi(J_\perp)\right|\left.\psi(J_\perp')\right>$ at
$\alpha=0.8$ ($J_\parallel=0.663$). Open symbols in various colors show the
$M=7$ coherent state results for several values of $J_\perp$ and $J_\perp'\in[0.377,0.880]$ 
(i.e. $\Delta\in[0.12/a,0.28/a]$). 
Solid curves represent the exact analytical result (\ref{luk1}). In the main panel, the 
improved renormalization group estimate (\ref{tkren}) was used for $T_K$. In the
inset, the less accurate value $T_K=1/\xi_\parallel$ was used.
\label{f7c}}
\end{center}
\end{figure}

Another interesting quantity for which an exact analytical result has recently been
obtained~\cite{luk}, and which gives some indirect information on the Kondo
cloud, is the overlap $\left<\psi(J_\perp)\right|\left.\psi(J_\perp
')\right>$ of two Kondo ground states with different transverse exchange
couplings, as a function of the Kondo temperature ratio $T_K(J_\perp)/T_K(J_\perp')$. 
Here $\left|\psi(J_\perp)\right>$ and $\left|\psi(J_\perp')\right>$ denote the
full many-body ground states obtained at different $J_\perp$ but for the same 
$J_\parallel$, while $T_K(J_\perp)$ and $T_K(J_\perp')$ are the associated Kondo 
temperatures. With the definition $z={\rm ln}[T_K(J_\perp')/T_K(J_\perp)]$, the exact 
result reads:
\begin{align}
&\left<\psi(J_\perp)\right|\left.\psi(J_\perp ')\right>=\frac{1}{1-\alpha}
\frac{\sinh[(1-\alpha)z/2]}{\sinh(z/2)}g_\alpha(z),\nonumber\\
&g_\alpha(z)=\exp\left[\int_0^\infty\frac{dt}{t}\frac{\sin^2(zt/\pi)}{\sinh(2t)\cosh(t)}
\frac{\sinh\left(\frac{\alpha t}{1-\alpha}\right)}{\sinh\left(\frac{t}{1-\alpha}\right)}\right].\label{luk1}
\end{align}
When comparing our numerical results to this analytical formula, a subtle issue
arises. The Kondo temperature is certainly related to the inverse of the size of
the screening cloud, but the exact relation may well involve an $\mathcal O(1)$
factor that is $J_\perp$ dependent. We have therefore tried various definitions
of the Kondo temperature. For small to moderate $\alpha$, we find that
$T_K\propto 1/\xi_\parallel$, as defined in Eq.~(\ref{asymptotics}), with a 
$J_\perp$-independent proportionality constant, works well. 
Indeed, in the Silbey-Harris regime, this correspondence is exact (see below). 
For larger $\alpha$ however, this definition seems to
incur a systematic error. For $J_\parallel$ and $J_\perp$ relatively small, as
is exemplified by the data we collected at $\alpha=0.8$, we have found it better
to estimate the Kondo scale from direct integration of the standard
weak-coupling in $J_\perp$ and $J_\parallel$ poor man's scaling equations:
\begin{equation}
\frac{dJ_\perp}{dl}=\frac{1}{\pi} J_\parallel J_\perp,~~~
\frac{dJ_\parallel}{dl}=\frac{1}{\pi} J_\perp^2,
\label{poorfloweq}
\end{equation} 
which lead to an expression of the Kondo scale for the
spin-anisotropic Kondo model that is valid for $J_\parallel\ll1$ and 
$J_\perp\ll1$:
\begin{equation}
T_K=\frac{1}{a} \exp\left[-\frac{\pi}{\sqrt{J_\perp^2-J_\parallel^2}}\arctan\left(\frac{\sqrt{J_\perp^2-J_\parallel^2}}{J_\parallel}\right)\right].\label{tkren}
\end{equation} 

Let us first consider the regime of small $\alpha$. Here we have seen that
single coherent state results are already well-converged. In the single coherent state
approximation, the overlap is given by:
\begin{align}
&\left<\psi(J_\perp)\right|\left.\psi(J_\perp ')\right>
=\exp\frac{\alpha}{2}\left[1-\frac{\bar z}{2}\coth(\bar z/2)\right],\label{rpa}
\end{align}
where $\bar z=\ln\,[\Delta_R(J_\perp')/\Delta_R(J_\perp)]$.
For small $\alpha$, Ref.~\onlinecite{luk} quotes the ``semi-classical'' result, 
\begin{align}
&\left<\psi(J_\perp)\right|\left.\psi(J_\perp ')\right>
=1+\frac{\alpha}{2}\left[1-\frac{z}{2}\coth(z/2)\right],\label{luk2}
\end{align}
which can be obtained by expanding the exact result (\ref{luk1})
to first order in $\alpha$. Since $z\coth(z/2)$ grows linearly in $z$ for large $z$, the semi-classical result
can at best only be valid for $z$ sufficiently smaller than $1/\alpha$.
Referring back to the single coherent state result (\ref{shxi})
for the correlation lengths, we see that if we make the identification 
$T_K\propto 1/\xi_\parallel$, the single coherent state approximation 
(\ref{rpa}) is nothing but a resummation of the small $\alpha$ result 
in Ref.~\onlinecite{luk}, in which $\ln \left<\psi(J_\perp)\right|\left.\psi(J_\perp ')\right>$
is calculated to second order in the impurity interaction, and then exponentiated.
In Figure~\ref{f7} we compare the single coherent state approximation to the 
exact result (\ref{luk1}). we see that, unlike the semiclassical formula~(\ref{luk2}), 
the single coherent state approximation remains valid up to large ratios of the 
Kondo temperatures, because the ground state is well captured for any value of
the Kondo temperature (provided $\alpha$ is small enough).

We now turn to the regime of larger dissipation, and in Figure \ref{f7b} we show 
multiple coherent state results for the Kondo overlap at $\alpha=0.3$ and 
$\alpha=0.6$, again using the (now approximate) identification 
$T_K\propto 1/\xi_\parallel$, together with the exact result (\ref{luk1}). Data points of
the same color were obtained by keeping $J_\perp$ fixed and varying
$J_\perp'<J_\perp$. If two data points have different colors, they correspond to
distinct $J_\perp$ {\em and} $J_\perp'$. It is therefore already nontrivial that
the differently colored points fall on the same curve, 
confirming the universality predicted by the exact analytical expression~(\ref{luk1}). Generally, 
good agreement with the analytical result (\ref{luk1}) is seen, with only a small 
systematic error pushing the numerical curves slightly below the analytical
ones, which increases as $\alpha$ increases. 
A small part of the error is likely due to an error of a few percent in the numerical
estimation of $\xi_\parallel$, and to $T_K$ being only an order of magnitude or
so less than the ultraviolet scale $1/a$ at large $J_\perp$. These errors would
be reduced if we could use more coherent states and smaller $J_\perp$. However, as
pointed out earlier, the main discrepancy comes from the chosen definition of 
the Kondo temperature in Ref.\,\onlinecite{luk}, which is not exactly equivalent to
$1/\xi_\parallel$ at larger $\alpha$. Indeed, Figure \ref{f7c} shows the multiple coherent 
state results for the overlap at $\alpha=0.8$, using both the naive identification
$T_K\propto 1/\xi_\parallel$ (inset) and the renormalization improved 
estimate~(\ref{tkren}). The latter choice leads clearly to a substantial reduction of 
the error. 
Our conclusion is that the actual Kondo overlap 
$\left<\psi(J_\perp)\right|\left.\psi(J_\perp ')\right>$ is calculated quite
accurately in the coherent state expansion, and that the (modest) observed errors are 
associated with the extraction procedure of the Kondo temperature from the size of 
the cloud.

\subsection{Comparison to the exact longitudinal forward-component of the
cloud at the Toulouse point}

\begin{figure}
\begin{center}
\includegraphics[width=\columnwidth]{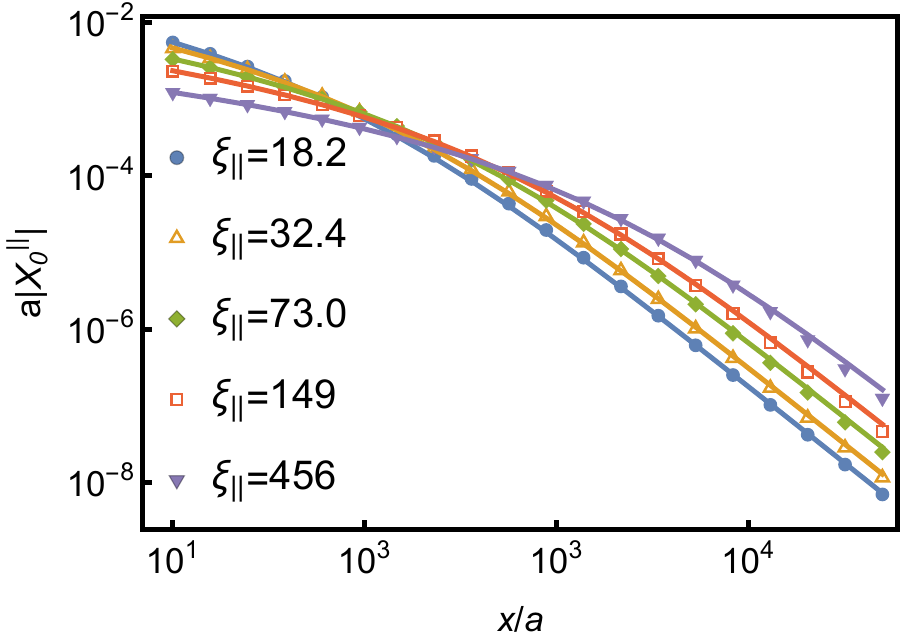}
\caption{(color online) The longitudinal $0k_F$ component of the screening cloud at the
Toulouse point $\alpha=1/2$. Five different values of $\Delta$ ranging from $0.02/a$ to
$0.1/a$ were used. The symbols correspond to results obtained with the coherent
state expansion (truncated at $M=7$ terms) for a choice of discrete
positions, while the solid lines represent the exact result (\ref{xtoul}). 
Also indicated for each curve is the associated Kondo length $\xi_\parallel$, ranging 
from $18.2 a$ at $\Delta=0.1/a$ to $456 a$ at $\Delta=0.02/a$.\label{ftoul}}
\end{center}
\end{figure}

In this subsection provide strong evidence that the coherent state
expansion~(\ref{superpose}) captures the full spatial structure
of the Kondo cloud, using an exact analytical result at the so-called Toulouse
point. Indeed, at $J_\parallel=2\pi(1-1/\sqrt{2})$ or equivalently, $\alpha=1/2$,
the Kondo model is equivalent to a fermionic non-interacting resonant level 
model [\onlinecite{Zarand}].  An exact result for
the longitudinal $0k_F$ component of the screening cloud can be obtained
following the route set out in Refs. \onlinecite{Posske1} and
\onlinecite{Posske2}. We review this calculation in Appendix \ref{appd}. The
result is given by the simple formula
\begin{equation}
X_0^\parallel(x)=-\frac{\sqrt{2}}{\pi^2}\frac{J_\perp^2}{4\pi a}F\left(-\frac{J_\perp^2|x|}{4\pi a}\right)^2,\label{xtoul}
\end{equation}
with $F$ as defined in Eq.~(\ref{efl}). From the point of view of the coherent
state expansion, there is nothing special about the Toulouse point. In the
coherent state expansion, $X_0^\parallel$ is expressed as a linear combination
of $F$ functions with different position dependent arguments, cf. Eq.~(\ref{abcd}), 
and an infinite number of terms are required to approximate $F^2$
exactly. In Figure \ref{ftoul}, we compare the exact expression to results
obtained with the coherent state expansion, truncated at $M=7$ terms. For
clarity, we show here results of the coherent state expansion only for a discrete 
set of $x$ values, because the complete curves would have completely covered the exact 
result.
We clearly find near perfect agreement between the coherent state expansion and the
exact result. Since the coherent state expansion does not exploit any special
features of Toulouse point, we expect similar accuracy at a similar cost ($M=7$)
to be achievable at other values of $\alpha$. 

\subsection{Detailed analysis of the screening cloud}

Having thoroughly established the accuracy of the coherent state approximation
for the Kondo ground state, we now proceed to investigate the physical features 
of the Kondo screening cloud, both for isotropic and anisotropic regimes.

\subsubsection{Summary of known results}

We first briefly recall the available analytical results regarding the
isotropic screening cloud. 
In the isotropic case ($J_\perp=J_\parallel$), the transverse components of the
screening cloud equal twice the longitudinal components, i.e.
\begin{equation}
X_0^\parallel=X_0^\perp/2\equiv X_0,~~X_{2k_F}^\parallel=X_{2k_F}^\perp/2\equiv X_{2k_F}.
\end{equation}
For $x\gg a$, both $X_0(x)$ and $X_{2k_F}(x)$ are expected to be universal scaling 
functions
\begin{equation}
\xi X_k(x)=\tilde X_k(\tilde x),~~\tilde x=x/\xi,~~k=0,\,2k_F.\label{eqisoscaling}
\end{equation}
Here $\tilde X_k$ is independent of $J$, and all parameter dependence is
contained in the Kondo length $\xi$. The Kondo length is expected to be
inversely proportional to the Kondo temperature, but the exact relation may
contain a $J$-dependent proportionality factor of order unity, as we
discussed previously. The following asymptotic results for $\tilde X_0$ and 
$\tilde X_{2kF}$ have been derived analytically~\cite{Affleck1}:
\begin{align}
&\tilde X_0(\tilde x)\propto-\frac{1}{\tilde x(\ln \tilde x)^2},
~~\tilde X_{2k_F}(\tilde x)\propto\frac{1}{\tilde x\ln \tilde x},
~~~\mathrm{for}~\tilde x\ll 1\nonumber\\
&\tilde X_0(\tilde x)\simeq\tilde X_{2k_F}(\tilde x)\simeq 
-\frac{1}{\tilde x^2}~~~
\mathrm{for}~\tilde x\gg 1.\label{eqisoasym}
\end{align}
The regime of small $\tilde x$ is perturbatively accessible with a calculation using 
renormalization group techniques, while the large $\tilde x$ regime is treated using
Fermi liquid theory.
Note that at small $\tilde x$, the $2k_F$-oscillatory component of the cloud
dominates the $0k_F$ component slightly. This implies that the total
correlation function $X(x) = X_0(x)+\cos(2k_Fx) X_{2kF}$ oscillates between positive
and negative values, with wavelength $\pi/k_F$ at small $x$. In other words,
close to the impurity, the spin correlations between the electron gas and the
impurity alternate between being ferromagnetic and being antiferromagnetic, on
the scale $\pi/k_F$. The crossover from slower than $1/\tilde x$ decay inside
the cloud to $1/\tilde x^2$ decay at large $x$ is expected to occur at $x\sim
\xi$, i.e. $\tilde x\sim 1$. Our goal in the next paragraph is to confirm these
results for the isotropic cloud, and to compute the full crossover curve from
the coherent state expansion. In a second step, we will examine how the cloud
correlations change when the Kondo couplings are not isotropic 
($J_\perp\not = J_\parallel$). 

\subsubsection{Kondo cloud in the isotropic case}
\label{secisocase}

\begin{figure}
\begin{center}
\includegraphics[width=\columnwidth]{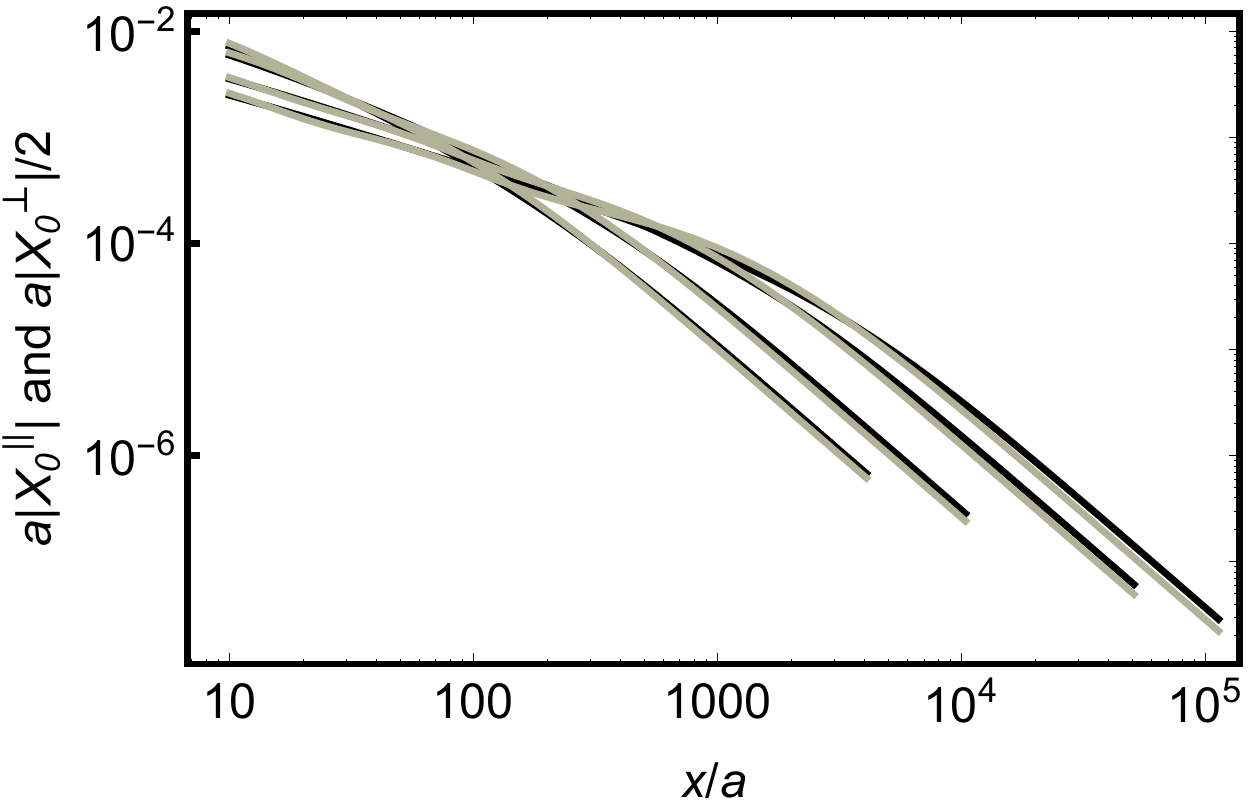}\\
\includegraphics[width=\columnwidth]{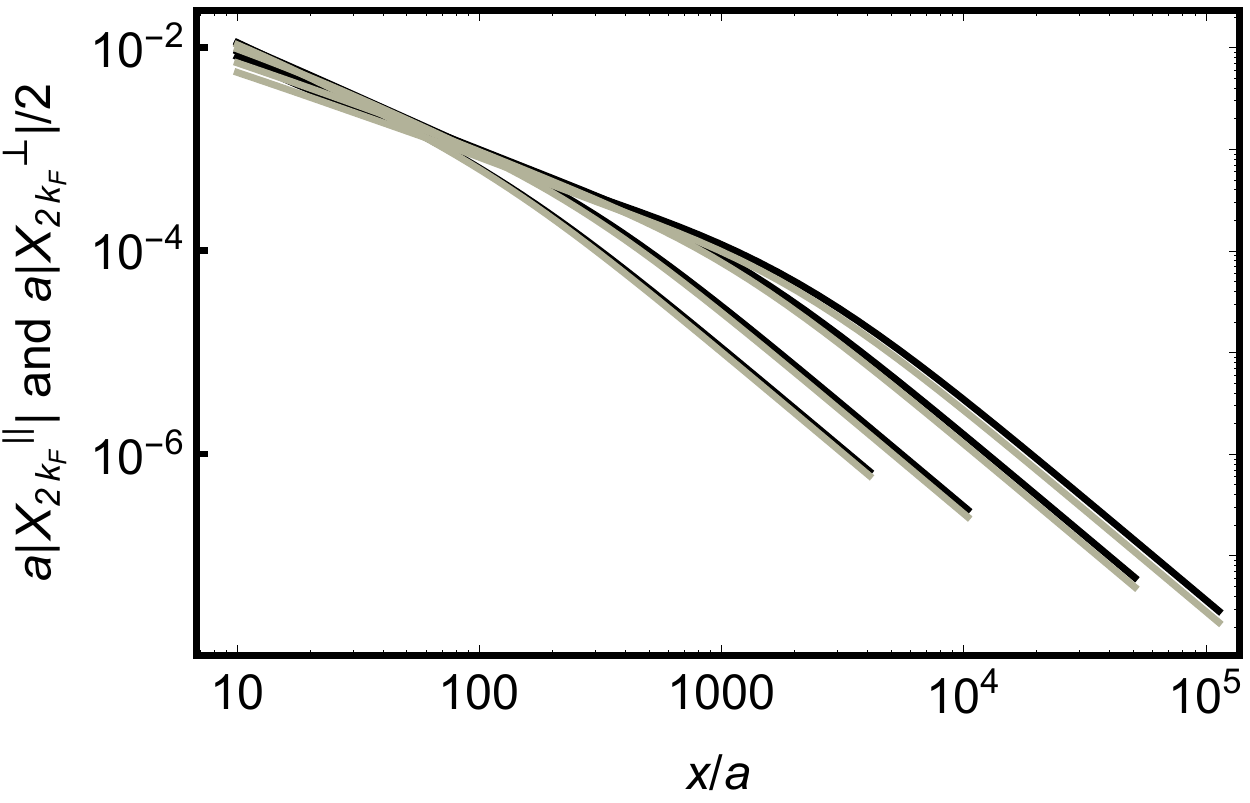}
\caption{The four components of the isotropic screening cloud: the top panel 
represents the longitudinal (black) and transverse (gray) $0k_F$ components, while the bottom
panel represents the longitudinal (black) and transverse (gray) $2k_F$ components.
The calculation was performed for the choice of Kondo couplings
$J=0.42,\,0.49,\,0.66,$ and $0.84$, which correspond to 
$\alpha=0.87,\,0.85,\,0.8,$ and $0.75$.
The various curves can be identified from the fact that the crossover to 
faster $1/x^2$ decay occurs on an increasing length scale $\xi$ as $\alpha$ is
increased (or equivalently $J$ is decreased).
For $\alpha=0.75$ convergence was achieved with $M=5$ coherent states, while for the 
other values of $\alpha$, $M=7$ was required.\label{fi1}}
\end{center}
\end{figure}

As a first step, we present results in Figure~\ref{fi1} for the screening cloud
calculated for several parameters on the isotropic line $J_\perp=J_\parallel$, or
equivalently $\alpha=(1-a\Delta/2)^2$. The correlation functions are plotted in
units of $1/a$ on the vertical axis and in units of $a$ on the horizontal axis.
When considering the small $\tilde x$ regime numerically, one must remember that
$x$ has to remain sufficiently larger than the short distance ultraviolet
cut-off $a$, since the limits $a\to 0$ and $x\to 0$ do not commute. This
restriction is implied whenever we consider the asymptotic $\tilde x \to 0$
limit. 
The transverse and longitudinal components are plotted in the same panel,
and we expect $X_k^\parallel=X_k^\perp/2$ for both the $k=0k_F$ and $k=2k_F$
components due to strict spin-isotropy. The numerical results of Figure~\ref{fi1} 
reveal this isotropy to a high degree, and this is a non-trivial check of our
method, since rotational symmetry in the bosonic model is only emergent. It 
is violated by the ultraviolet regularization (see Appendix~\ref{appmap}). 
This fact also explains the small differences that are still visible between the 
transverse and longitudinal components. 

\begin{figure}
\begin{center}
\includegraphics[width=\columnwidth]{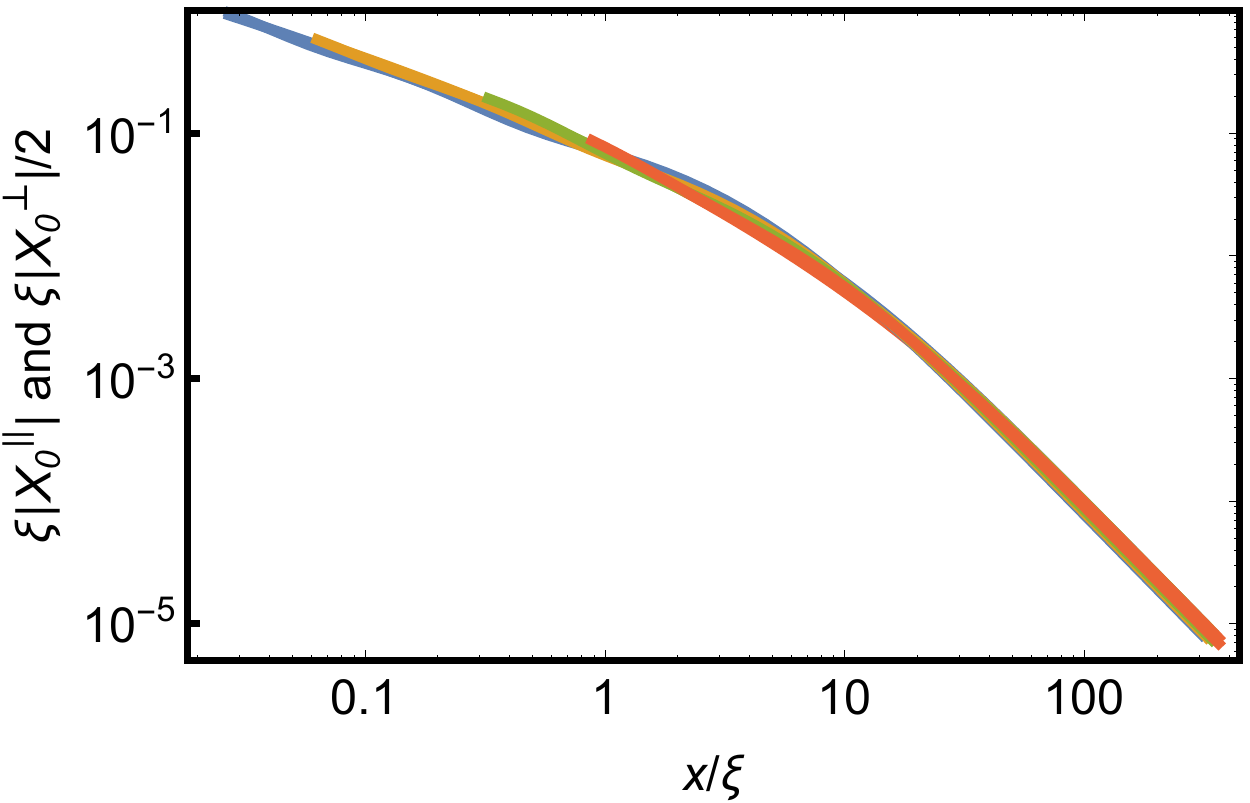}\\
\includegraphics[width=\columnwidth]{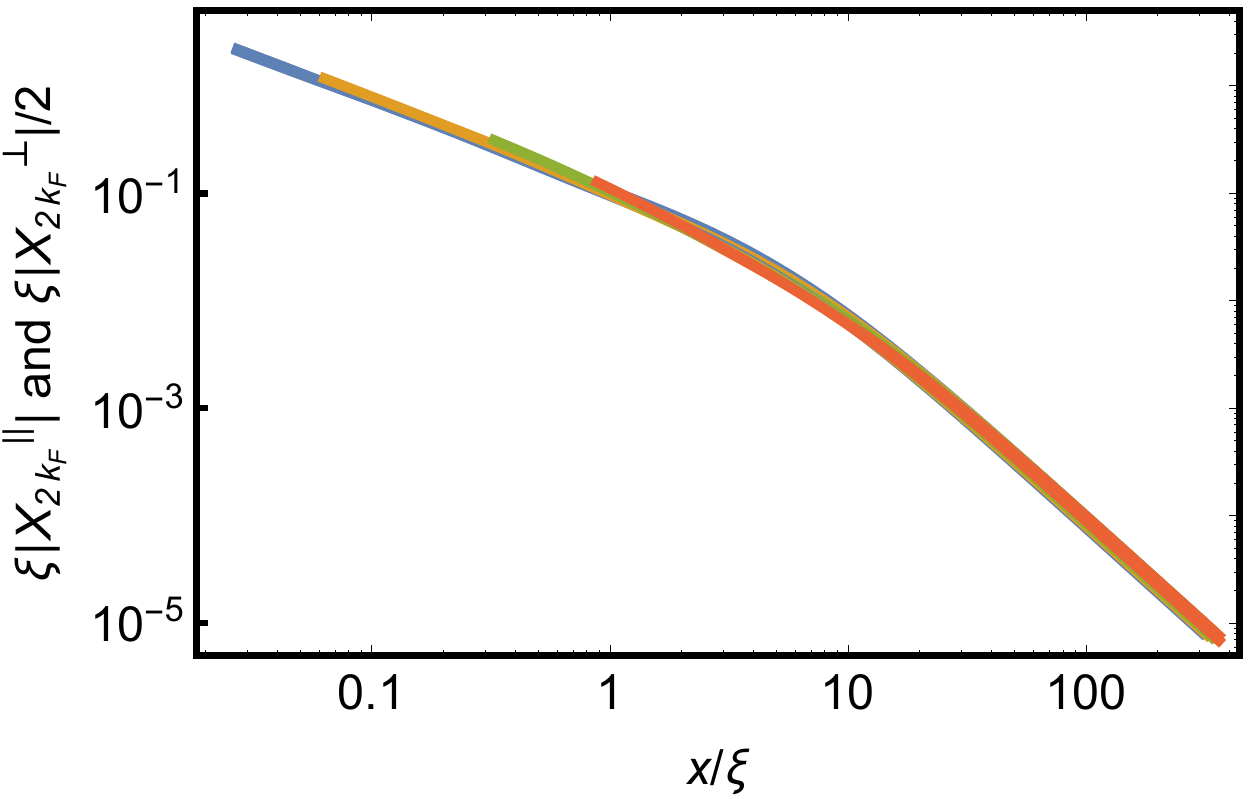}
\caption{(color online) The isotropic correlation functions of Figure \ref{fi1}, in rescaled
units $1/\xi$ on the vertical axis and $\xi$ on the horizontal axis. The
Kondo length $\xi=\xi_\parallel$ was estimated from Eq.~(\ref{ea1}).\label{fi2}}
\end{center}
\end{figure}

In Figure~\ref{fi2} we plot the same data as in Figure~\ref{fi1}, but now in
rescaled units, according to (\ref{eqisoscaling}). We ignored small differences
between the large distance behavior of the transverse and longitudinal
components, and scaled all components with $\xi=\xi_\parallel$, where the Kondo length
$\xi_\parallel$ was calculated using Eq.~(\ref{ea1}). We clearly see that it is
possible to scale correlation functions calculated at different $J$ onto
universal curves. As expected, we observe that the cross-over from slower than
$1/x$ decay inside the cloud to $1/x^2$ decay at large $x$ occurs around
$x\sim\xi$. The precise behavior of the Kondo length $\xi$ as a function of $J$ will be
analyzed further in the next subsection. At this point we note that it varies
from $11 a$ at $J=0.84$ to $366 a$ at $J=0.42$. Such a large variation in the
spatial scale implies that the observed scaling is non-trivial and
reflects the universality of the Kondo problem.

\begin{figure}[t]
\begin{center}
\includegraphics[width=\columnwidth]{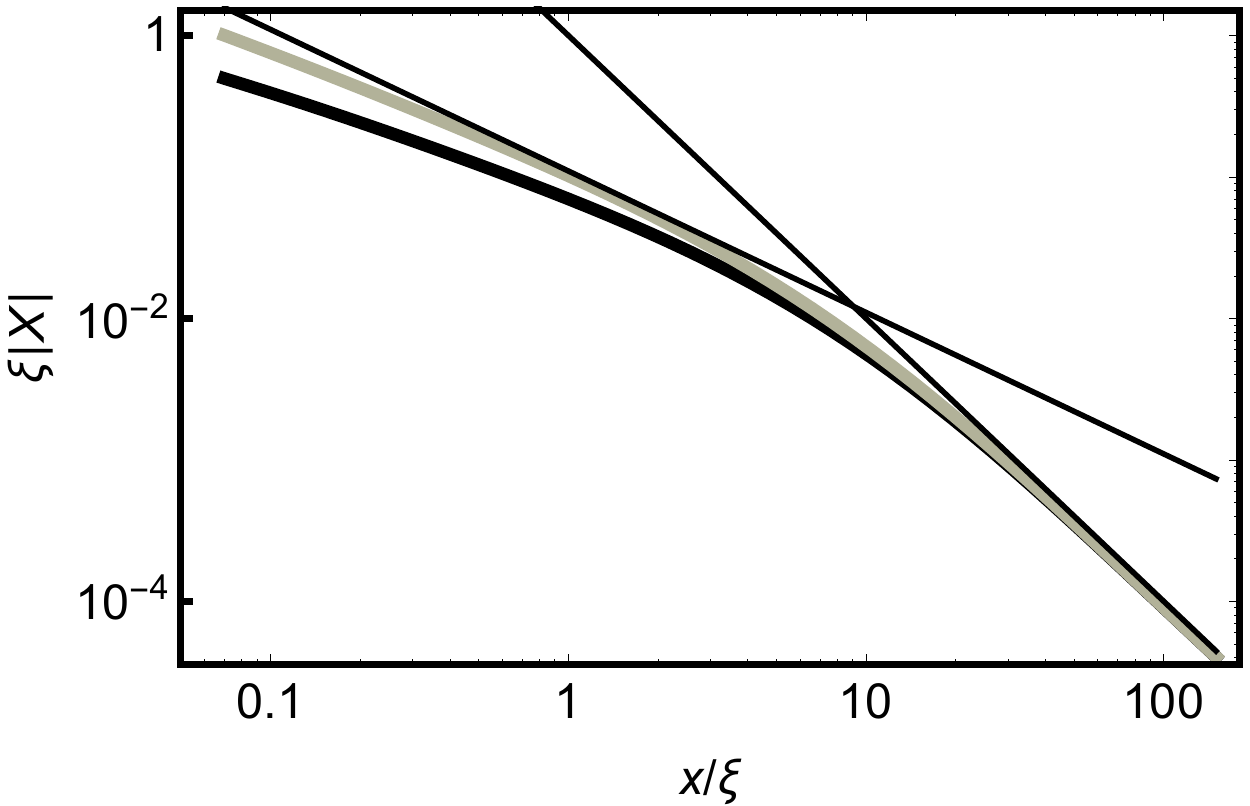}
\caption{Universal scaling curves on the isotropic line: the black curve
represents the longitudinal component $\tilde X_0$, and the gray curve
represents the transverse component $\tilde X_{2k_F}$. These two single 
curves were obtained by fitting high order polynomials through the rescaled data 
set of Figure~\ref{fi2}. Thinner straight lines indicate the pure power 
laws $X(x)\propto1/x$ and $X(x)\propto 1/x^2$. \label{fi3}}
\end{center}
\end{figure}

In Figure~\ref{fi3} we compare the universal scaling functions for the $0k_F$
and the $2k_F$ components of the cloud, by plotting them on top of each other.
We determined the single universal scaling functions by fitting high order polynomials
through the scaled data set of Figure~\ref{fi2}. We clearly see that 
the $2k_F$ component dominates the $0k_F$ component at small $\tilde x$, 
consistent with the known small $x$ asymptotics of Eq.~\ref{eqisoasym}.

\subsubsection{Kondo cloud in the anisotropic case}

\begin{figure}[t]
\begin{center}
\includegraphics[width=\columnwidth]{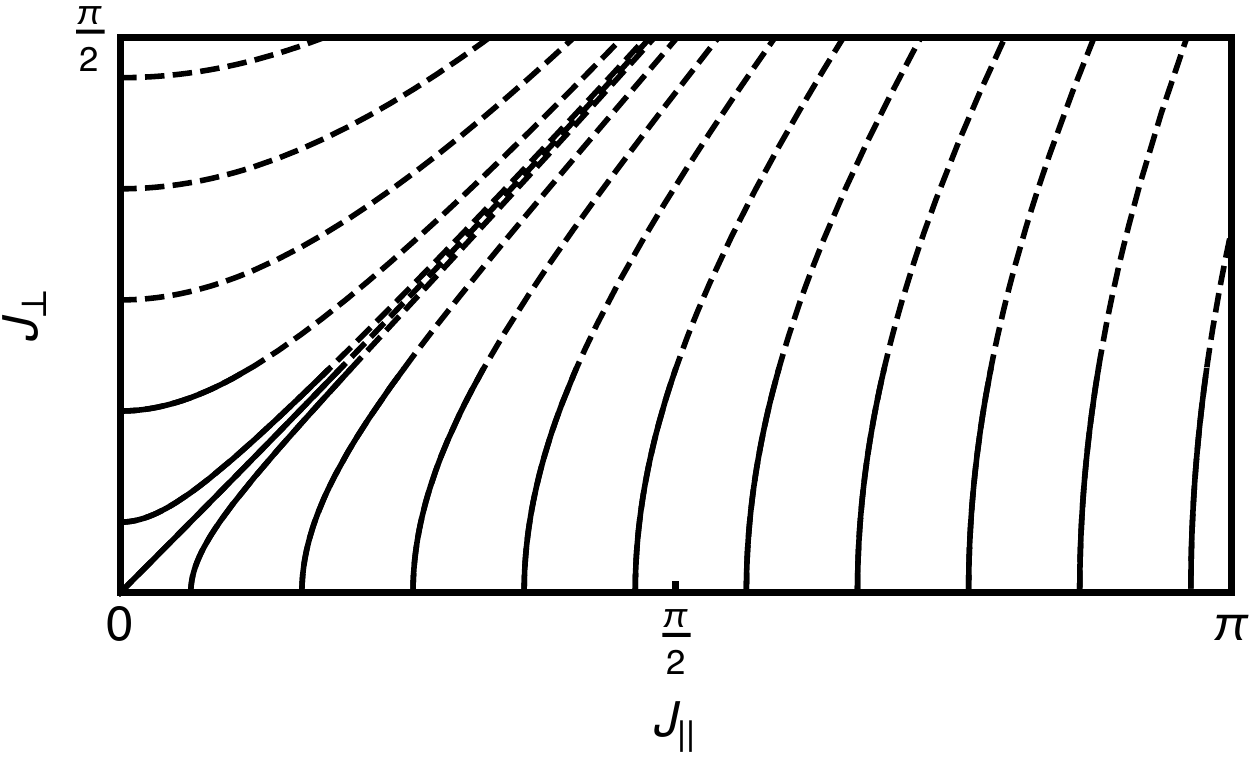}
\caption{Scaling trajectories that derive from the improved poor man's scaling equations
(\ref{floweq}), that are correct to leading order in $J_\perp$. 
In order to emphasize that only the small $J_\perp$ part of trajectories are 
to be trusted, dashed lines are used for $J_\perp/2\pi>0.1$.
 \label{figflow}}
\end{center}
\end{figure}

Having confirmed that the cloud displays the expected universal scaling for
isotropic couplings, we move on to the general anisotropic case. The
existence of two independent couplings $J_\parallel$ and $J_\perp$, or
equivalently $\alpha$ and $\Delta$, implies that the universal scaling picture
is less straightforward. To guide our investigation, let us review known
results obtained by an improved poor man's scaling argument, applicable
with greater generality in the anisotropic case. (Standard poor man's 
equations~(\ref{poorfloweq}) can only be trusted when both Kondo exchange
couplings are small.)\cite{Chakravarty} 
In the language of the spin-boson model, it is known that to leading order 
in $\Delta$, but arbitrary $\alpha$, increasing the short-distance scale $a$ by $a\,dl$ is
approximately equivalent to changing $\alpha$ to $\alpha+d\alpha$ and $\Delta$ to $\Delta+d\Delta$,
where 
\begin{equation}
\frac{d\alpha}{dl}=-\left(a\Delta\right)^2\alpha,~~~\frac{d}{dl}(a\Delta)=(1-\alpha)(a\Delta).
\label{floweq}
\end{equation} 
Integration of these flow equations yields scaling trajectories
\begin{equation}
\frac{(a\Delta)^2}{2}-\alpha+\ln(\alpha)=\mbox{constant},
\end{equation}
or equivalently, in the language of the Kondo model
\begin{align}
J_\perp^2&-J_\parallel^2+(2\pi)^2\left[\ln\left(1-\frac{J_\parallel}{2\pi}\right)+\frac{J_\parallel}{2\pi}+\frac{1}{2}\left(\frac{J_\parallel}{2\pi}\right)^2\right]\nonumber\\
&=\mbox{constant}.
\end{align}
A few of these trajectories in the $J_\parallel$ - $J_\perp$ plane are plotted
in Figure \ref{figflow}. We stress that the standard weak coupling
renormalization equations~(\ref{poorfloweq}) cannot be trusted in the regime
where $J_\parallel\gtrsim1$, and that is why we have to work
non-perturbatively in $\alpha$.
It is also important to remember that these trajectories are
only meaningful at $J_\perp$ sufficiently smaller than $\pi$. Always bearing
this proviso in mind, the following statement holds: screening clouds associated
with two Kondo Hamiltonians, whose parameters lie on the same scaling
trajectory, can be scaled onto the same universal line shape.

\begin{figure}[t]
\begin{center}
\includegraphics[width=\columnwidth]{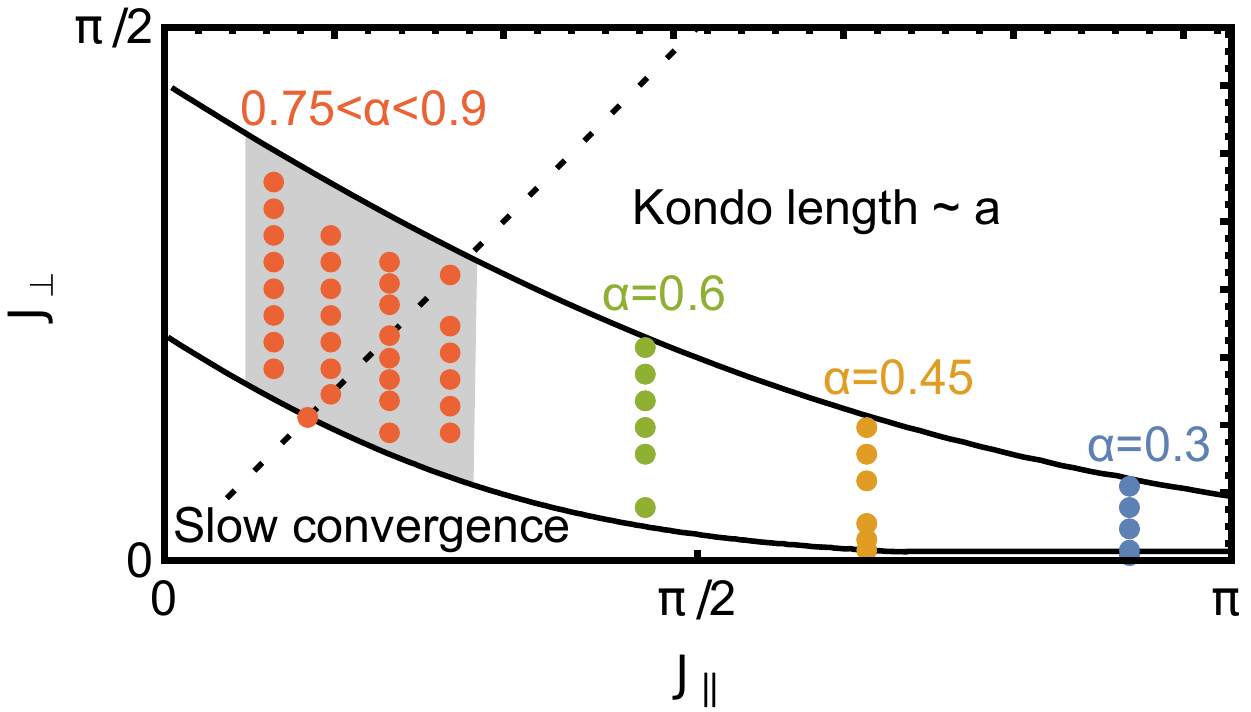}
\caption{(color online) Map of the parameter space of the Kondo model, where dots indicate points 
where we have collected data to investigate the scaling behavior of the 
screening cloud. 
Below the lower solid curve, $T_K\lesssim 10^{-3}/a$ and we generally find that more than $M=7$ coherent
states are required for a converged result. Above the upper solid curve, $T_K\gtrsim 1/a$ and
the screening cloud is poorly resolved because of non-universal ultraviolet effects.
At the points included in the shaded region, we find that the screening cloud is nearly isotropic down to distances deep inside the cloud.\label{fa1}}
\end{center}
\end{figure}

\begin{figure*}[t]
\begin{center}
\includegraphics[width=.95\columnwidth]{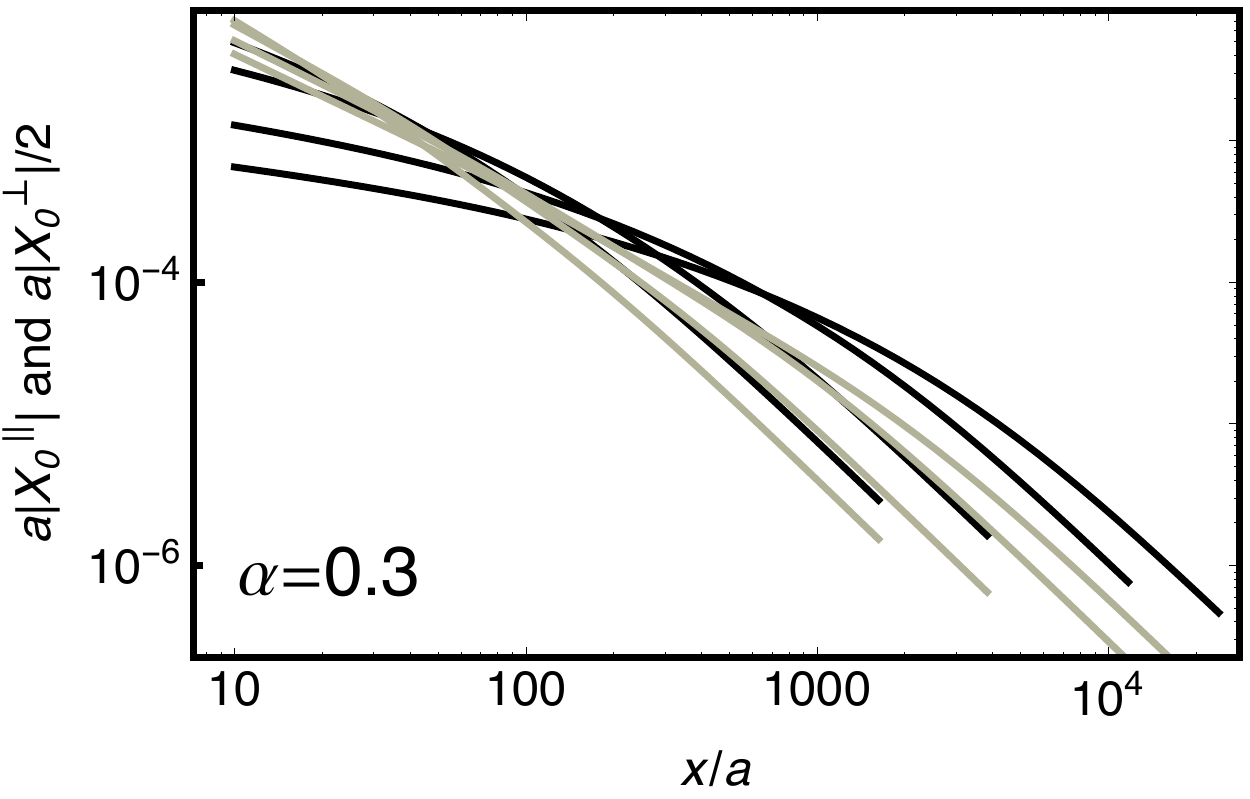}~~
\includegraphics[width=.95\columnwidth]{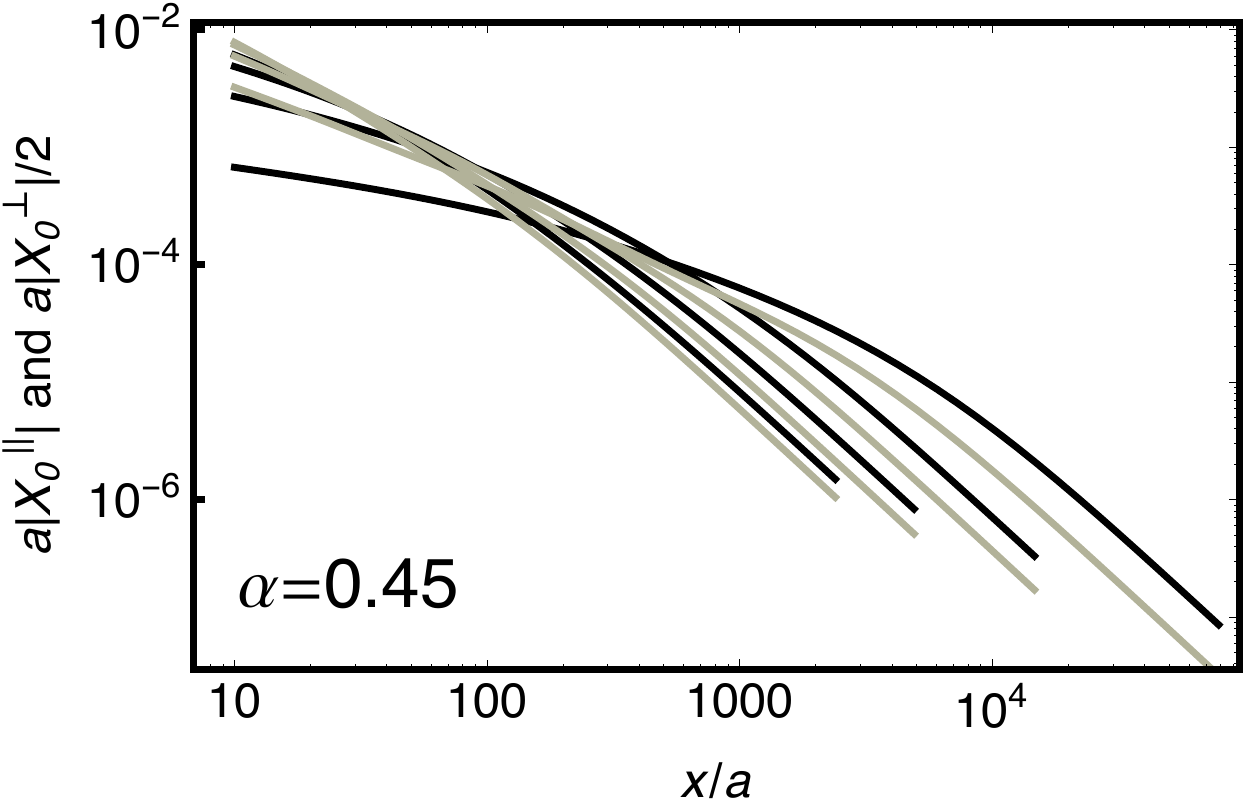}
\includegraphics[width=.95\columnwidth]{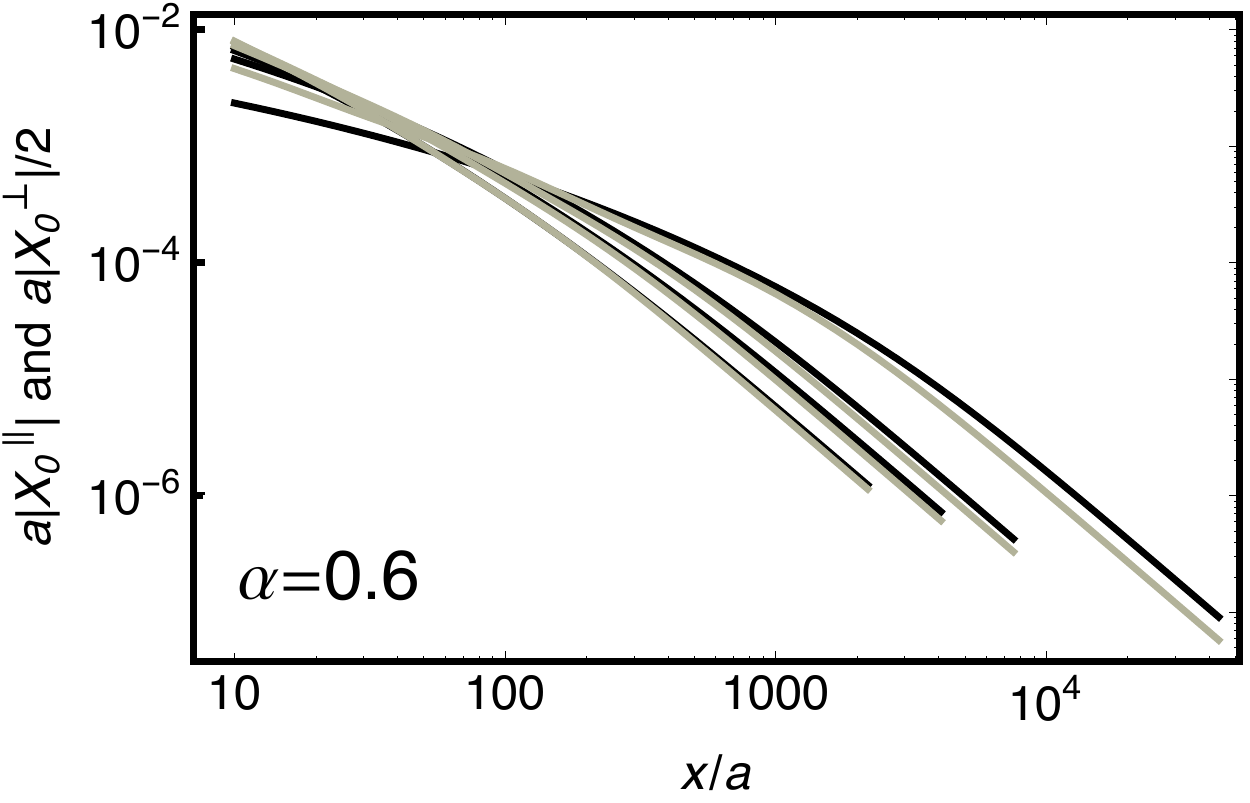}~~
\includegraphics[width=.95\columnwidth]{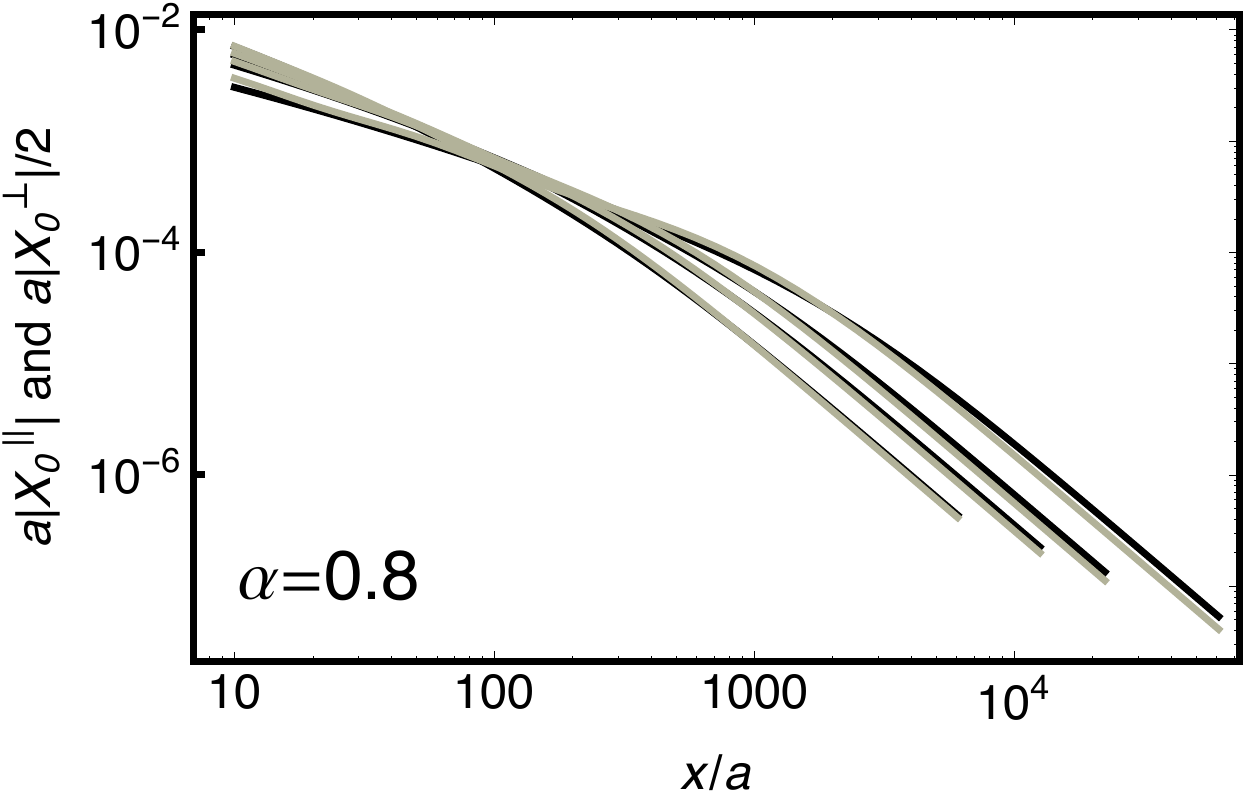}
\caption{Onset of spin-isotropy in the $0k_F$ component of the Kondo cloud,
black curves corresponding to the longitudinal correlator $a|X_0^\parallel|$ and
gray curves to the transverse correlator $a|X_0^\perp|/2$. 
Top left panel is $\alpha=0.3$, top right panel is $\alpha=0.45$, bottom left 
panel is $\alpha=0.6$, and bottom right panel is $\alpha=0.8$. 
In each panel, different curves of the same shade correspond 
to a few selected values of $\Delta$. \label{roadtoiso}
}
\end{center}
\end{figure*}

\begin{figure}[t]
\begin{center}
\includegraphics[width=\columnwidth]{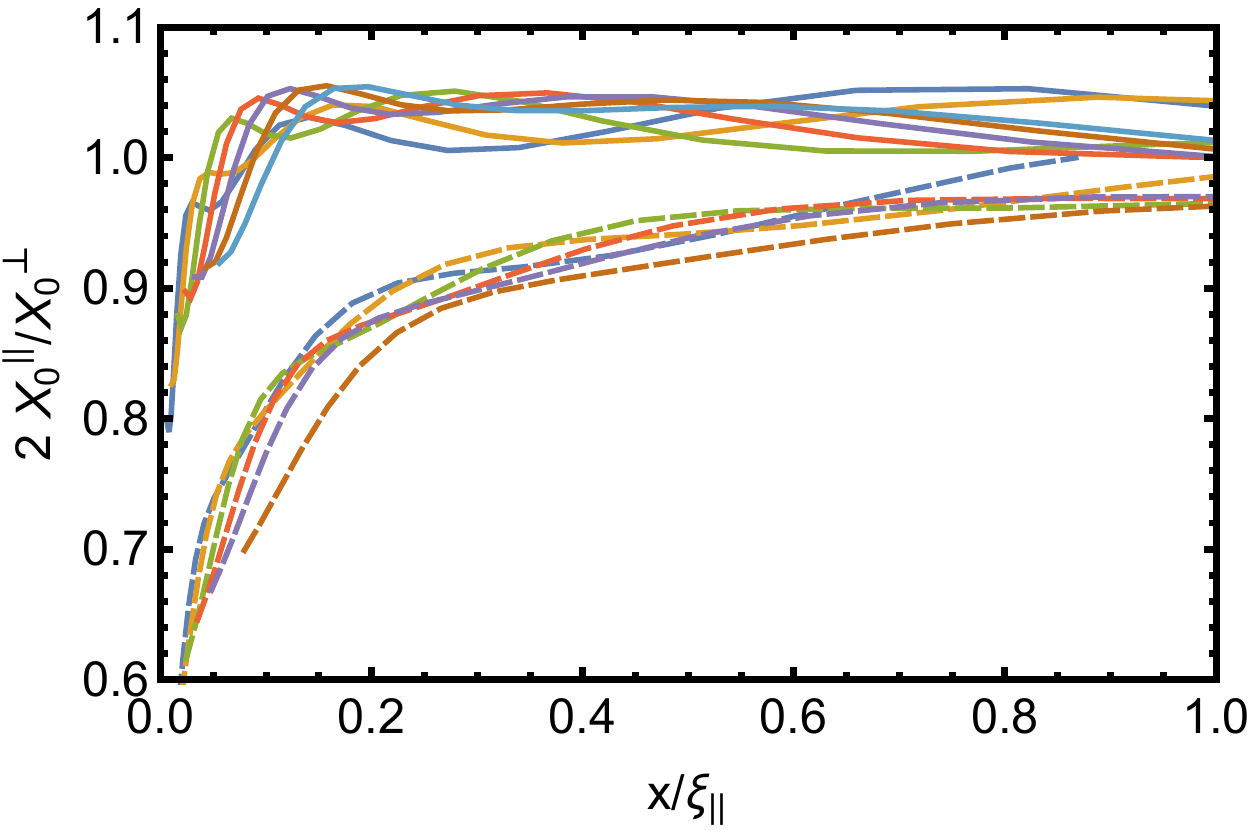}
\caption{(color online) 
The ratio $2X_0^\parallel/X_0^\perp$ of longitudinal and transverse
cloud components plotted against the scaled distance $x/\xi_\parallel$. 
Perfect isotropy corresponds to $2X_0^\parallel/X_0^\perp=1$. 
Solid curves correspond to $\alpha=0.85$ and various values of $\Delta$ ranging from $0.156/a$ to $0.305/a$.
Dashed curves correspond to $\alpha=0.75$ and various values of $\Delta$ ranging from $0.120/a$ to $0.268/a$. 
The Kondo length $\xi_\parallel$ was 
calculated for each curve from Eq.~(\ref{ea1}). For the set of data at 
$\alpha=0.85$, $\xi_\parallel$ varies from $18.0 a$ to $158 a$, while for 
$\alpha=0.75$, $\xi_\parallel$ varies from $12.6 a$ to $108 a$.
\label{figratio}}
\end{center}
\end{figure}

\begin{figure*}[tbh]
\begin{center}
\includegraphics[width=.95\columnwidth]{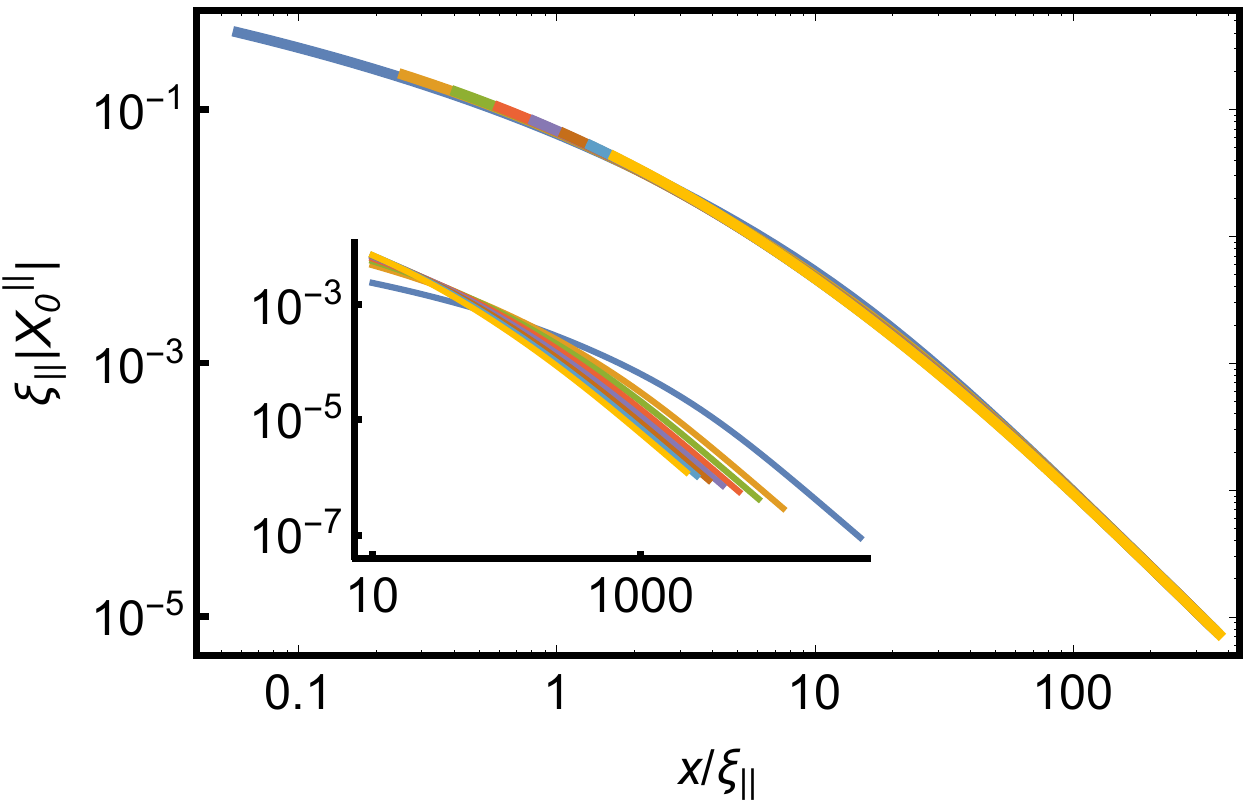}~~
\includegraphics[width=.95\columnwidth]{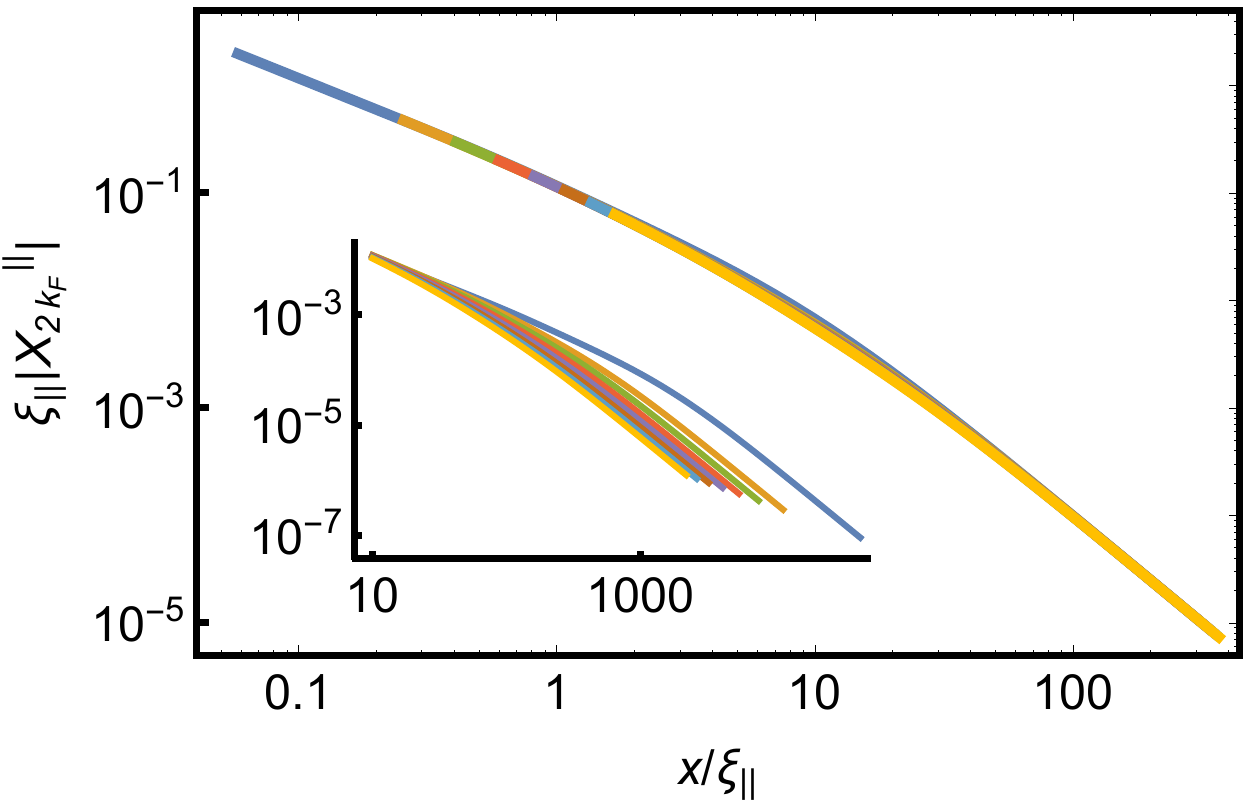}
\includegraphics[width=.95\columnwidth]{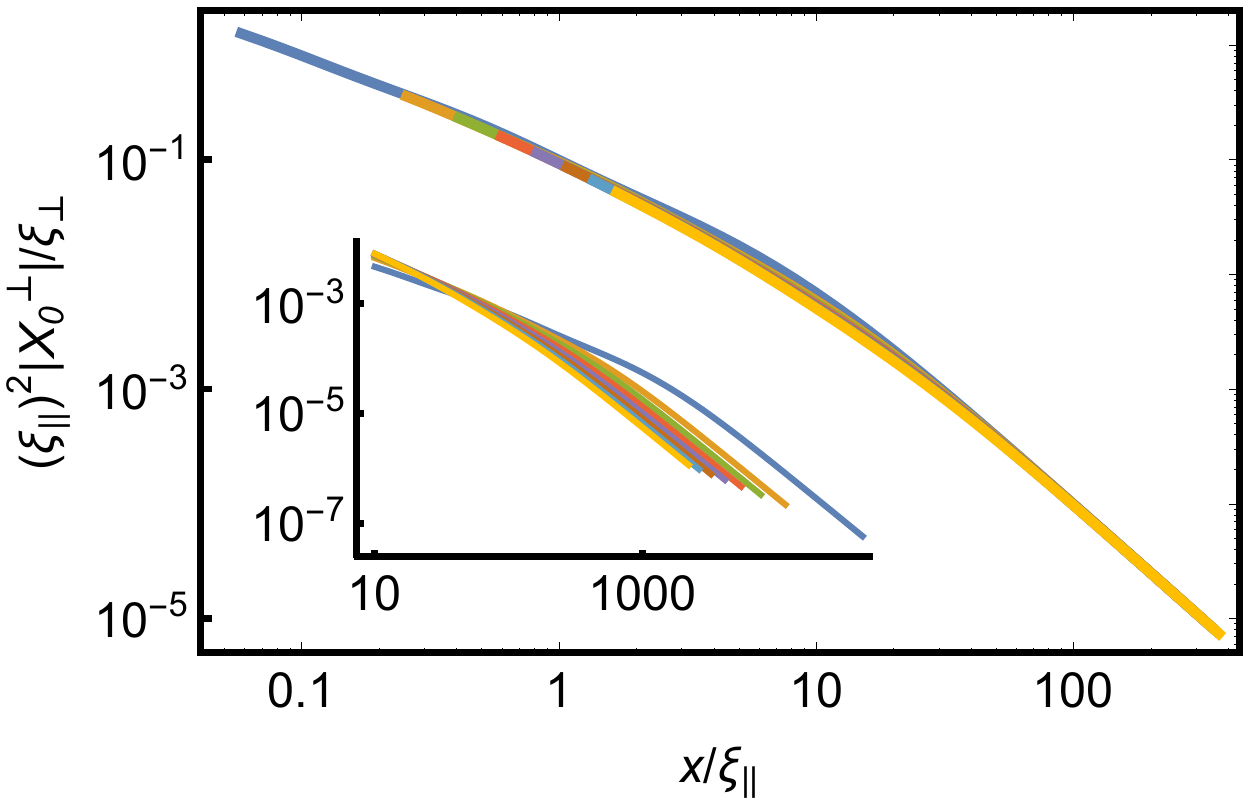}~~
\includegraphics[width=.95\columnwidth]{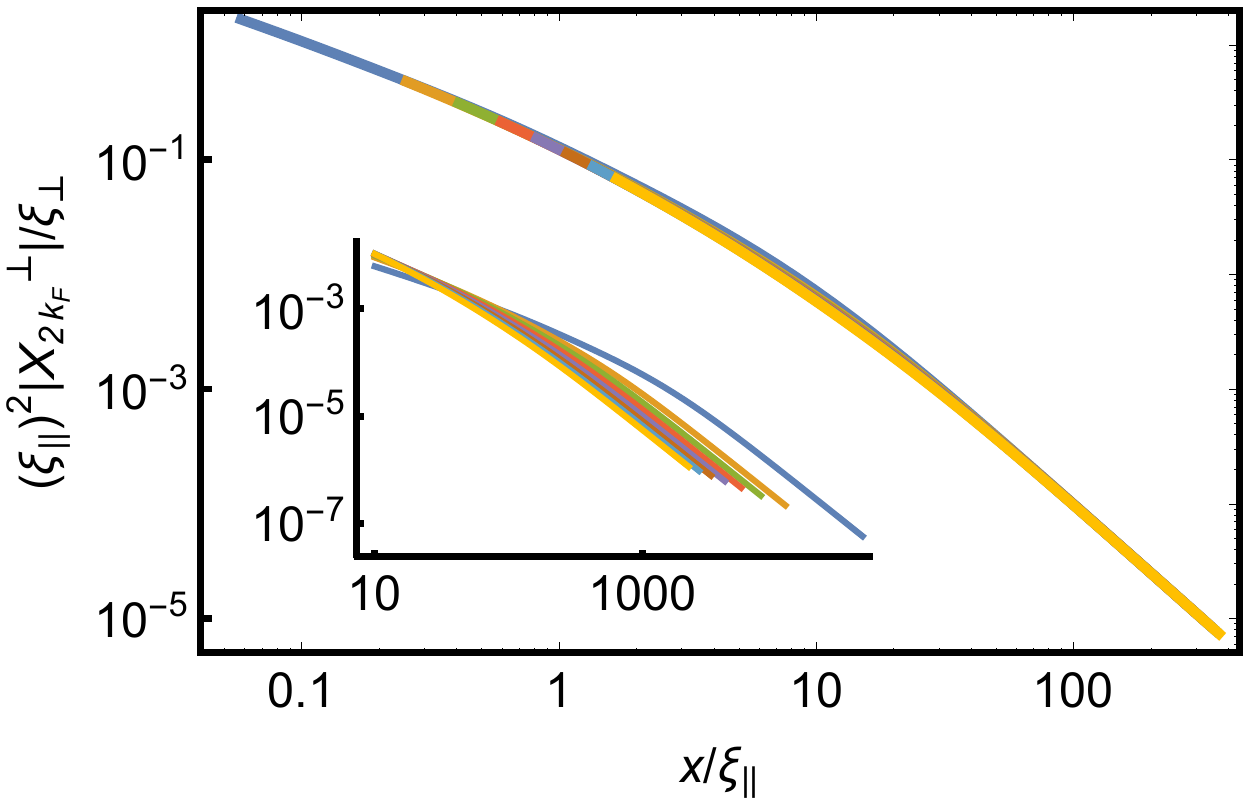}
\caption{Scaling curves for the four component of the Kondo cloud computed
at fixed $\alpha=0.6$ ($J_\parallel=1.42$). Raw data is shown in the insets, with
different curves correspond to eight different values of $\Delta$ collected 
in the interval $\Delta\in[0.05/a,0.25/a]$ (i.e. $J_\perp\in
[0.157,0.785]$). Excellent scaling behavior is obtained for all components.  \label{f8}}
\end{center}
\end{figure*}

\begin{figure}[t]
\begin{center}
\includegraphics[width=\columnwidth]{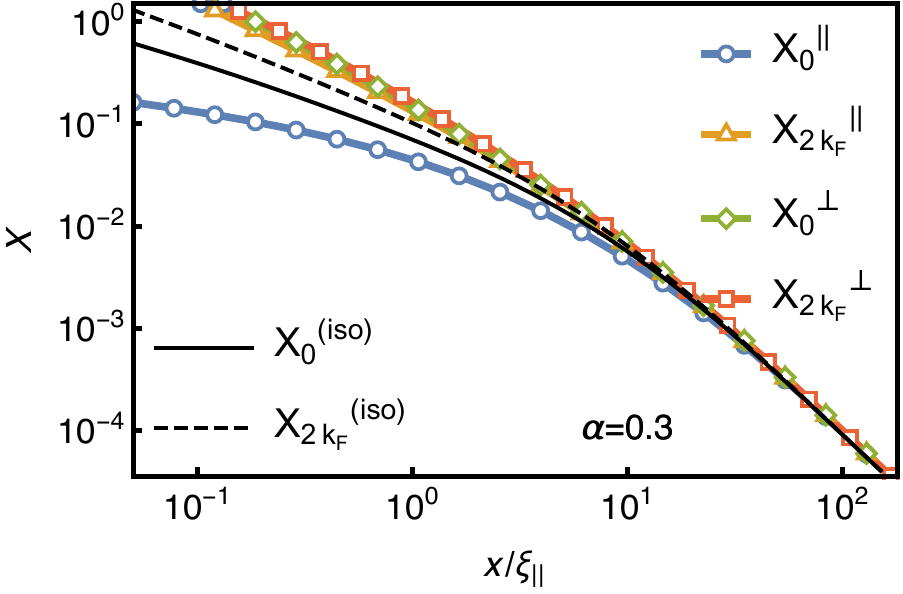}
\includegraphics[width=\columnwidth]{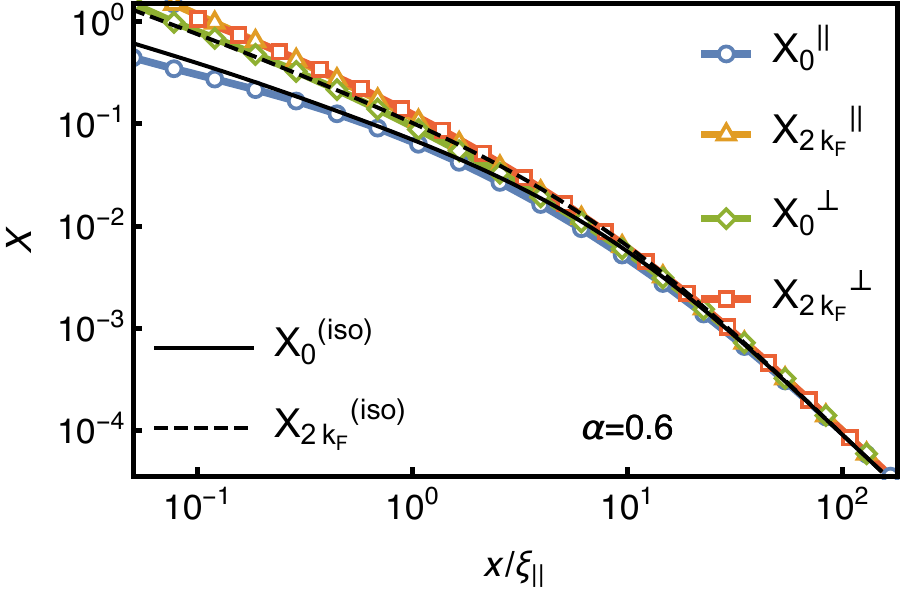}
\caption{(color online) Comparison between the universal line shapes of the scaled 
cloud (symbols in color), at $\alpha=0.3$ (top panel) and at $\alpha=0.6$
(bottom panel), to the universal isotropic line shapes (full and dashed black
lines). The curves were
obtained by fitting data sets such as those displayed in the main panels of
Fig.~\ref{f8} to polynomials of high degree. On the vertical axis, the longitudinal components are plotted in units of $1/\xi_\parallel$, while the 
transverse components are plotted in units of $\xi_\perp/\xi_\parallel^2$. The isotropic curves are those of Figure \ref{fi3}.\label{f3comp}}
\end{center}
\end{figure}

\begin{figure}[t]
\begin{center}
\includegraphics[width=\columnwidth]{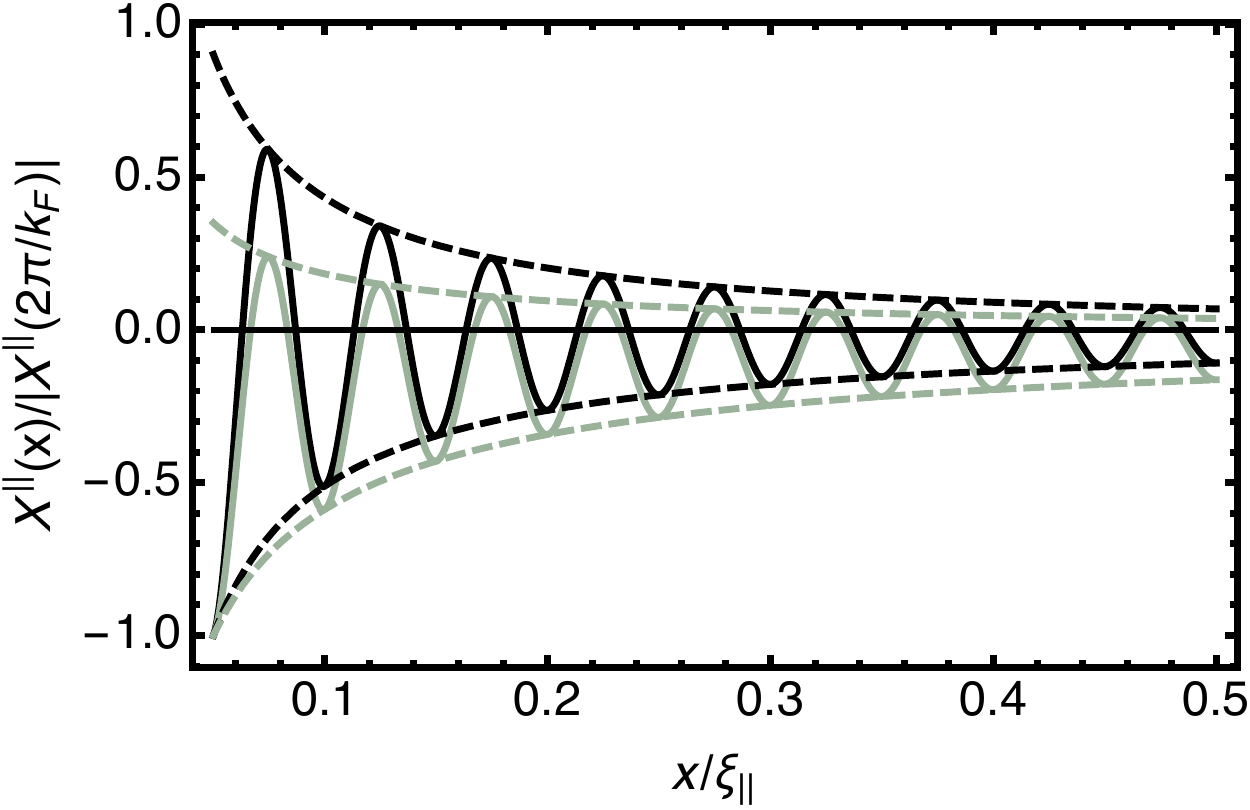}
\caption{The full longitudinal cloud
$X^\parallel(x)=X^\parallel_0(x)+\cos(2k_Fx)X^\parallel_{2k_F}(x)$ at $\alpha=0.3$ (black)
and on the isotropic line (gray), as shown by full lines (dashed lines are the
respective envelope functions $X^\parallel_0(x)\pm X^\parallel_{2k_F}(x)$).
Both curves were normalized using their magnitude at $x=2\pi/k_F$. 
For the purpose of showing the oscillations, we took $k_F=20\pi/\xi_\perp$.\label{figfriedel}}
\end{center}
\end{figure}

Below, our work will be guided by two qualitative features of Figure
\ref{figflow}. The first is that the flow trajectories intersect the
$J_\parallel$-axis at $90^\circ$. For large $J_\parallel$, the parts of the
trajectories that may reasonably be trusted, run nearly parallel to the
$J_\perp$ axis. Going beyond the small $J_\perp$ regime of the figure, we know
that, at the Toulouse point $J_\parallel=2\pi(1-1/\sqrt{2})$, the scaling
trajectory runs exactly parallel to the $J_\perp$ axis. (See Appendix
\ref{appd}.) 
Universality at fixed $J_\parallel$ is also
an obvious feature of the exact analytical expression for the Kondo
overlap~(\ref{luk1}).
For $J\gtrsim0.45\pi$ ($\alpha \leq 0.6$), we will therefore try to scale 
clouds at fixed $J_\parallel$, but different $J_\perp$, onto each other.

The second pertinent feature of Figure \ref{figflow} is that for $J_\perp,
J_\parallel \lesssim \pi/2$, trajectories are approximately hyperbolas
$J_\parallel^2-J_\perp^2=c$. In other words, the isotropic line
$J_\parallel=J_\perp$ is an attractor for the renormalization flow.  
This implies that at sufficiently large distances $x>x_{\rm iso}$, the cloud must
tend to the isotropic cloud.
However, for $x<x_{\rm iso}$, the poor man's scaling picture of Figure
\ref{figflow} indicates pronounced anisotropic behavior. This short
distance region corresponds to the scales that have to be integrated out for an
anisotropic point $(J_\parallel,J_\perp)$ to flow close to the isotropic line.
Note, furthermore that the above discussion only applies to points that are in
the general vicinity of the isotropic line, and are associated with a Kondo
temperature sufficiently lower than $1/a$. If a point is too far from the
isotropic line to start with, or if the initial Kondo temperature is too large,
the size of the cloud renormalizes down to the ultraviolet scale, before the
renormalized couplings become isotropic. At this point the notion of flow
trajectories in a two-dimensional parameter space breaks down, and Kondo physics
mixes with ultraviolet physics. 
Based on these observations, we will compare the cloud calculated at $J_\parallel, J_\perp \lesssim \pi/3$ (i.e. reasonably close to the isotropic line and with a decent Kondo length)
with the isotropic cloud.

We start the presentation of our results in the regime of small $\alpha$ (large
$J_\parallel$). This regime does not overlap with the isotropic Kondo regime,
because the size of the screening cloud flows to the short distance cut-off $a$
before the renormalized couplings come close to the isotropic line. 
For small $\alpha$, the single coherent state (Silbey-Harris) approximation is
accurate (see Figure \ref{f3}), and we have derived simple analytical expressions 
(\ref{shcloud}) for the
cloud correlation functions, that reveal the following universality. By
appropriately rescaling the four cloud components and the coordinate $x$, clouds
calculated at different $J_\perp\ll \pi$ but fixed $J_\parallel$ (i.e. fixed
$\alpha$) have the same line shape: For the longitudinal components
$X_j^\parallel$, $j\in\{0,2k_F\}$ one defines 
\begin{equation}
\tilde X_j^\parallel(\alpha,\tilde x)=\xi_\parallel X_j(x),~~~\tilde x=x/\xi_\parallel,\label{ansatz1}
\end{equation}
and finds from the Silbey-Harris result the simple analytical formulas:
\begin{align}
\tilde X_0^\parallel(\alpha,\tilde x)&=-{\rm 
Re}\,F\left(\frac{i\sqrt{\alpha}\tilde{x}}{\pi}\right),\nonumber\\
\tilde X_{2k_F}^\parallel(\alpha,\tilde x)&=-\frac{1}{2\pi \tilde x}\sin\left[2\sqrt{\alpha}{\rm Im\,}F\left(\frac{i\sqrt{\alpha}\tilde{x}}{\pi}\right)\right].\label{uncomps1}
\end{align} 
In the small $\alpha$ regime, the two transverse components $X_j^\perp$, $j\in\{0,2k_F\}$, are equal. One defines
\begin{equation}
\tilde X_j^\perp(\alpha,\tilde x)=\xi_\parallel^2 X_j(x)/\xi_\perp,~~~\tilde x=x/\xi_\parallel,\label{ansatz2}
\end{equation}
and finds the analytical expression:
\begin{equation}
\tilde X_j^\perp(\alpha,\tilde x)=-\frac{2}{\tilde x^2}\exp\left[ -2\sqrt{\alpha}{\rm Re\,}F\left(\frac{i\sqrt{\alpha}\tilde{x}}{\pi}\right)\right].\label{uncomps2}
\end{equation}
The expression for $\tilde X_0^\parallel(x)$ could suggest a further rescaling 
$\bar x=\sqrt{\alpha} \tilde x$ that would lead to an $\alpha$ independent
line shape for $X_0^\parallel$. However, one cannot simultaneously get rid of 
the $\alpha$ dependence in the other components of the cloud. Furthermore, it 
turns out that $\tilde X^\parallel (\alpha,z/\sqrt{\alpha})$ is no longer 
$\alpha$-independent beyond $\alpha\gtrsim 0.4$ (as shown by our numerical
simulations below).

The two pertinent facts to emerge from this discussion are the following.
Firstly, whereas the magnitudes of the longitudinal and transverse components of
the cloud have to be rescaled by different inverse lengths $1/\xi_\parallel$ and
$\xi_\perp/(\xi_\parallel)^2$, the coordinate $x$ is always rescaled by $\xi_\parallel$. In
other words, the longitudinal and transverse clouds have different
characteristic magnitudes $1/\xi_\parallel$ and $\xi_\perp/(\xi_\parallel)^2$, but the same
characteristic size $\xi_\parallel$. Secondly, the universal scaling curves depend in
general on $\alpha$. Thus, as expected from poor man's scaling, clouds at
the same $J_\parallel$ but different $J_\perp$ have the same shape, whereas
clouds at different $J_\parallel$ in general have different shapes.

The differences in scaling behavior between the small $\alpha$ and isotropic
clouds are thrown into sharp relief when one considers the small $x$
asymptotic form of the cloud correlators. At $\tilde x\ll 1$, the small
$\alpha$ results (\ref{uncomps1}) and (\ref{uncomps2}) lead to asymptotic
behavior:
\begin{align}
&\tilde X^\perp_0=\tilde X^\perp_{2k_F} \sim -\tilde x^{-2(1-\sqrt{\alpha})},\nonumber\\
&\tilde X_0^\parallel\sim \ln \tilde x,~~~\tilde X_{2k_F}^\parallel \sim - \tilde x^{-1}.\label{small1}
\end{align} 
The $\tilde X_0^\parallel$ component diverges much more slowly in the $\tilde
x\to 0$ limit than is the case at isotropic couplings ($\ln \tilde x$ vs. $-1/\tilde x(\ln\tilde x)^2$), whereas the $\tilde
X_{2k_F}^\parallel$ component diverges more rapidly than in the isotropic case ($-1/\tilde x$ vs. $1/\tilde x\ln \tilde x$).
Thus, the alternation between ferro- and antiferromagnetic correlations close
to the impurity is enhanced in the longitudinal direction. This is not
surprising, as small $\alpha$ implies $J_\parallel\gg J_\perp$. Furthermore we
see that the small $x$ asymptotic behavior of the transverse cloud is explicitly
$\alpha$-dependent, displaying power-law behavior with an exponent
$-2(1-\sqrt{\alpha})=-J_\parallel/\pi$, implying a divergence as $\tilde x\to 0$
that is closer to $1/x^2$, than to $1/x$.

The next obvious question is how the small $\alpha$ scaling picture evolves as
$\alpha$ increases. In order to investigate this, we calculated the cloud
correlators at the points in parameter space shown in Figure~\ref{fa1}.
The most obvious feature of the data, is that the cloud becomes more and more
isotropic as $\alpha$ increases. This is revealed in Figure~\ref{roadtoiso},
where we plot $X_{0}^\parallel(x)$ and $X_{0}^\perp(x)/2$ on top of each other
for respectively $\alpha=0.3$, $\alpha=0.45$, $\alpha=0.6$, and $\alpha=0.8$, in
separate panels. Each panel contains curves for several values of $J_\perp$. At
$\alpha=0.3$, the transverse and longitudinal clouds are very different. As
$\alpha$ increases, the differences become smaller. By the time we reach
$\alpha=0.8$, the cloud is isotropic to a high degree of accuracy for
all values of $J_\perp$ considered.
The behavior of the $X_{2k_F}^\parallel(x)$ and $X_{2k_F}^\perp(x)$ components of 
the cloud (not shown) present very similar behavior. 
We find similarly isotropic clouds at all the points in Figure \ref{fa1} inside 
the shaded region.

Poor man's scaling suggests that for anisotropic exchange couplings, there
is a scale $x_{\rm iso}$ below which isotropy breaks down.  To investigate this,
we plot the ratio $2 X_0^{\parallel}/X_0^\perp$ as a function of
$x/\xi_\parallel$ (Figure \ref{figratio}).  We do so for the data collected at $\alpha=0.75$ and for
$\alpha=0.85$. In both cases we see that $2 X_0^{\parallel}/X_0^\perp$ reaches a
plateau that is within 10\% of unity. This is the same degree of isotropy as we
obtain on the isotropic line itself.  In both cases, the plateau is reached at a
distance that is a small fraction of the Kondo length $\xi_\perp$. In other
words, isotropy extends deep inside the cloud.  The ratio $x_{\rm
iso}/\xi_\parallel$ between the scale where isotropy sets in, and the Kondo
length, is more or less constant at fixed $\alpha$, while the Kondo length
itself varies by an order of magnitude.  From the figure one can read off that
$x_{\rm iso}/\xi_\parallel\sim 0.2$ for $\alpha=0.75$ and $x_{\rm
iso}/\xi_\parallel\sim 0.05$
for $\alpha=0.85$.

Next, we investigate the precise scaling behavior of the data. For $\alpha\leq0.6$, we take our cue 
from the small $\alpha$ results we obtained above, together with the
understanding from poor man's scaling, and we thus scale correlation functions 
according to the Ansatz (\ref{ansatz1}) and (\ref{ansatz2}).
In Fig.~\ref{f8}, we show raw and scaled results for $\alpha=0.6$. Similar
figures for $\alpha=0.3$ and $0.45$ can be found in the supplementary material.\cite{supmat}
For all values of $\alpha$ that we considered, the data nicely scale onto single
universal curves. We conclude that for practical purposes, scaling Ansatz of 
the form (\ref{ansatz1}) and (\ref{ansatz2}) hold for all values of $\alpha<0.6$.

These universal scaling curves are $\alpha$-dependent and differ from the
isotropic scaling curves. This can clearly be seen in Figure \ref{f3comp}, where
we compare the universal line shapes of the scaled cloud at $\alpha=0.3$ (top
panel) and at $\alpha=0.6$ (bottom panel) to the universal isotropic line shapes
obtained in Sec. \ref{secisocase}. Although the single coherent state (Silbey-Harris) 
approximation underestimates the Kondo length $\xi_\parallel$ by $50\%$
at $\alpha=0.3$, the simple equations~(\ref{uncomps1}-\ref{uncomps2}) still predict
the universal line shape of the screening cloud very well.
Inside the cloud ($x<\xi_\parallel$), these universal curves are very different
from the universal isotropic curves. As $\alpha$ increases, deviations from the
Silbey-Harris line-shape set in, and scaling curves start to resemble the
isotropic curves more closely, as can be seen for $\alpha=0.6$ in the bottom panel 
of Figure \ref{f3comp}.

A prominent feature of the anisotropic scaling curves is that, at small
distances, there is a stronger suppression of the $X_{0}^\parallel$ component
over the $X_{2k_F}^\parallel$ component, than in the isotropic case. As a
result, the alternation from ferro- to antiferromagnetic correlations
between the impurity and the electron gas inside the cloud is enhanced in the
longitudinal direction. 
This is illustrated in Figure~\ref{figfriedel}, where we compare the full longitudinal cloud
$X^\parallel(x)=X^\parallel_0(x)+\cos(2k_Fx)X^\parallel_{2k_F}(x)$ calculated
at $\alpha=0.3$ with the result on the isotropic line.

\begin{figure}[t]
\begin{center}
\includegraphics[width=\columnwidth]{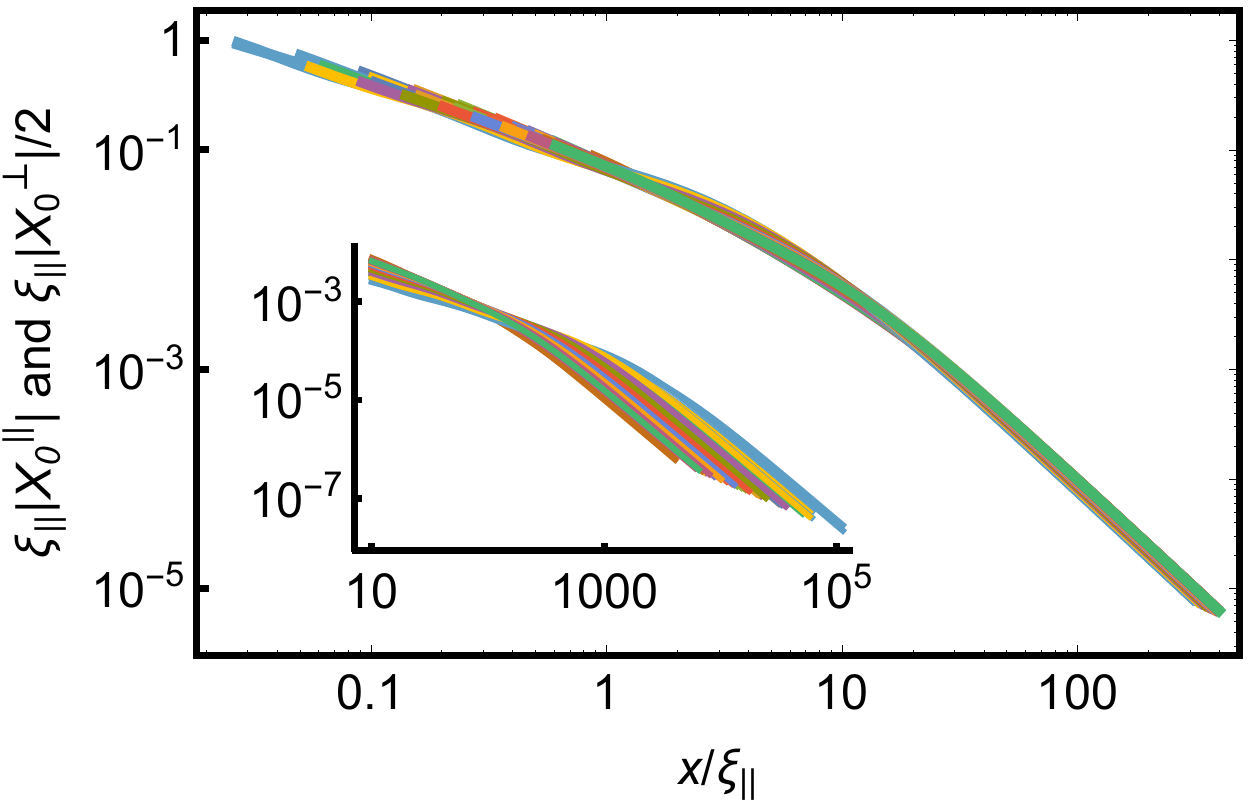}
\includegraphics[width=\columnwidth]{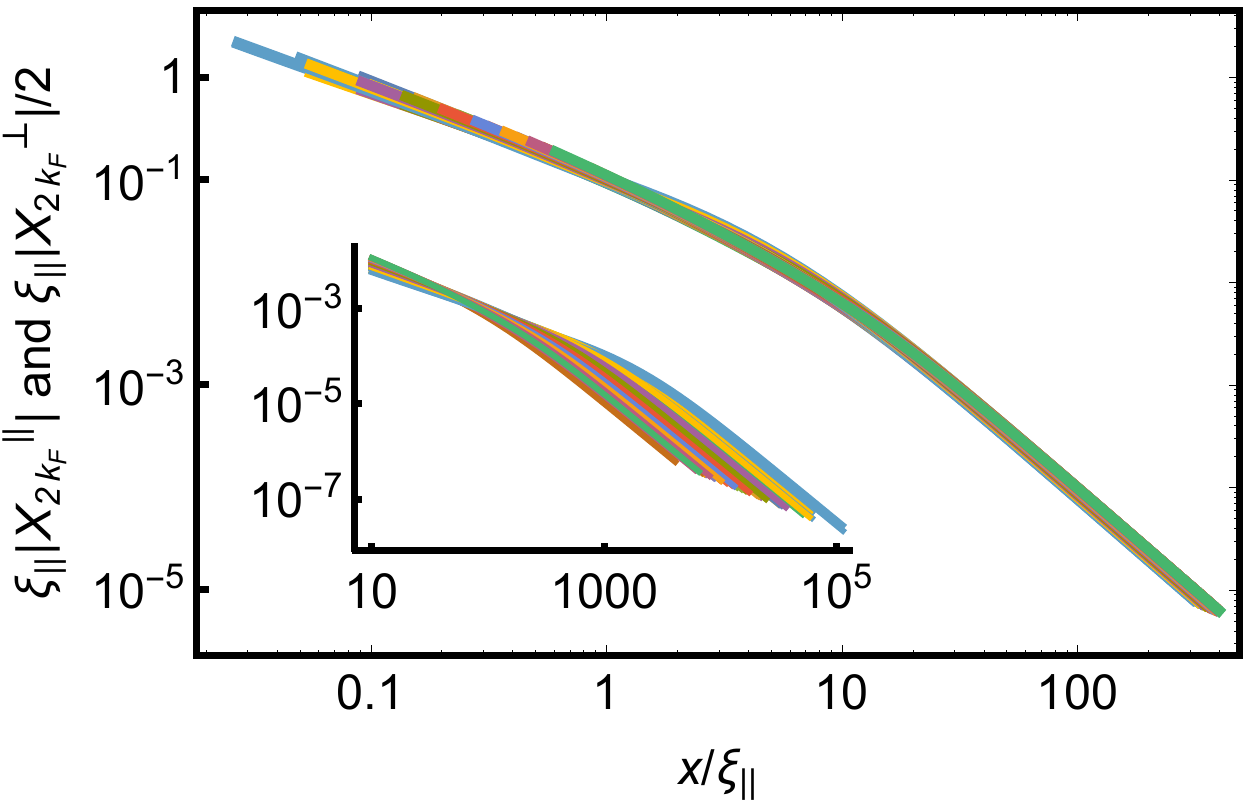}
\caption{(color online) The screening cloud calculated at all points inside the shaded region ($0.75\leq \alpha\leq0.9$) in Figure \ref{fa1}.
Insets show the raw data, while the main panels show data in units of $\xi_\parallel$ on the horizontal axis and units of
$1/\xi_\parallel$ on the vertical axis. The top panel contains the transverse and longitudinal $0 k_F$ components plotted together. 
The bottom panel contains the transverse and longitudinal $2k_F$ components.\label{fall}}
\end{center}
\end{figure}

\begin{figure}[t]
\begin{center}
\includegraphics[width=\columnwidth]{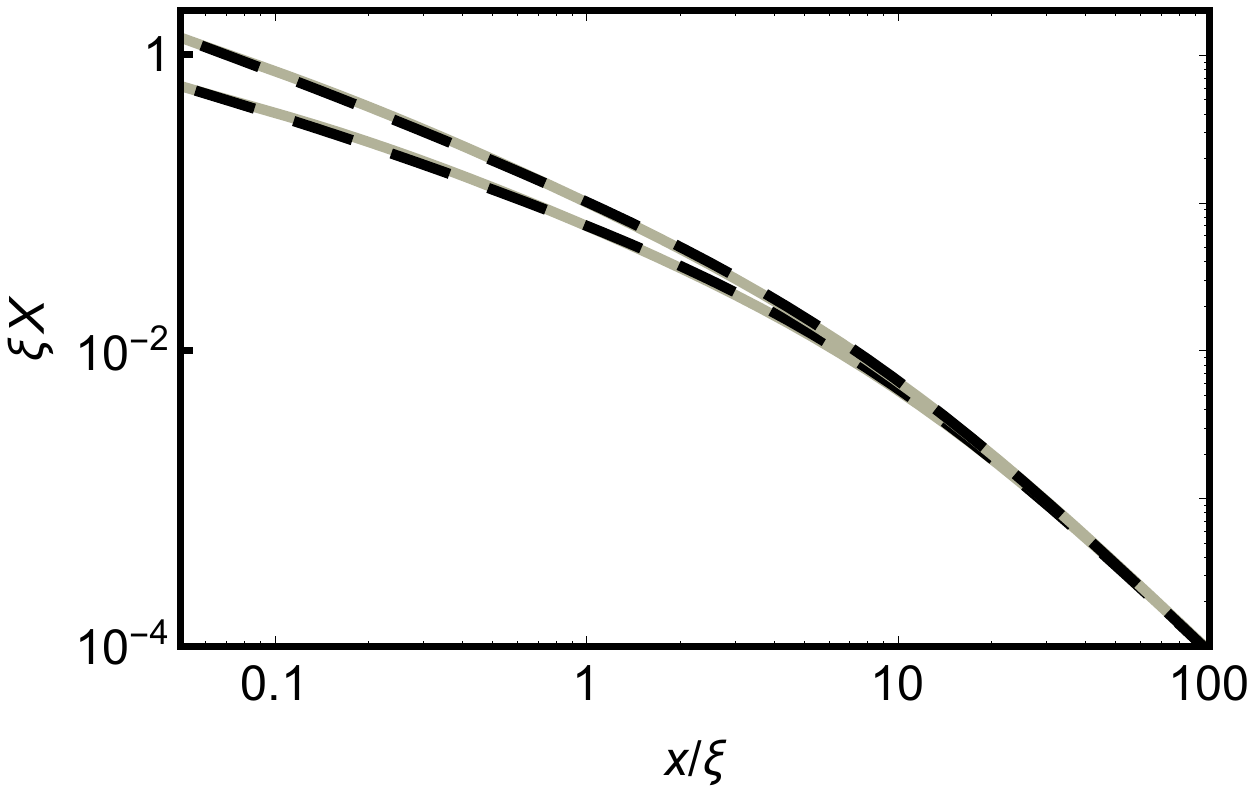}\\
\caption{The solid gray curves represent the (isotropic) universal scaling curves inferred using all the data of Figure \ref{fall}, i.e. clouds
calculated at all the points inside the shaded region in Figure \ref{fa1}. 
These curves were obtained as the best fit of a high order polinomial through the scaled data of Figure \ref{fall}.
The dashed black curves are the universal scaling curves inferred using only
clouds calculated at perfectly isotropic couplings (See Figure \ref{fi3}).
\label{fallviso}}
\end{center}  
\end{figure}

To complete the scaling analysis, we consider the screening cloud at
$0.75\leq\alpha\leq0.9$ (the shaded region in Figure \ref{fa1}). In this region,
we have seen that the cloud is approximately isotropic. In Figure \ref{fall}, we
therefore plot the transverse and longitudinal components of the cloud on the
same graph. Insets show raw results in units of $a$ on the horizontal and $1/a$ on
the vertical axes. The main panels show data in rescaled units of
$\xi_\parallel$ on the horizontal and $1/\xi_\parallel$ on the vertical axes. We
see that the scaled data collapse very well onto single scaling curves.  As
before, the Kondo length $\xi_\parallel$ was calculated using Eq.~(\ref{ea1}),
i.e. no fitting parameters were used to obtain the high degree of collapse. It
is instructive to  compare the main panels of Figure \ref{fall} for all the data
collected, to Figure \ref{fi2}, that shows scaled data only for the subset of
points that lie on the isotropic line $J_\perp=J_\parallel$. We see the same
degree of collapse onto single curves in both figures. In other words, within
the numerical accuracy of our calculation, the universal line shape of the
screening cloud is the same for all points in the shaded region of Figure
\ref{fa1}. To further confirm this conclusion, we fitted a high order polynomial
through the complete scaled data sets of Figure \ref{fall}. In Figure
\ref{fallviso} we compare this fit to the universal scaling curves we obtained
previously by only considering the cloud on the isotropic line (Figure
\ref{fi3}).  We see nearly perfect agreement. Of course, the screening cloud
at an anisotropic value of the exchange couplings, $J_\perp\not= J_\parallel$,
only follows this universal line shape for distances larger than $x_{\rm iso}$,
the scale at which isotropy sets in. However, as we have seen in Figure \ref{figratio},
there is a sizable region around the isotropic line where isotropy already sets in
deep inside the cloud.

\begin{figure}[tbh]
\begin{center}
\includegraphics[width=\columnwidth]{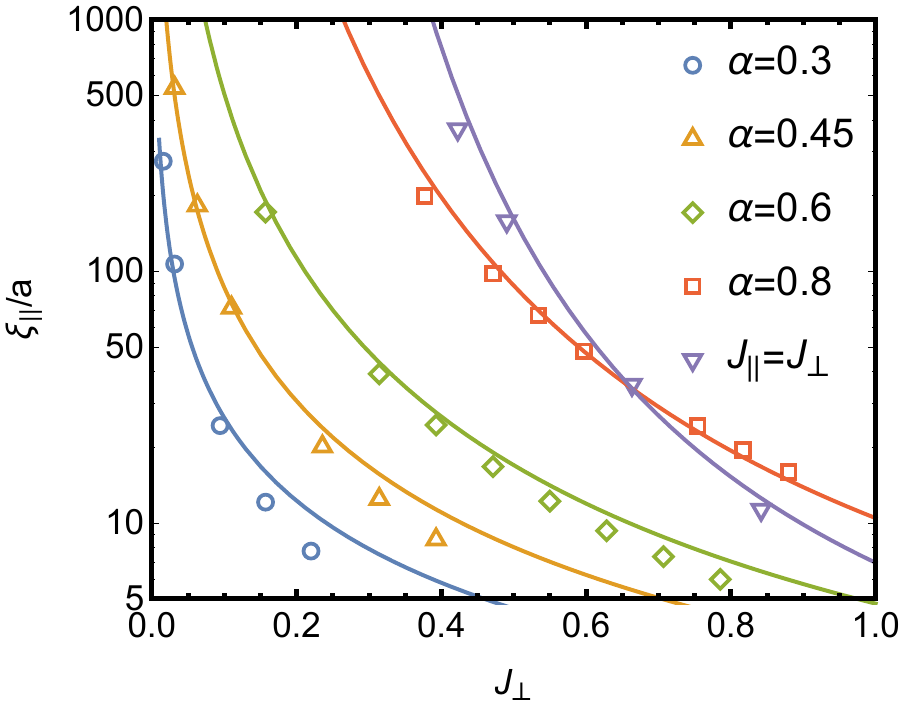}
\caption{(color online) The Kondo length $\xi_\parallel$ as a function of
$J_\perp$ for several $\alpha$ values. The solid lines
shows $\xi=c/T_K$, with $T_K$ the standard poor man's scaling estimate (\ref{tkren})
and $c$ a fitting parameter. The same value of $c=0.3$ we used for all five 
curves.\label{fklen}}
\end{center}
\end{figure}

Finally, we plot in Figure~\ref{fklen} the Kondo length $\xi_\parallel$, which
corresponds to the characteristic size of the cloud, for some of the data we have
collected. The large range over which $\xi_\parallel$ varies for each value of
$\alpha$, confirms that the scaling behavior we see is non-trivial. We compare
the calculated value of $\xi_\parallel$ to $\xi\propto 1/T_K$, with $T_K$ the
standard poor man's scaling estimate (\ref{tkren}) of the Kondo temperature.
With a logarithmic scale on the vertical axis, which hides mismatches by factors of order
unity, we find good agreement for the isotropic and the $\alpha=0.8$ data sets.
That these sets yield the best agreement is expected, as they are closer to the
point $J_\perp=J_\parallel=0$ in parameter space than the other sets, and the
version of poor man's scaling that leads to the estimate (\ref{tkren}) for $T_K$
assumes small exchange couplings. 
 
\section{Summary of results and conclusion}

\label{secdisc}

This paper has provided a very extensive study of the Kondo screening cloud for
a wide range of parameter values, including the spin-isotropic and strongly
spin-anisotropic regimes. Methodologically, we derived simple but controlled
analytical expressions in the case of strong anisotropy, and developed an
original and very powerful numerical technique to tackle the problem in
its complete generality. This allowed us to investigate 
the $0k_F$ and $2k_F$ component of the Kondo cloud correlator separately, 
both for the longitudinal and transverse response. In addition, we have examined, again both
analytically and numerically, the structure of Kondo overlaps introduced in
recent works.\cite{luk}

Our main results concern the universal scaling of correlations between the
impurity spin and the electron spin density in the anisotropic Kondo model. They
can be summarized as follows. At large $x$, transverse correlators equal
$X_k^{\perp}(x)=-\xi_\perp/x^2$ while longitudinal correlators equal
$X_k^\parallel(x)=-\xi_\parallel/x^2$ (here $k=0,\,2k_F$ refer to respectively
the forward or backscattering component of the correlator). In general, the two
emergent length scales $\xi_\perp$ and $\xi_\parallel$ are not equal, and we find
the following universal scaling behavior:
\begin{align}
&X_k^\perp(x)=\frac{\xi_\perp}{\xi_\parallel^2}\tilde
X_k^\perp(\alpha,x/\xi_\parallel),\nonumber\\
&X_k^\parallel(x)=\frac{1}{\xi_\parallel}\tilde
X_k^\parallel(\alpha,x/\xi_\parallel).
\end{align}
The scaling curves are $J_\perp$-independent, but remain $\alpha$-dependent (and
therefore $J_\parallel$-dependent). As can be seen from the scaling equations,
the two lengths $\xi_\perp$ and $\xi_\parallel$ have fundamentally different
roles. The length $\xi_\parallel$ sets the size of the screening cloud, whereas
the ratio $\xi_\perp/\xi_\parallel$ sets the relative magnitude of transverse
correlations, compared to longitudinal correlations.

As $J_\parallel$ decreases, the explicit $J_\parallel$ dependence of the scaling curves
becomes weaker and weaker. For $J_\parallel\ll 1$, we find no discernible
$J_\parallel$-dependence anymore, and the universal scaling curves become those
of the isotropic model. In this region of parameter space, we find that isotropy
already sets in deep inside the cloud.
We managed to calculate the universal scaling curves down to distances
$x/\xi=0.1$ or less. This far inside the cloud, the backscattering ($2k_F$)
component of the longitudinal correlator dominates the forward scattering
($0k_F$) component. Inside the cloud, correlations between the impurity spin and
the electron spin density therefore alternate with wave vector $2k_F$ between
being ferro- and antiferromagnetic. The effect is more pronounced in the regime
$J_\parallel\gg J_\perp$, but we can still clearly resolve it when the cloud
becomes isotropic.

In previous numerical studies, results were obtained
using the numerical renormalization group (NRG). This required solving a
two-impurity problem, where the position of the second (fictitious) impurity
represents the point $x$ where the correlator is evaluated. An independent 
two channel Kondo problem must be solved by NRG at every position where the correlator 
is evaluated, making the calculation computationally expensive. In addition, it proved
difficult to resolve in NRG the dominance of the backscattering components of the cloud 
over the forward scattering components inside the cloud. 

We have shown that the coherent state expansion, the method we employed
in this Article, complements previous numerical renormalization group studies,
and even offers simple analytical insights in the regime of strong
spin-anisotropy.
Thanks to a technical improvement that dramatically reduces the number of 
variational parameters, it offers excellent accuracy at a reduced computational 
cost, and does not require one to discretize the bath degrees of
freedom.
Yet, we note that the numerical renormalization group was successfully used to 
study finite temperature \cite{Borda} and time-dependent\cite{Lechtenberg} features 
of the Kondo cloud. We envisage future work to extend the coherent state expansion 
into these domains as well.
 
\appendix
\section{Mapping between the Kondo and Spin-Boson Models}
\label{appmap}

In this Appendix, we review the mapping from the Kondo Hamiltonian (\ref{h2}) to
the spin-boson Hamiltonian (\ref{sbham}). The first step is to express the
(even) fermionic degrees of freedom in terms of bosonic ones $b_{q\sigma}$,
$q=2\pi n/L$, $n\in\{1,\,2,\,3,\,\ldots\}$ with
$[b_{q\sigma},b_{q'\sigma'}^\dagger]=\delta_{qq'}\delta_{\sigma\sigma'}$. For
this purpose, we invoke the bosonization identities
\begin{equation}
b_{q\sigma}=\sqrt{\frac{2\pi}{Lq}}\int_{-L/2}^{L/2} dx\, e^{-iqx}\psi^\dagger_\sigma(x)
\psi_\sigma(x),
\end{equation}
and
\begin{align}
&\psi^\dagger_\sigma(x)=\frac{F^\dagger_\sigma}{\sqrt{2\pi a}} 
e^{-i\phi_\sigma(x)},\nonumber\\
&\phi_\sigma(x)=i\sum_{q>0}\sqrt{\frac{2\pi}{qL}}
e^{-(a/2+ix)q}b_{q\sigma}^\dagger+\mbox{h.c.} .\label{phixdef}
\end{align}
Here $F_\sigma^\dagger$ denotes the unitary Klein factor operator associated
with increasing the number of $\sigma$ electrons by one. 

We separate spin and charge degrees of freedom by defining:
\begin{equation}
a_q=(b_{q\uparrow}+b_{q\downarrow})/\sqrt{2},~~~
b_q=(b_{q\uparrow}-b_{q\downarrow})/\sqrt{2}.
\end{equation}
We also define new SU(2) operators, in which fermion number and impurity degrees
of freedom are mixed.
\begin{equation}
s^-=F^\dagger_\uparrow F_\downarrow\sigma^-,~~~
s^+=(s^-)^\dagger,~~~s_z=\sigma_z.
\end{equation} 
 
In the bosonic representation, the Hamiltonian reads
\begin{align}
&H_0=\sum_{q>0} q(a_q^\dagger a_q+b_q^\dagger b_q),\nonumber\\
&H_\parallel=\frac{J_\parallel}{2}\sum_{q>0}\sqrt{\frac{q}{\pi L}}
e^{-aq/2}(b_q^\dagger+b_q)s_z,\nonumber\\
&H_\perp=\frac{J_\perp}{2\pi a}\left[e^{-\sqrt{2}i\phi} s^- 
+ e^{\sqrt{2}i\phi} s^+\right],\label{bosrep}
\end{align} 
where 
\begin{align}
\phi&=\frac{\phi_\uparrow(0)-\phi_\downarrow(0)}{\sqrt{2}}
=i\sum_{q>0}\sqrt{\frac{2\pi}{Lq}} e^{-aq/2}(b_q^\dagger - b_q).\label{appphidef}
\end{align}
We note that spin ($b_q$) and charge ($a_q$) degrees of freedom are decoupled,
and that only spin degrees of freedom couple to the impurity. We restrict the
Hamiltonian to the vacuum of the charge sector and drop the $a^\dagger a$ terms
in future expressions.

Finally, we bring the Hamiltonian into the standard form of the (unbiased)
spin-boson model, via a unitary transform
\begin{equation}
U=\exp(-i s_z\phi/\sqrt{2}).\label{uni}
\end{equation} 
Under the action of $U$, 
\begin{align}
&Ub_qU^\dagger=b_q-\sqrt{\frac{\pi}{Lq}}e^{-aq/2}s_z,~~~
Us^-U^\dagger=e^{\sqrt{2} i \phi}s^-,
\end{align}
so that $UHU^\dagger=H_{\rm SB}+\mbox{const}$, where $H_{\rm SB}$ 
is given by (\ref{sbham}).

\section{Correlation functions in the bosonic representation}
\label{appcloud}
In this Appendix, we transform the fermionic cloud correlators (\ref{clouddef})
onto their bosonic counterparts (\ref{bosoncloud}). To do so, we represent the
electron spin density in terms of bosonic degrees of freedom, and subsequently
perform the unitary transformation of (\ref{uni}) on these operators. To this
end, it is useful to note from (\ref{psi}) the following identities:
\begin{align}
&\frac{1}{\sqrt{2}}\left[\tilde \psi_\sigma(x)+\tilde \psi_\sigma(-x)\right]
\simeq e^{ik_F x}\psi_\sigma(x)+ e^{-ik_F x}\psi_\sigma(-x),\nonumber\\
&\frac{1}{\sqrt{2}}\left[\tilde \psi_\sigma(x)-\tilde \psi_\sigma(-x)\right]
\simeq e^{ik_F x}\bar \psi_\sigma(x)- e^{-ik_F x}\bar \psi_\sigma(-x).\label{slow}
\end{align}
The `approximately equal' signs in these equations indicate that the operators
on the right hand side represent versions of the point operators on the left,
that have been broadened by an amount $\sim a$, inversely proportional to the
ultraviolet cut-off. Now consider the expectation value $X^\perp(x)$, which,
from (\ref{clouddef}), can also be written as 
\begin{equation}
X^\perp(x)=4{\rm Re}\left<\sigma^-\tilde\psi^\dagger_\uparrow(x)\tilde\psi_\downarrow(x)\right>_{\rm K}.
\end{equation}
Because the ground state has good spatial parity, $X^\perp(x)$ is an even function, and we may write
\begin{equation}
X^\perp(x)=2{\rm Re}\left<\sigma^-\left[\tilde\psi^\dagger_\uparrow(x)\tilde\psi_\downarrow(x)
+\tilde\psi^\dagger_\uparrow(-x)\tilde\psi_\downarrow(-x)\right]\right>_{\rm K}.
\end{equation}
The last expression is manipulated to become
\begin{widetext}
\begin{align}
&X^\perp(x)={\rm Re}\left\{\left<\sigma^-\left[\tilde\psi^\dagger_\uparrow(x)+\tilde\psi^\dagger_\uparrow(-x)
\right]\left[\tilde\psi_\downarrow(x)+\tilde\psi_\downarrow(-x)\right]\right>_{\rm K}
+\underbrace{\left<\sigma^-\left[\tilde\psi^\dagger_\uparrow(x)-\tilde\psi^\dagger_\uparrow(-x)
\right]\left[\tilde\psi_\downarrow(x)-\tilde\psi_\downarrow(-x)\right]\right>_{\rm
K}}_{=0}\label{xp}\right\}.
\end{align}
The term marked with the underbrace is zero because of the following reason:
$\tilde\psi^\dagger_\uparrow(x)-\tilde\psi^\dagger_\uparrow(-x)$ creates an
electron in an odd parity single particle orbital, and these orbitals do not
couple to the impurity, so that:
\begin{equation}
\left<\sigma^-\left[\tilde\psi^\dagger_\uparrow(x)-\tilde\psi^\dagger_\uparrow(-x)
\right]\left[\tilde\psi_\downarrow(x)-\tilde\psi_\downarrow(-x)\right]\right>_{\rm K}
=\left<\sigma^-\right>_{\rm K}\left<\left[\tilde\psi^\dagger_\uparrow(x)-\tilde\psi^\dagger_\uparrow(-x)
\right]\left[\tilde\psi_\downarrow(x)-\tilde\psi_\downarrow(-x)\right]\right>_{\rm K}.
\end{equation}
Since the antiferromagnetic Kondo ground state is a singlet, $\left<\sigma^-\right>_{\rm K}=0$.
Substituting from (\ref{slow}) into the first line of (\ref{xp}), we obtain
\begin{align}
&X^\perp(x)=2{\rm Re}\left<\sigma^-\left[\psi^\dagger_\uparrow(x)\psi_\downarrow(x)
+\psi^\dagger_\uparrow(-x)\psi_\downarrow(-x)
+e^{-2ik_Fx}\psi^\dagger_\uparrow(x)\psi_\downarrow(-x)
+e^{2ik_Fx}\psi^\dagger_\uparrow(-x)\psi_\downarrow(x)\right]\right>_{\rm K}.
\end{align}
Performing a similar calculation for $X^\parallel(x)$, we find
\begin{align}
&X^\parallel(x)=\frac{1}{2}\sum_\sigma{\rm sgn}
(\sigma)\left<\sigma_z\left[\psi^\dagger_\sigma(x)\psi_\sigma(x)
+\psi^\dagger_\sigma(-x)\psi_\sigma(-x)+e^{-2ik_Fx}\psi^\dagger_\sigma(x)\psi_\sigma(-x)
+e^{2ik_Fx}\psi^\dagger_\sigma(-x)\psi_\sigma(x)\right]\right>_{\rm K},
\end{align}
\end{widetext}
where ${\rm sgn}(\uparrow)=+$ and ${\rm sgn}(\downarrow)=-$.
We finally manipulate the above expressions as follows:
\begin{enumerate}
\item The fermion operators are bosonized.
\item We apply the unitary transformation that maps the Kondo Hamiltonian onto
the spin-boson model.  In the resulting expressions, expectation values are with
respect to the ground state of the spin-boson model, indicated without a
subscript.
\item We normal order the bosonic operators. In the resulting expressions, the
operators $a_q^\dagger$ and $a_q$, are replaced by zeros, since $a_q$
annihilates the ground state. 
\end{enumerate}
Finally, we take the large system limit, which allows us to make replacements
such as $1-e^{-(a+ix)2\pi/L}\to2\pi(a+ix)/L$. The resulting expressions are
those presented in the main text (\ref{bosoncloud}).

\section{Contour integral for cloud calculation}
\label{appci}
The integrals involved in calculating the cloud are of the form:
\begin{equation}
\int_0^\infty dq e^{-(a\pm ix)q}r(q),
\end{equation}
where the function $r(q)$ goes to zero as least as fast as $1/q$ for large $q$. Furthermore,
the only non-analyticities that $r(q)$ possesses in the complex $q$ plane are (first order) poles
at $\omega_n$, $n=1,\,\ldots,\,M$. These integrals can be performed by extending the method
employed in Sec.~\ref{secen}, where the energy was calculated.

\begin{figure}
\begin{center}
\includegraphics[width=.6\columnwidth]{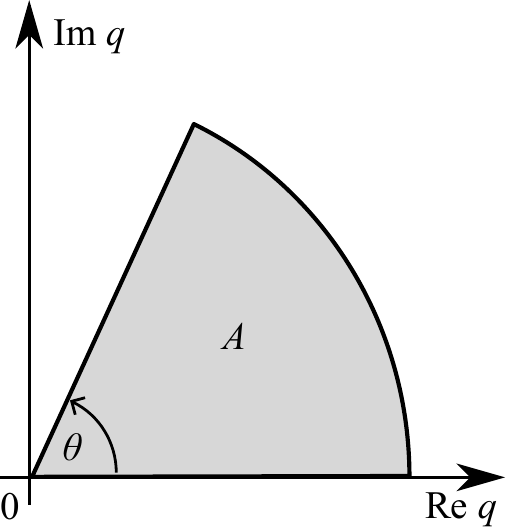}
\caption{Integration contour used in Appendix \ref{appci}.\label{con2}}
\end{center}
\end{figure}

Define $z=a\pm ix$ and $\theta=-{\rm arg}(z)$. Integrating around the shaded area 
$A$, whose azimuthal boundary
is understood to lie at $|q|\to\infty$ in Figure \ref{con2}, we obtain
\begin{align}
&\int_0^\infty dq\,e^{-zq}r(q)=e^{i\theta}\int_0^\infty dp\,e^{-|z|p}r(pe^{i\theta})\nonumber\\
&+2\pi i\sum_{\omega_n\in A}{\rm Res}\left[e^{-zq}r(q),\omega_n\right].
\end{align}

The integral on the right hand side on the first line is of precisely the same form as those considered
in Sec.~\ref{secen}. We also note that the summation on the second line vanishes if there are
no poles $\omega_n$ with positive real parts. Using the results of
Sec.~\ref{secen}, we obtain:
\begin{align}
&\int_0^\infty dq\,e^{-zq}r(q)=\sum_{m=1}^N{\rm Res}\left[e^{-zq}E_1(-zq)r(q),\omega_n\right]\nonumber\\
&+2\pi i\sum_{\omega_n\in A}{\rm Res}\left[e^{-zq}r(q),\omega_n\right].
\end{align}

\section{Exact results at the Toulouse point}
\label{appd}
In this appendix, we review the mapping between the Kondo model at the Toulouse
point $\alpha=1/2$ (i.e. $J_\parallel=2\pi(1-1/\sqrt{2})$), and the
non-interacting fermionic  resonant level model, following Ref.
\onlinecite{Zarand}. We also derive an exact expression for the longitudinal
$0k_F$ component of the screening cloud. The derivation takes the same route as
that set out in Refs. \onlinecite{Posske1} and \onlinecite{Posske2}.

The starting point is the bosonic representation (\ref{bosrep}) of the 
Kondo hamiltonian. The fermionic non-interacting resonant level $H_{\rm rlm}$ is obtained, by applying the unitary
transform $H_{\rm rlm}=U' H (U')^\dagger$, where
\begin{equation}
U'=\exp\left\{i\sigma_z\left[\frac{\pi}{2} N_--\left(1-\frac{1}{\sqrt{2}}\right)\frac{\phi}{\sqrt{2}}\right]\right\}.
\end{equation}
The bosonic field $\phi$ is defined in Eq.~(\ref{appphidef}), and the operator
\begin{equation}
N_-=(N_\uparrow-N_\downarrow)/2,
\end{equation}
is the $z$-component of the total spin of the electron gas. New degrees of freedom
\begin{eqnarray}
\bar\psi(x)^\dagger&=&\exp\left[-\frac{2\pi i}{L}\left(N_--\frac{1}{2}\right)x-i\phi(x)\right]\frac{F_\uparrow^\dagger F_\downarrow}{\sqrt{2\pi a}}\nonumber\\
&=&\frac{1}{\sqrt{L}}\sum_k e^{-ikx}e^{-a|k|/2}\bar c_k^\dagger\nonumber\\
d^\dagger&=&\exp\left[i\pi\left(N_-\frac{1}{2}\right)\right]\sigma^+,
\end{eqnarray}
emerge, that obey the usual fermionic commutation relations. Here 
\begin{equation}
\phi(x)=\frac{\phi_\uparrow(x)-\phi_\downarrow(x)}{\sqrt{2}},
\end{equation}
with $\phi_{\uparrow(\downarrow)}(x)$ defined in Eq.~(\ref{phixdef}),
and $F_{\uparrow(\downarrow)}^\dagger$ denote Klein factors.
In terms of these new fermions, $H_{\rm rlm}$ reads
\begin{equation}
H_{\rm rlm}=\sum_{q} q \bar c_q^\dagger \bar c_q+\frac{J_\perp}{\sqrt{2\pi a}}\left[\bar \psi(x)^\dagger d+
d^\dagger\bar\psi(x)\right].
\end{equation}

At the Toulouse point, the unitary transformation $U'$ maps the ground state of the Kondo Hamiltonian onto the ground state of
$H_{\rm rlm}$. Since $H_{\rm rlm}$ is a quadratic Hamiltonian, ground state correlation functions can be calculated straight-forwardly
for transformed observables, provided they are expressed simply in terms of the
new fermionic degrees of freedom. Here we focus on the longitudinal $0k_F$
component $X_0^\parallel$ for which this is easily accomplished.

Applying the unitary mapping to the $z$-components of the impurity spin and the electron spin density operators, one finds
\begin{align}
X_0^\parallel(x)=&\sqrt{2}\left<\left(d^\dagger d-\frac{1}{2}\right)\left[\bar \psi(x)^\dagger\bar\psi(x)+ \bar \psi(-x)^\dagger\bar\psi(-x)\right]\right>\nonumber\\
&-(2-\sqrt{2})\frac{a}{x^2+a^2},
\end{align}
where the expectation value on the right hand side of the first line is with respect to the ground state of $H_{\rm rlm}$.
This expression is further simplified using Wick's theorem and noting that
the occupation probability for the resonant level is one half, i.e. $\left<d^\dagger d\right>=1/2$.
In terms of the original degrees of freedom, this corresponds to
$\left<\sigma_z\right>=0$ which is guaranteed by the singlet nature of the Kondo
ground state. Working directly with $H_{\rm rlm}$, this follows from
particle-hole symmetry. (Note that the energy of the resonant level co-incides
with the Fermi energy.) 
To be more precise, $H_{\rm rlm}$ is invariant under the combined action of
particle-hole and spatial inversion.  Owing to the same symmetry,
$\left<d^\dagger\bar \psi(x)\right>=\left<d^\dagger\bar \psi(-x)\right>^*$.
Putting everything together, one obtains 
\begin{equation}
X_0^\parallel(x)=-2\sqrt{2}\left|\left<d^\dagger \bar \psi(x)\right>\right|^2
-(2-\sqrt{2})\frac{a}{x^2+a^2}.\label{toulcloud1}   
\end{equation}
The expectation value $\left<d^\dagger \bar \psi(x)\right>$ can be calculated from the known single particle Green's functions of the resonant level model.
In the thermodynamic limit ($L\to\infty$) the exact result is
\begin{align}
&\left<d^\dagger \bar \psi(x)\right>\nonumber\\
&=\frac{J_\perp}{\sqrt{2\pi a}}\int_{-\infty}^0\frac{d\omega}{2\pi}e^{i\omega x}\left[\frac{\theta_R(x)}{\omega-\Sigma(\omega)}
-\frac{\theta_R(-x)}{\omega-\Sigma(\omega)^*}\right],\label{cor1}
\end{align}
where $\theta_R(x)$ is a regularized step function
\begin{align}
&\theta_R(x)=i\int_{-\infty}^\infty \frac{dk}{2\pi}\frac{e^{i(k-\omega)x-a|k|}}{\omega^+-k}\nonumber\\
&=e^{\omega a}\left[\theta(x)+\frac{i\Gamma(0,\omega(ix+a))}{2\pi}\right]-e^{-\omega a}\frac{i\Gamma(0,\omega(ix-a))}{2 \pi},
\end{align} 
with $\omega^+=\omega+i0^+$, and $\Sigma(\omega)$ is the retarded resonant level self-energy
\begin{align}
&\Sigma(\omega)=\frac{J_\perp^2}{4\pi^2 a}\int_{-\infty}^\infty dk\,\frac{e^{-a|k|}}{\omega^+-k},\nonumber\\
&=\frac{J_\perp^2}{4\pi^2 a}\left[e^{a\omega}\Gamma(0,a\omega^+)-e^{-a\omega}\Gamma(0,-a\omega^+)\right].
\end{align}

In general, the integral (\ref{cor1}) has to be evaluated numerically. However, for $|x|\gg a$, the following approximations are valid, which yields an analytical result.
As a function of (negative) $\omega$, the regularized step function $\theta_R(x)$ is suppressed as $e^{\omega a}$ for $x>0$.
As a result, the integrand of (\ref{cor1}) is strongly suppressed for $-\omega\gtrsim 1/a$. 
As a function of $x$, the regularized step function is smoothed at a scale $|x|\lesssim a$.
For $-\omega< 1/a$ and $|x|\gg a$, the regularized step function can be
replaced with the sharp step function $\theta(x)$. For $x\gg a$, the rapidly oscillating
factor $e^{i\omega x}$ then cuts off the integral at $\omega\sim 1/x\ll 1/x$, so that the self-energy $\Sigma(\omega)$, that varies on the scale of $1/a$, can be evaluated at
$\omega=0$
\begin{equation}
\Sigma(0)=-\frac{iJ_\perp^2}{4\pi a}.
\end{equation}
For $|x|\gg a$, one then finds
\begin{align}
\left<d^\dagger \bar \psi(x)\right>
&=\frac{J_\perp}{\sqrt{2\pi a}}\int_{-\infty}^0\frac{d\omega}{2\pi}\frac{e^{i\omega |x|}}{\omega+i\frac{J_\perp^2}{4\pi a}}\nonumber\\
&=-\frac{J_\perp}{\sqrt{2\pi a}}\frac{F\left(-\frac{J_\perp^2|x|}{4\pi a}\right)}{2\pi},
\end{align}
where $F(z)$ is defined in Eq.~(\ref{efl}). We substitute this into
(\ref{toulcloud1}) and also drop the second term in that equation, which is
valid for $|x|\gg a$, to obtain
\begin{equation}
X_0^\parallel(x)=-\frac{\sqrt{2}}{\pi^2}\frac{J_\perp^2}{4\pi a}F\left(-\frac{J_\perp^2|x|}{4\pi a}\right)^2.
\end{equation}
The Kondo length $\xi_\parallel$ is defined as in (\ref{asymptotics}) in the main text, i.e. $\xi_\parallel=\lim_{x\to\infty}x^2X_0^\parallel(x)$.
This yields 
\begin{equation}
\xi_\parallel=\frac{\sqrt{2}}{\pi^2}\frac{4\pi a}{J_\perp^2},
\end{equation}
and a universal scaling function (cf. Eq.~(\ref{ansatz1}))
\begin{equation}
\tilde X^\parallel_0(\alpha=1/2,\tilde x)=-\frac{2}{\pi^4}F\left(-\frac{\sqrt{2}\tilde x}{\pi^2}\right)^2.
\end{equation}

\acknowledgments
This work is based on research supported in part 
by the National Research Foundation of South Africa (Grant Number 90657).


\setcounter{figure}{0}
\setcounter{table}{0}
\setcounter{equation}{0}
\setcounter{section}{1}
\onecolumngrid

\pagebreak

\global\long\def\theequation{S\arabic{equation}}
\global\long\def\thefigure{S\arabic{figure}}

\vspace{1.0cm}
\begin{center}
{\bf \large Supplementary information for
``Universal spatial correlations in the anisotropic Kondo screening cloud:
analytical insights and numerically exact results from a coherent state expansion''}
\vspace{0.5cm}
\end{center}

\section*{Scaling analysis of the anisotropic cloud}
\label{appscalefigs}

\begin{figure}[tbh]
\begin{center}
\includegraphics[width=.4\textwidth]{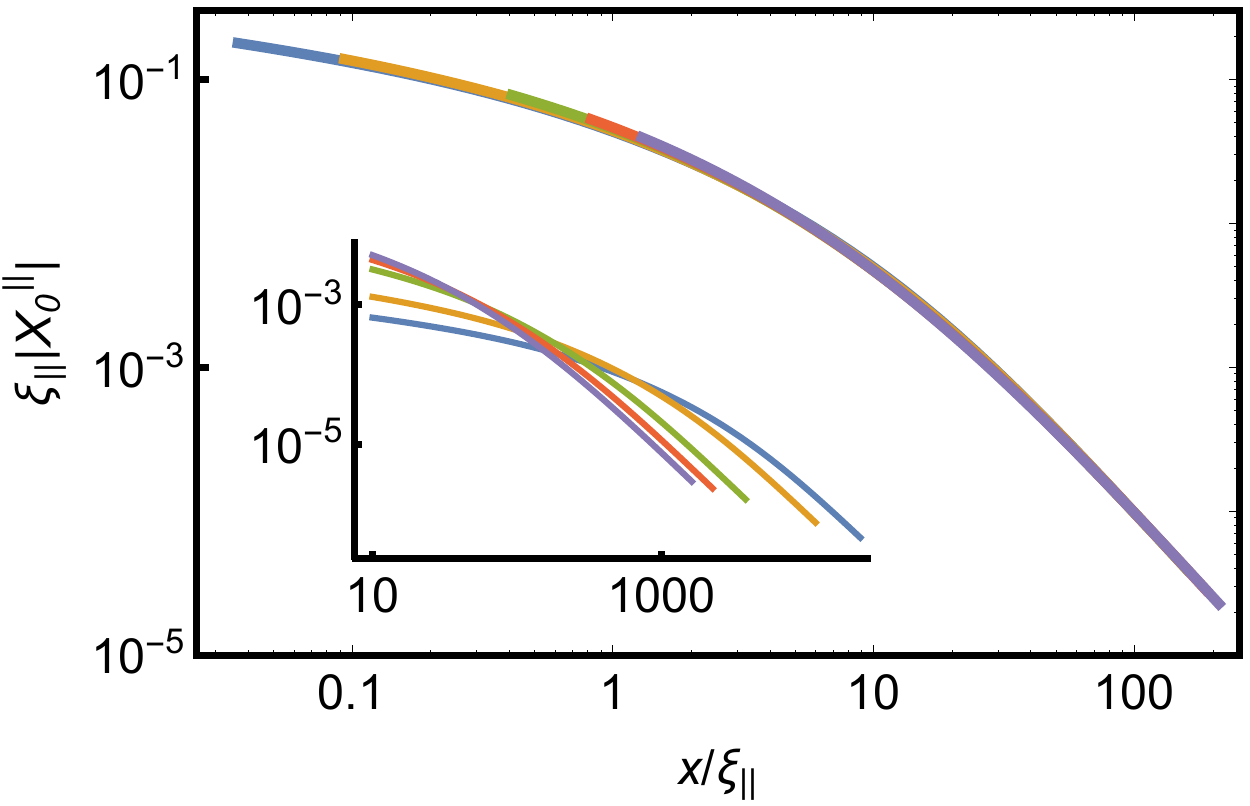}~~
\includegraphics[width=.4\textwidth]{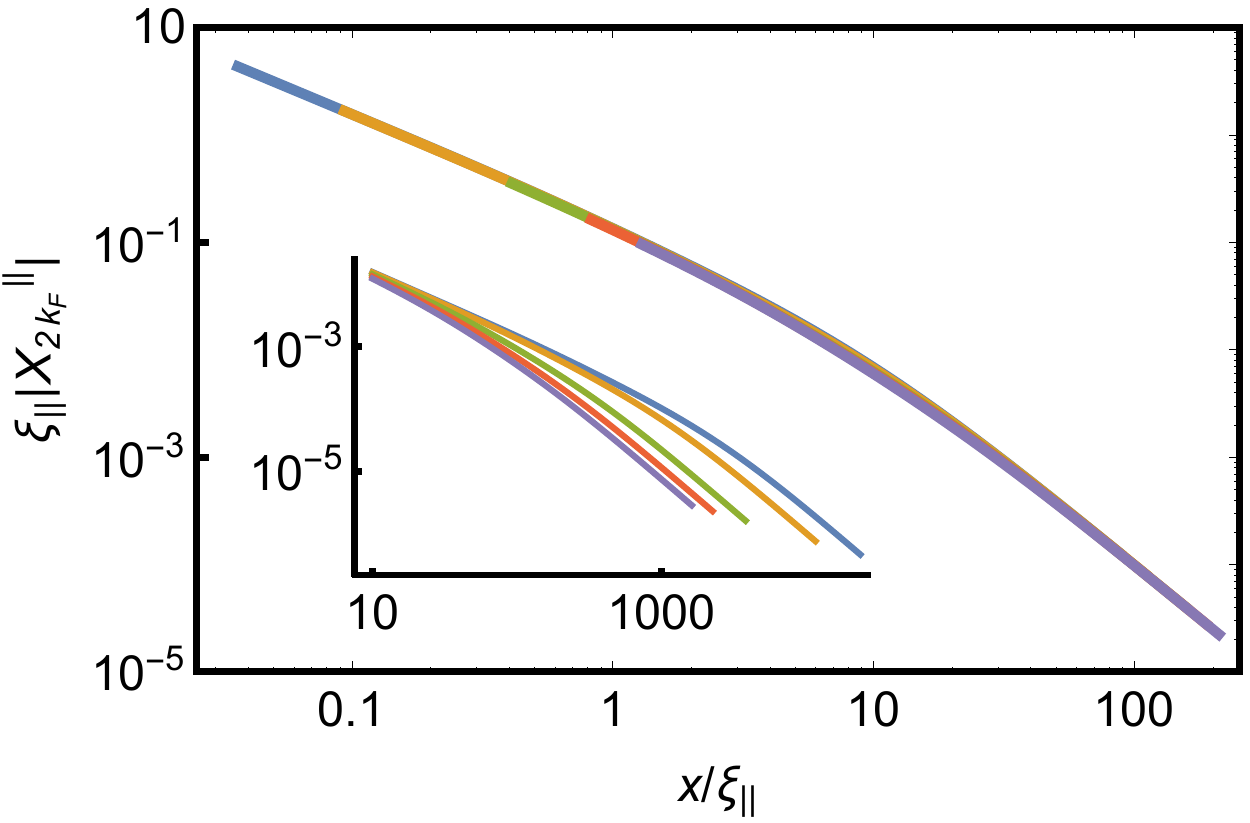}
\includegraphics[width=.4\textwidth]{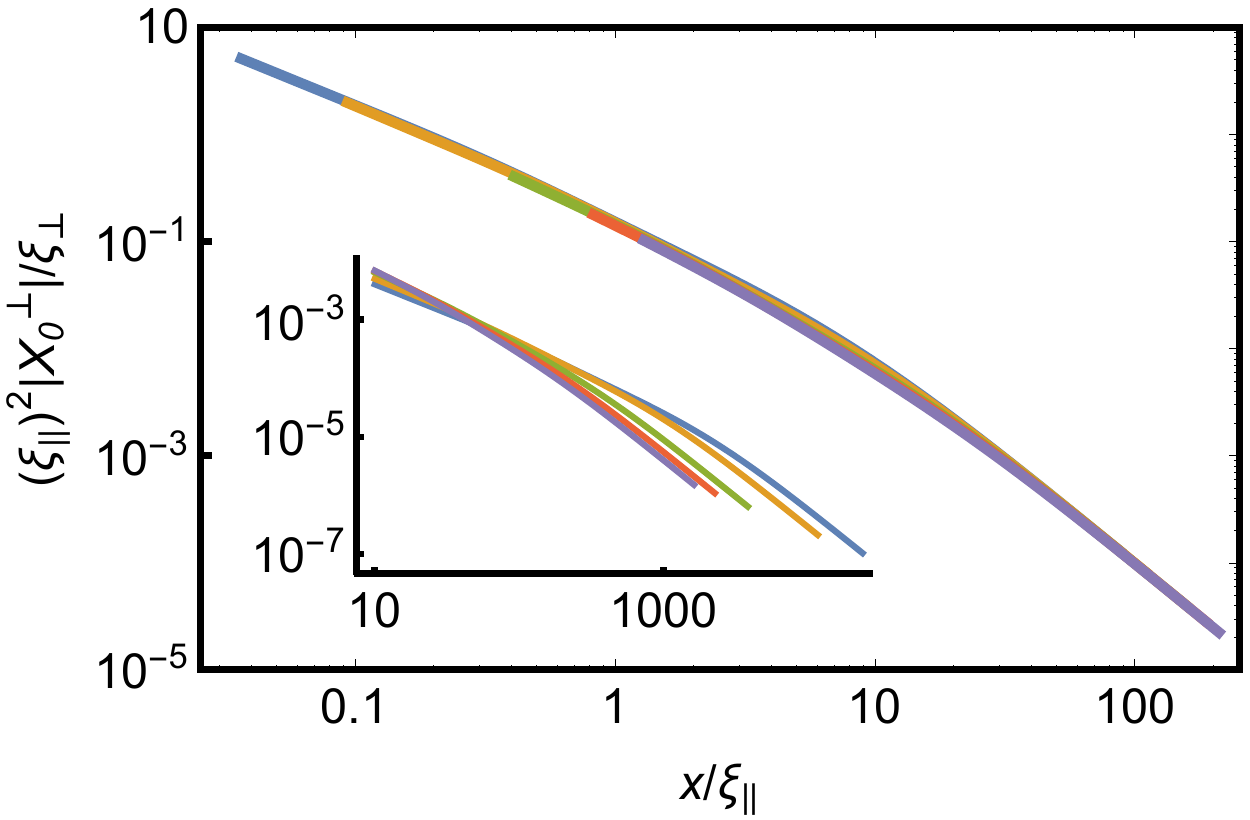}~~
\includegraphics[width=.4\textwidth]{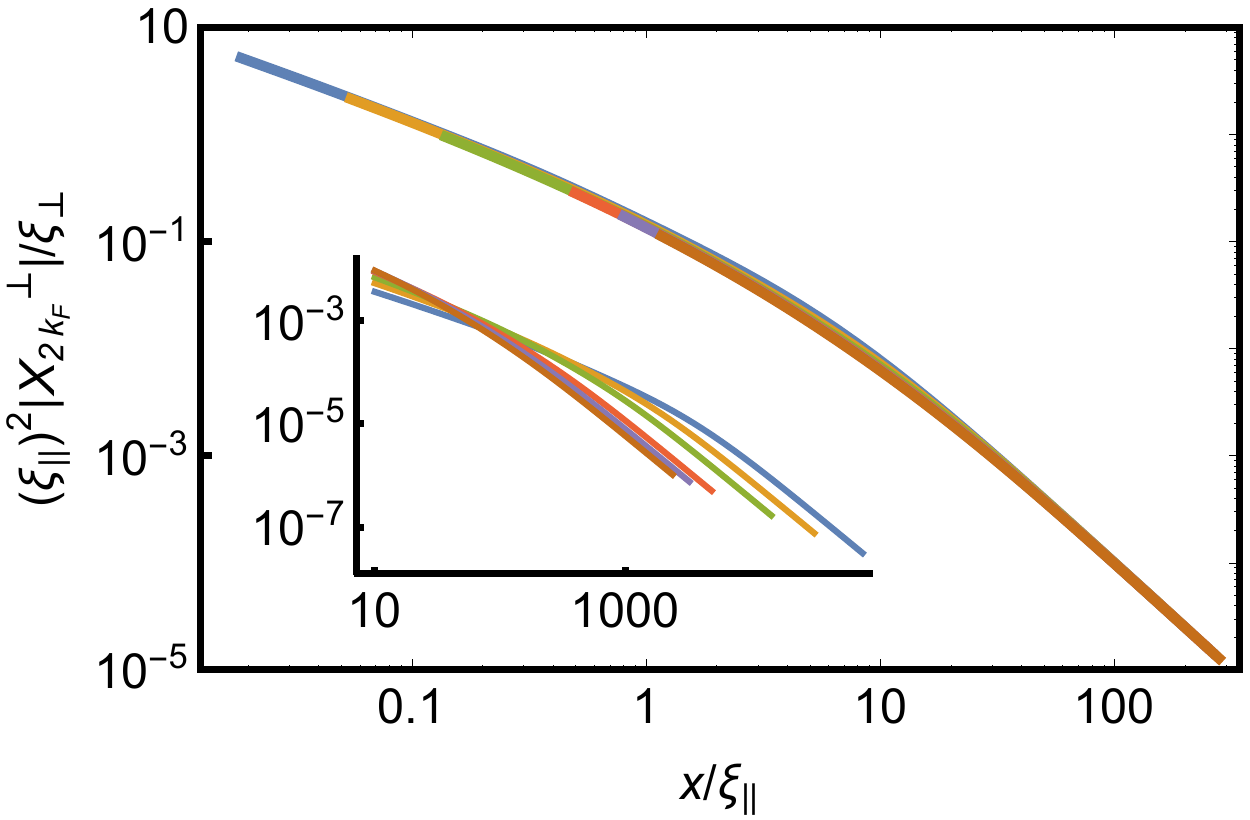}
\caption{Scaling curves for the four components of the Kondo cloud computed
at fixed $\alpha=0.3$ ($J_\parallel=2.84$). Raw data is shown in the insets, with
different curves correspond to five different values of $\Delta$ collected 
in the interval $\Delta\in[0.005/a,0.07/a]$ (i.e. $J_\perp\in
[0.0157,0.220]$). \label{fapp1}}
\end{center}
\end{figure}

\begin{figure}[tbh]
\begin{center}
\includegraphics[width=.4\textwidth]{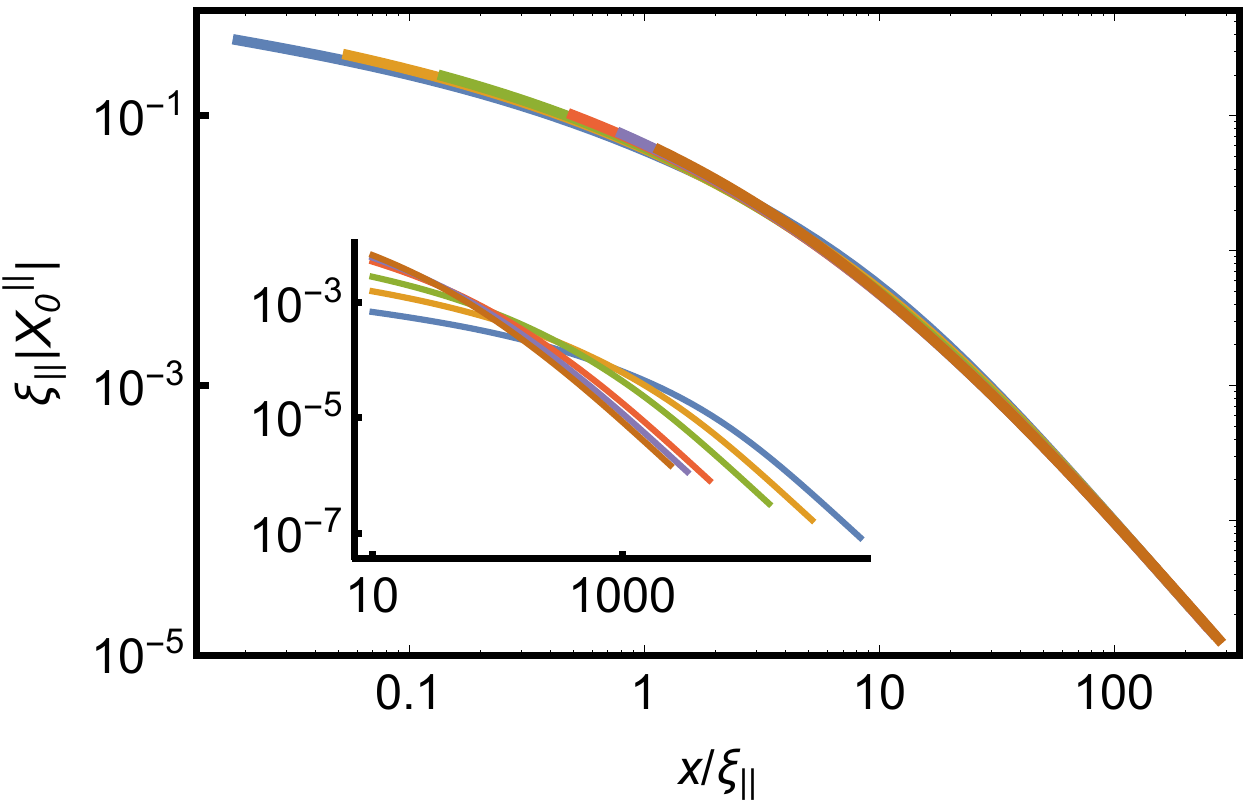}~~
\includegraphics[width=.4\textwidth]{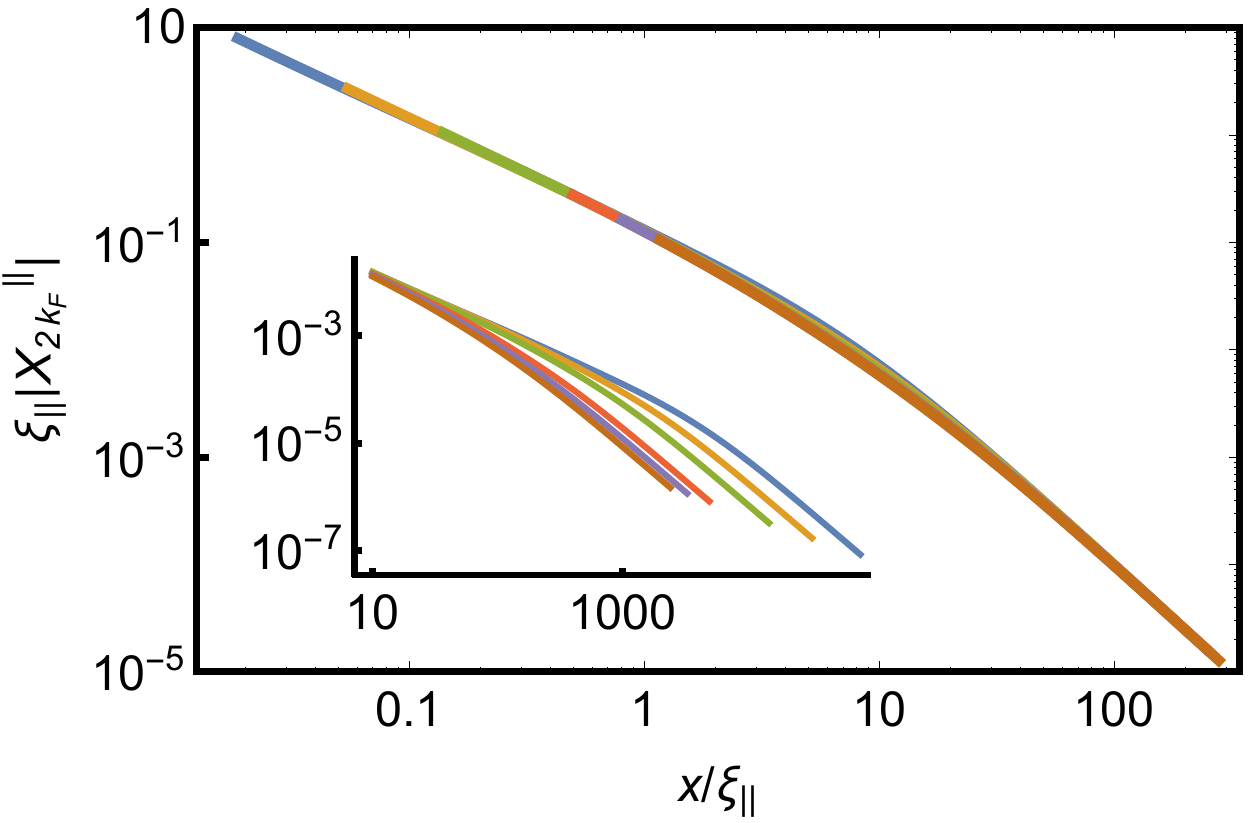}
\includegraphics[width=.4\textwidth]{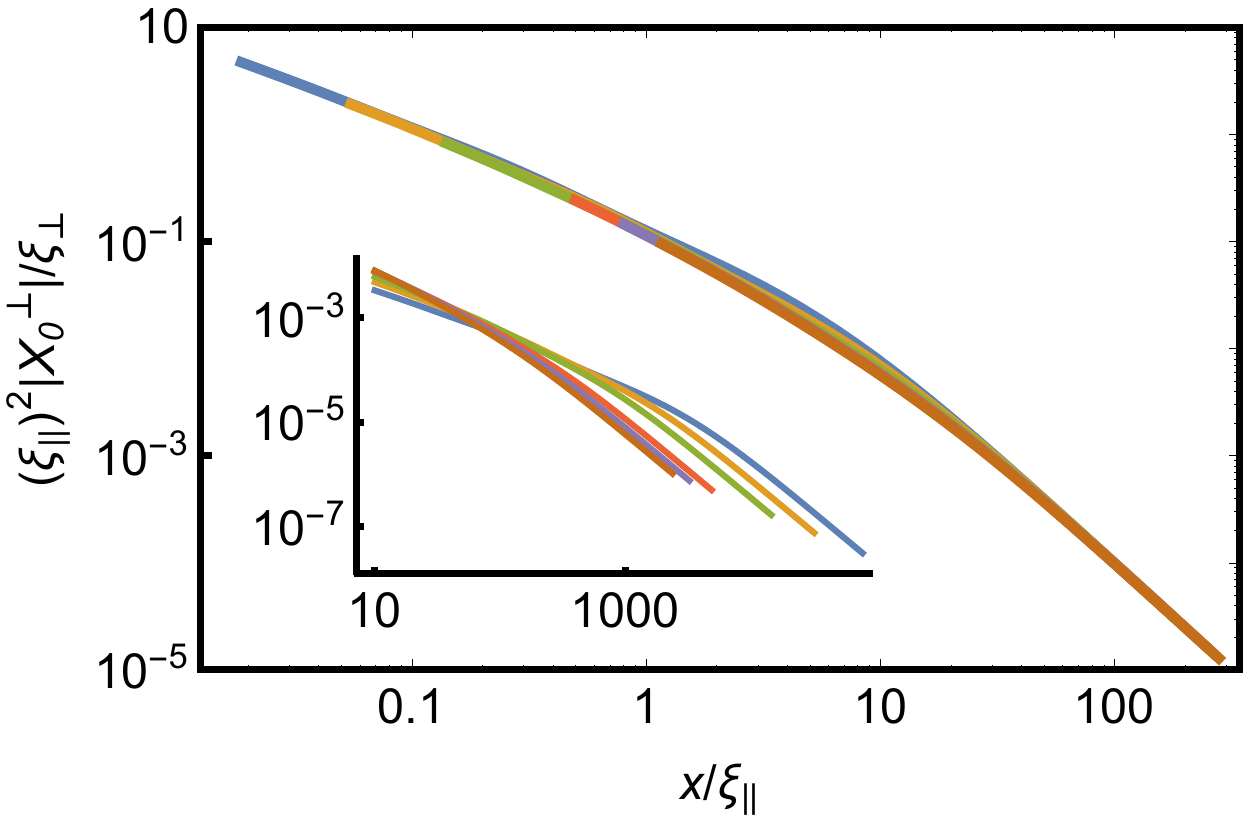}~~
\includegraphics[width=.4\textwidth]{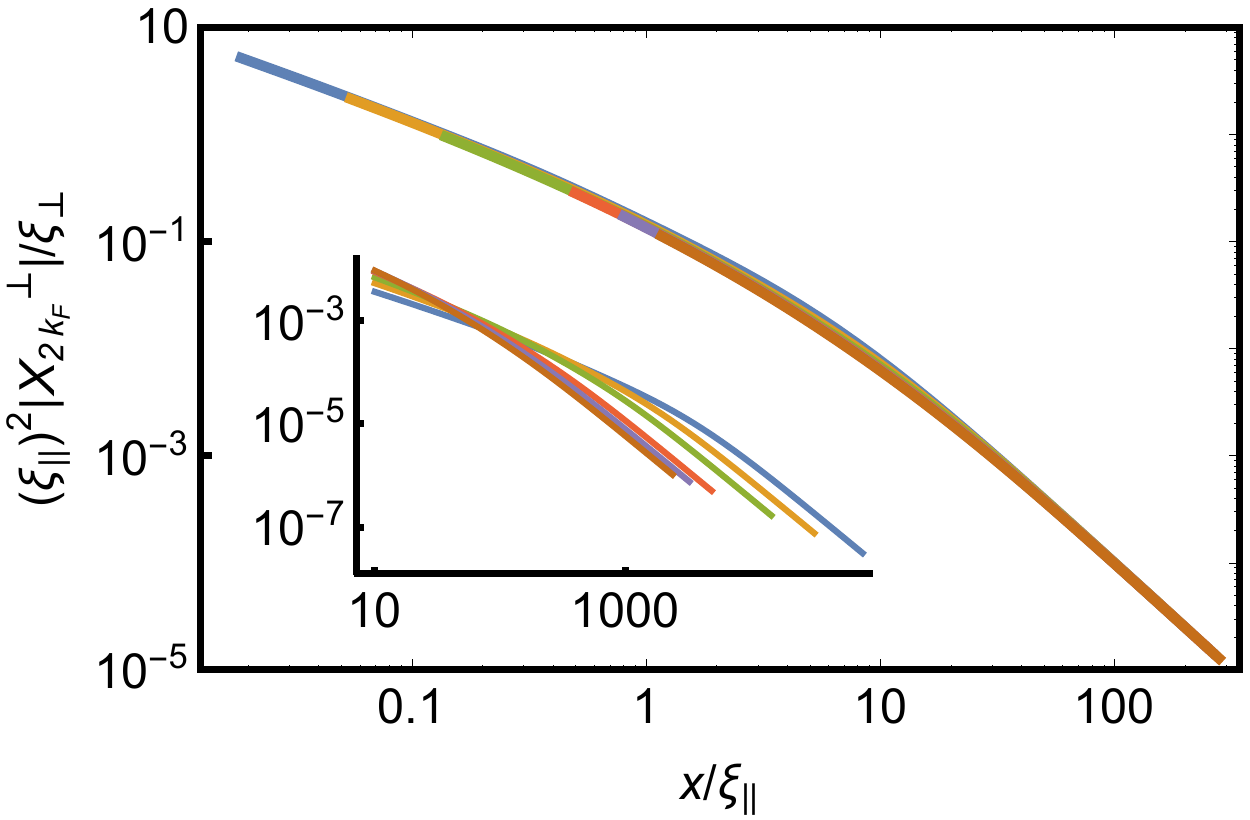}
\caption{Scaling curves for the four components of the Kondo cloud computed
at fixed $\alpha=0.45$ ($J_\parallel=2.07$). Raw data is shown in the insets, with
different curves correspond to six different values of $\Delta$ collected 
in the interval $\Delta\in[0.01/a,0.125/a]$ (i.e.  $J_\perp\in
[0.0314,0.393]$).\label{fapp2}}
\end{center}
\end{figure}

\begin{figure}[tbh]
\begin{center}
\includegraphics[width=.4\textwidth]{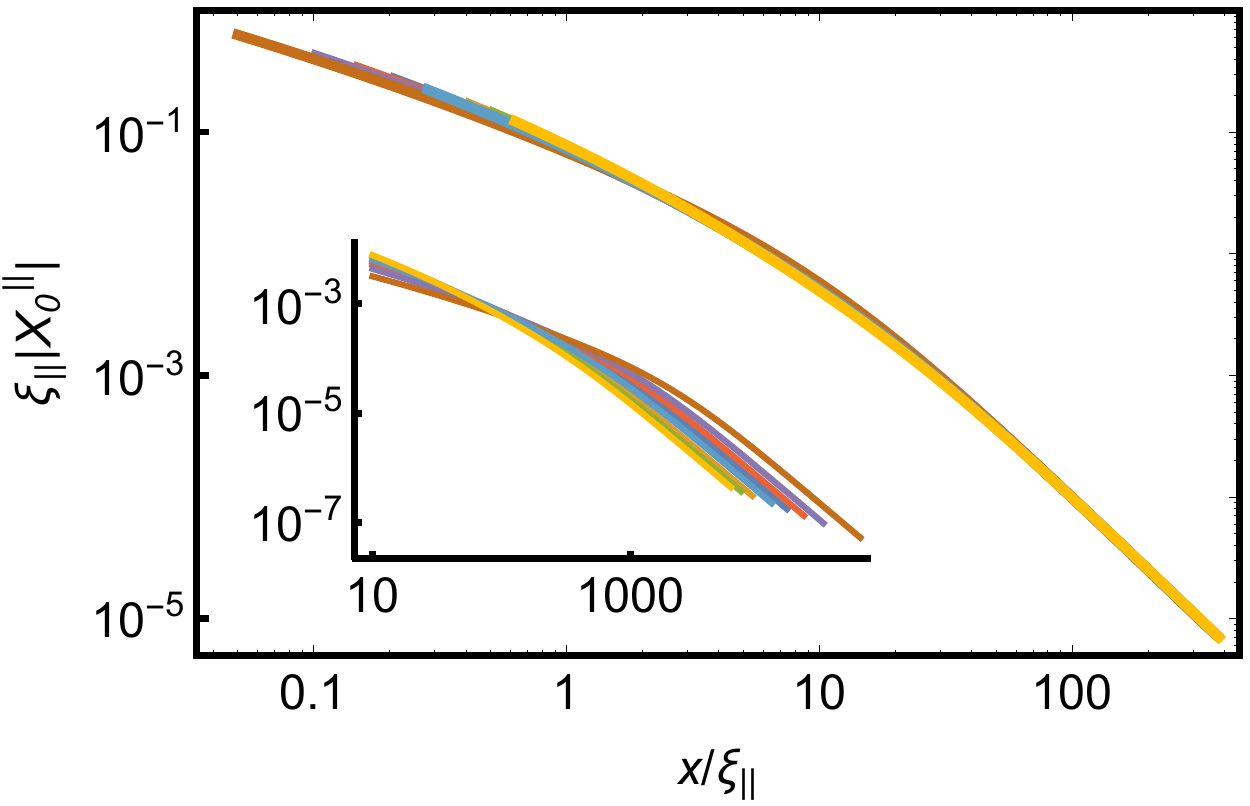}~~
\includegraphics[width=.4\textwidth]{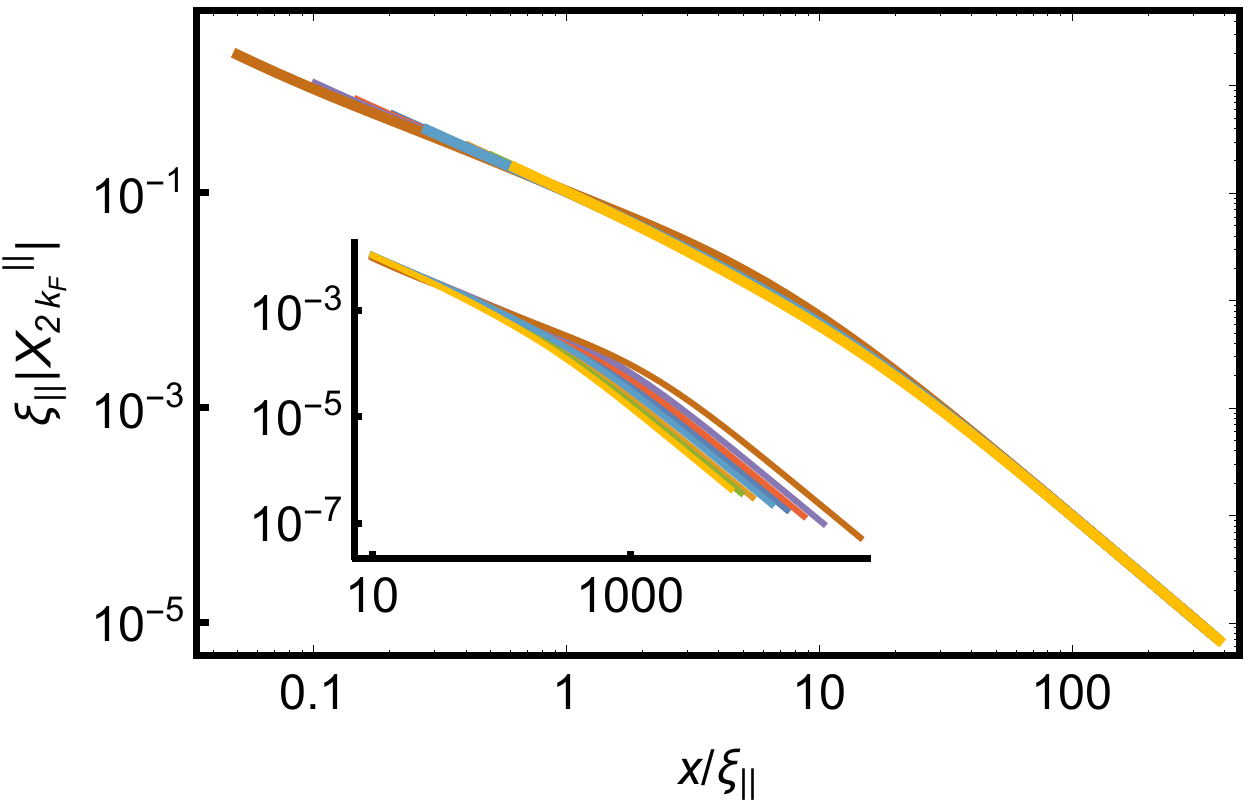}
\includegraphics[width=.4\textwidth]{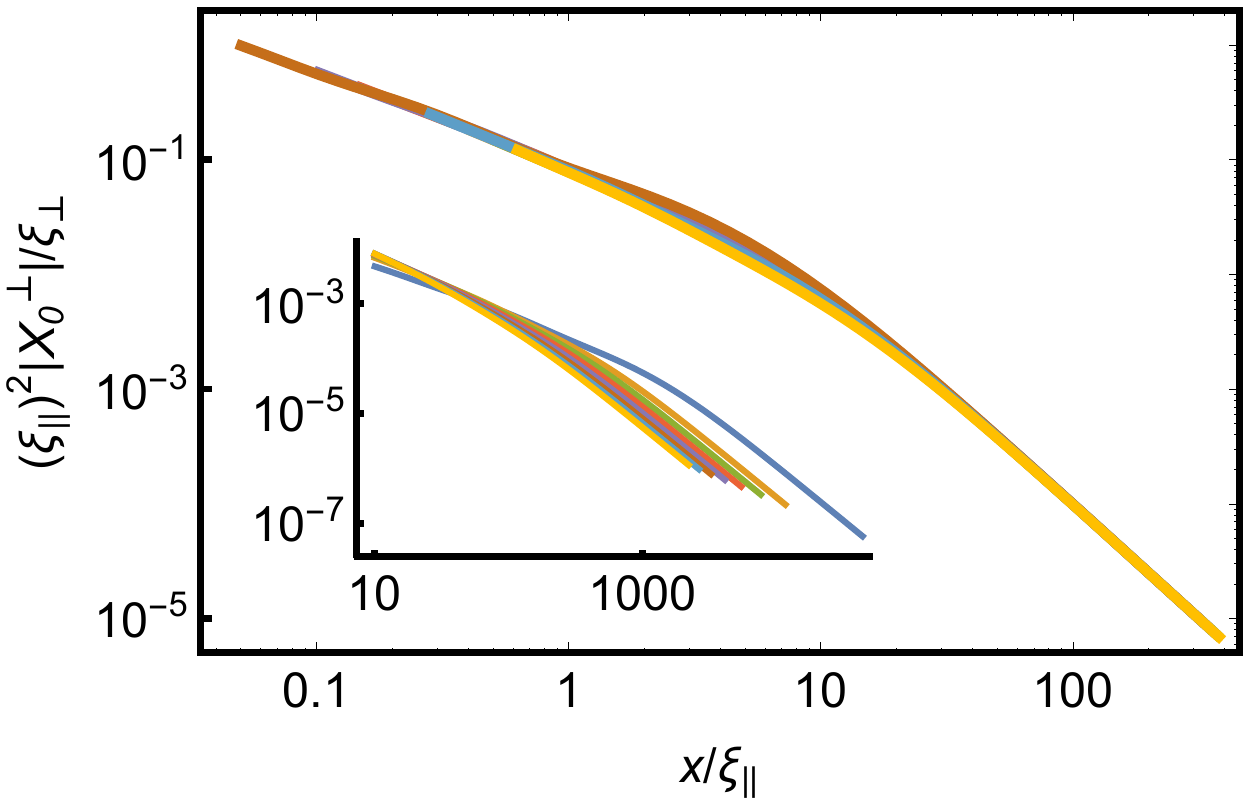}~~
\includegraphics[width=.4\textwidth]{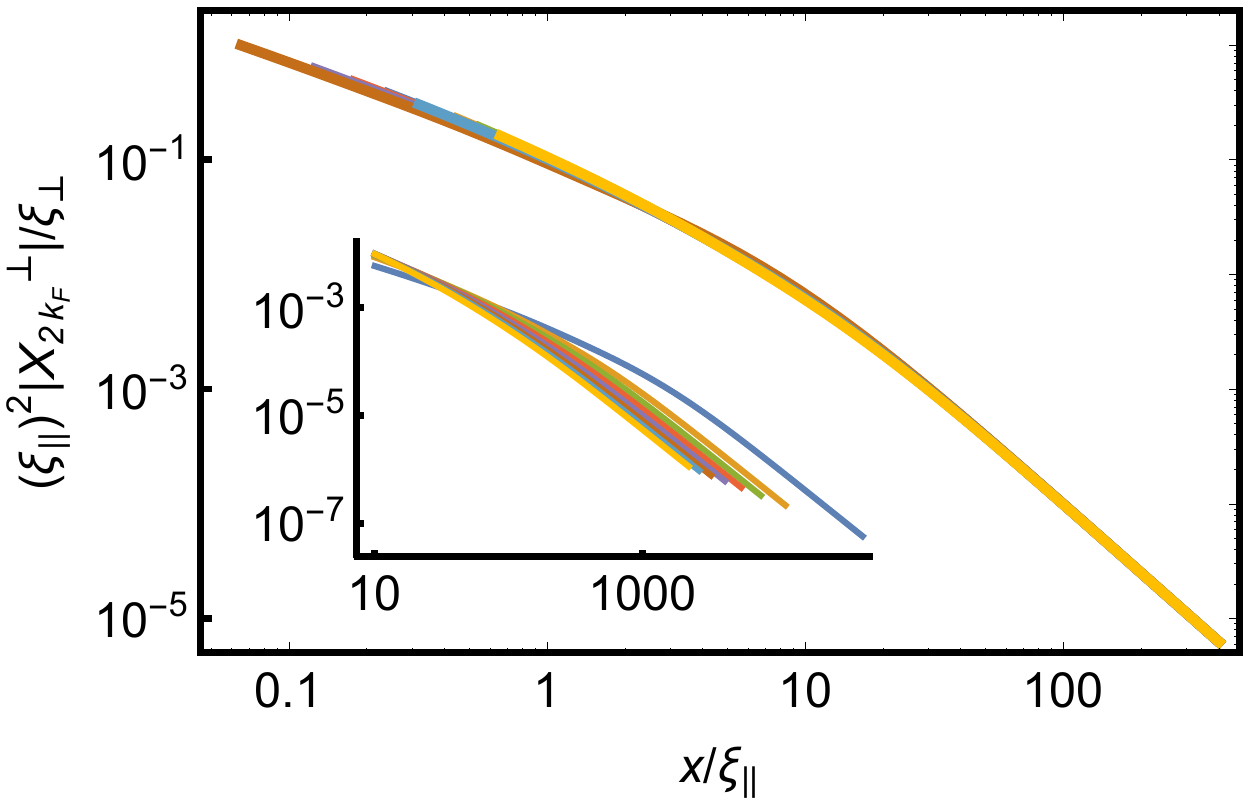}
\caption{Scaling curves for the four component of the Kondo cloud computed
at fixed $\alpha=0.8$ ($J_\parallel=0.663$). Raw data is shown in the insets, with
different curves correspond to eight different values of $\Delta$ collected 
in the interval $\Delta\in[0.12/a,0.28/a]$ (i.e. $J_\perp\in
[0.377,0.880]$).\label{fapp3}}
\end{center}
\end{figure}

In the main text, we showed extensive data at $\alpha=0.6$ (i.e. $J_\parallel=1.42$)
demonstrating that cloud correlation functions calculated at different values of
$\Delta=J_\perp/\pi a$ scaled onto the same universal functions (Figure
\ref{f8}). Here, we show for completeness data that demonstrate scaling at $\alpha=0.3$ (Figure
\ref{fapp1}), $\alpha=0.45$ (Figure \ref{fapp2}), and $\alpha=0.8$ (Figure
\ref{fapp3}). 
 
\end{document}